\documentclass[english,11pt,oneside]{article}

\pdfoutput=1

\usepackage{amsmath}
\usepackage{amsfonts}
\usepackage{amssymb}
\usepackage{graphicx, rotating}
\usepackage{epstopdf}
\usepackage{epsfig}
\usepackage{latexsym}
\usepackage{multirow}
\usepackage{color}
\usepackage[dvipsnames]{xcolor}
\usepackage{slashed}
\usepackage[top=2.8 cm, bottom=2.8 cm, left=3 cm, right=3 cm]{geometry}
\usepackage{amstext} 
\usepackage{array}  
\usepackage[colorlinks]{hyperref}

\hypersetup{citecolor={blue}}

\newcolumntype{L}{>{$}l<{$}} 
\newcommand{\ba}{\begin{eqnarray}}
\newcommand{\ea}{\end{eqnarray}}
\newcommand{\no}{\nonumber}
\newcommand{\be}{\begin{equation}}
\newcommand{\ee}{\end{equation}}
\newcommand{\eq}[1]{Eq.~(\ref{#1})}

\newcommand{\LambdaL}{\Lambda_{\slashed{\rm L}}}

\usepackage{hyperref}
\hypersetup{
colorlinks = true,
 linkcolor  = black,
 citecolor={blue}
}

\begin{document}

\title{\bf The Bearable Compositeness of Leptons}

\author{Michele Frigerio$^a$, Marco Nardecchia$^{b,c}$, Javi Serra$^d$, Luca Vecchi$^e$\\ \\
{\small\emph{$^a$ Laboratoire Charles Coulomb (L2C), University of Montpellier, CNRS, Montpellier, France}}\\
{\small\emph{$^b$ INFN, Sezione di Trieste, Italy}}\\
{\small\emph{$^c$ Theoretical Physics Department, CERN, Geneva, Switzerland}}\\
{\small\emph{$^d$ Physik-Department, Technische Universit\"at M\"unchen, 85748 Garching, Germany}}\\
{\small\emph{$^e$ Theoretical Particle Physics Laboratory, Institute of Physics, EPFL, Lausanne, Switzerland }}\\
}
\date{}

\maketitle
\begin{abstract}
\noindent Partial compositeness as a theory of flavor in the lepton sector is assessed. We begin presenting the first systematic analysis of neutrino mass generation in this context, 
and identifying the distinctive mass textures. We then update the bounds from charged lepton flavor and CP violating observables. We put forward a $U(1)^3\times CP$ symmetry of the composite sector, in order to allow the new physics to be not far above the TeV scale.
This hypothesis effectively suppresses the new contributions to the electron EDM and $\mu\to e\gamma$, by far the most constraining observables, and results in a novel pattern of flavor violation and neutrino masses. 
The CP violation in the elementary-composite mixing is shown to induce a CKM phase of the correct size, as well as order-one phases in the PMNS matrix. We compare with the alternative possibility of introducing multiple scales of compositeness for leptons, that also allow to evade flavor and CP constraints.
Finally, we examine violations of lepton flavor universality in $B$-meson semi-leptonic decays.
The neutral-current anomalies can be accommodated, predicting strong correlations among different lepton flavors, with a few channels close to the experimental sensitivity.
\vspace{5cm}
\begin{flushright}
CERN-TH-2018-160
\end{flushright}
\end{abstract}

\newpage

\begingroup
\color{black} 
\tableofcontents
\endgroup 

\newpage

%%%%%%%%%%%%%%%%%%%%%%%%%%%%%%%%%%%%%%
\section{Introduction} 
%%%%%%%%%%%%%%%%%%%%%%%%%%%%%%%%%%%%%%

The absence of new physics at the LHC puts significant pressure on solutions of the hierarchy problem at the TeV scale. The message, already suggested by precision measurements at LEP, is that the new physics must lie at a high scale $m_* \gtrsim$ few TeV significantly larger than the Higgs mass $m_h$, leaving the hierarchy $m_*^2/m_h^2$ unexplained, i.e.~tuned. Nevertheless, {as long as $m_*$ is stabilized in the TeV ballpark, such a setting}  
offers a far more convincing picture compared to scenarios with no physics beyond the Standard Model (SM) up to the ultimate cutoff of particle physics, the Planck scale.

Still, even this point of view faces an important challenge (or an opportunity, in the absence of direct experimental probes of energies beyond the TeV), when one inspects processes that are very rare within the SM, such as baryon and lepton number violation, {flavor-changing neutral-current} transitions and electric dipole moments. Considering order one couplings between the SM and the new physics one finds that the proton lifetime and neutrino masses put $m_*$ in the $10^{11 - 13}$ TeV range. Flavor and CP violation instead suggest the new mass threshold has to be in excess of $10^{2 - 5}$ TeV.

Taking these numbers at face value, one should not expect new physics in the TeV ballpark. But the truth is that some of these constraints might be evadable if the new physics enjoys approximate global symmetries. The SM itself has global symmetries, appearing as accidents of the leading, dimension-four 
Lagrangian. 
A typical extension of the SM will have its own accidental symmetries, and the couplings between the two sectors might leave some combination of these unbroken. If this is the case then one can easily imagine new physics scenarios in which the most severe bounds, from proton decay and neutrino masses, are completely removed because a generalized baryon and lepton numbers is preserved.

To explain the suppression of flavor and CP violation is more challenging if, as we will do here, one wishes to rely on a framework in which the flavor structure of the SM is motivated by some dynamical mechanism. 
A very appealing flavor paradigm that achieves this goal goes under 
the name of Partial Compositeness (PC). This framework finds a natural implementation within models of Higgs compositeness, where indeed it first emerged~\cite{Kaplan:1991dc,Grossman:1999ra,Gherghetta:2000qt,Huber:2000ie,Agashe:2004cp}, but it is in fact more general and may be realized in other extensions of the SM. The defining assumption of PC is the existence of a new flavorful strongly-coupled sector (with or without global symmetries) that breaks the SM flavor symmetries via a mixing between the SM fermions and ``composite'' fermions of the new sector. Under quite reasonable assumptions this picture can explain the curious pattern of masses and mixings in the SM, and {\emph{predict}} the structure of flavor and CP violation beyond the SM. There is no hierarchy of parameters in the UV description: all hierarchies in the low energy theory derive from {{large renormalization group effects}} within the strong flavorful sector. In this sense, PC does not rely on the presence of a suitable flavor structure in the UV.

Even in the most unstructured realizations of PC (that we will refer to as anarchic scenarios) the bounds of order $m_*>10^{2-5}$ TeV from quark flavor violation are significantly relaxed down to a few tens of TeVs. The most severe constraints on the new physics scale then come from CP violation (in particular the electric dipole moments for the neutron and the electron) and lepton flavor violation (most notably $\mu\to e\gamma$). The main focus of this paper is precisely on these two crucial aspects, and their connection with neutrino masses and mixing. Leptonic observables in PC \cite{Agashe:2006iy,Agashe:2009tu,Csaki:2010aj,KerenZur:2012fr,Redi:2013pga} have not received as much attention as the analogous observables in the quark sector so far, partly because the latter, and in particular the top quark sector, is more likely to be probed by high energy colliders. Still, it is fair to say that leptonic observables appear to be quite problematic for vanilla realizations of PC, and a dedicated study is missing, especially for scenarios that retain the PC explanation of fermion masses and mixings. Furthermore, no systematic investigation of how neutrino masses arise in PC has been presented before, and in fact we will identify a number of interesting features for neutrino phenomenology that went unnoticed.

In section~\ref{sec:PC} we introduce the PC paradigm. Some of the basic elements are discussed in greater detail than in the existing literature. This care is not necessary when analyzing the so-called anarchic versions of PC, where the flavorful sector has no internal structure (no symmetries). However, our derivation turns out to be essential when studying the predictions in non-generic scenarios with additional flavor or CP symmetries. 
{Some general properties of the compositeness framework are collected in appendix~\ref{sec:ass}.}

In section~\ref{NeutrinoSection} we study Majorana neutrino masses within PC. We will prove that the neutrino mass texture in anarchic PC scenarios must belong to one of three classes, which we identify and contrast with neutrino data. We provide explicit examples for each of these classes and discuss their main differences. Some of these models are novel and their predictions could be investigated in detail e.g.~by embedding them in a warped extra dimension.
{In appendix~\ref{sec:textures} we corroborate the generality of our classification of the neutrino textures.}
Dirac neutrino masses are discussed in appendix~\ref{dirac}.

An analysis of the dominant constraints from charged lepton observables is presented in section~\ref{CLFV}. Here we first discuss the bounds in a rather model-independent fashion (our assumptions are clearly stated
and the derivation of the bounds is detailed in appendix \ref{sec:bounds}). Then, we concentrate on anarchic scenarios of PC. We will see that this picture is in tension with bounds on flavor and CP violation unless $m_* \gtrsim 100$ TeV. 
Technical aspects and a more comprehensive list of constraints are collected in appendix \ref{compApp}.
We are thus motivated to consider non-anarchic scenarios.

This is done in section~\ref{sec:supp}, where we identify two main realizations of PC that can simultaneously preserve its defining property (namely the ability to generate fermion mass hierarchies from a UV description with no large or small numbers), and relax the experimental constraints. The first is based on the assumption that the flavorful sector has a $U(1)^3 \times {CP}$ symmetry. In the second non-anarchic realization of PC the flavorful dynamics is characterized by several flavor-dependent mass scales, 
rather than a single one. We will see that in both cases it is possible to keep the new physics scale close to a few TeVs. The structure of neutrino masses is also analyzed in the two non-anarchic scenarios, and we find for instance that
the lightest neutrino mass may be naturally very small.

As a concrete application of our results,  in section~\ref{anoB} we discuss the long-standing anomaly in semi-leptonic, neutral-current $b\to s$ transitions. We identify the PC scenarios that can reproduce the excess while still being compatible with leptonic data. 
This is done by analyzing the correlations between different channels predicted by our flavor scenarios. We also discuss a tentative PC scenario where the bounds on flavor and CP violating transitions in the quark sector are compatible with the size of the anomaly.

We summarize and discuss perspectives in section~\ref{end}.

%%%%%%%%%%%%%%%%%%%%%%%%%%%%%%%%%%%%%%
\section{Flavor hierarchy from flavor anarchy}
\label{sec:PC}
%%%%%%%%%%%%%%%%%%%%%%%%%%%%%%%%%%%%%%

The basic assumption of PC is that the SM flavor symmetry is dominantly broken by a mixing between {``elementary''} fermions $\psi=\ell_L,e_R,q_L,u_R,d_R$, 
and fermionic {``composite''} operators $O^\psi$ of an exotic dynamics. Schematically, we postulate that at some UV cutoff $\Lambda_{\rm UV}$ we have: 
\ba\label{PC1}
\lambda^\psi_{ai}\overline{O}^\psi_a{\psi_i}+{\rm h.c.}~,
\ea
where $\lambda^\psi_{ai}$ is a coupling, $i=1,2,3$ the SM flavor index, $a=1,\cdots,n_a$ a flavor index in the new dynamics. More explicitly, the PC couplings for the lepton sector read $\lambda^\ell_{ai}\overline{O}^\ell_a{(\ell_L)_i}+\lambda^e_{ai}\overline{O}^e_a{(e_R)_i}+{\rm h.c.}$. In order to break the SM flavor symmetry completely and generate masses for all fermions one needs
$n_a\geq3$. In the following we will simplify our formulas taking $n_a=3$, commenting on the generalization to $n_a>3$ when useful.

We will never specify what the fundamental degrees of freedom that describe $O^\psi_a$ are, in an attempt to keep our analysis as general as possible, and simply refer to the sector the operators
belong to as ``composite sector'', ``strong dynamics'', or ``conformal field theory (CFT)'' in the following. The operator language breaks down at scales of order TeV or higher, where the composite sector confines and generates  massive resonances, possibly including a Nambu-Goldstone Higgs doublet $H$.
For simplicity we assume all couplings and masses of the resonances are controlled by two parameters~\cite{Giudice:2007fh}
\ba\label{g*m*}
g_*~,~m_*~,
\ea
where $g_*\in[1,4\pi]$.
Our extra assumptions on the strong sector are collected in appendix~\ref{sec:ass}.

What makes PC attractive is the possibility to dynamically generate a {\emph{flavor hierarchy}} from parameters $\lambda^\psi_{ai}$ that have no particular structure. 
The hierarchy is generated by renormalization group (RG) effects, provided the CFT violates the flavor symmetry associated to the index $a$ in the operators $O^\psi_a$. 
To see this, without loss of generality we can work in a basis in which $O^\psi_a$ have definite scaling dimensions $\Delta^\psi_a\equiv\Delta[O^\psi_a]$. 
The key assumption then reads $\Delta^\psi_a\neq\Delta^\psi_b$ for $a\neq b$.
At scales of the order of the CFT mass gap, $m_*$, the strong dynamics is integrated out, leaving an effective field theory (EFT) containing only the SM fields, see (\ref{NDA}) for details. The flavor-violating couplings, including the SM Yukawas, are therefore controlled by the parameters (\ref{PC1}) renormalized at the scale $m_*$. At leading order in the small CFT perturbations the RG evolution of $\lambda^\psi$ is given by $\mu\frac{d}{d\mu}\lambda^\psi_{ai}=(\Delta^\psi_a-5/2)\lambda^\psi_{ai}+{\cal O}(\lambda^3)$, and its solution for $m_*\ll\Lambda_{\rm UV}$ reads (no sum in $a$)
\ba\label{PC2}
\lambda^\psi_{ai}(m_*)=\varepsilon^\psi_a\lambda^\psi_{ai}~,
\ea
with $\varepsilon^\psi_a=\varepsilon^\psi_a(m_*/\Lambda_{\rm UV})$. The specific expression of $\varepsilon^\psi_a$ depends on whether the operator (\ref{PC1}) is irrelevant (that occurs when $\Delta^\psi_a>5/2$) or relevant ($\Delta^\psi_a<5/2$). In the former case $\varepsilon^\psi_a\simeq (m_*/\Lambda_{\rm UV})^{\Delta^\psi_a-5/2}$ is suppressed. In the latter case the coupling grows at lower scales and may reach a nontrivial IR fixed point.

We want to show now that the hierarchy encoded in $\varepsilon_a^\psi$ translates into a hierarchy of the flavor-violating couplings in the EFT. 
It is convenient to label the operators in such a way that $O_3^\psi$ is more relevant than $O_2^\psi$, that is more relevant than $O_1^\psi$, that is more relevant than all the others (if present), 
in formulas: $\Delta^\psi_1>\Delta^\psi_2 >\Delta^\psi_3$ and $\varepsilon^\psi_{1}<\varepsilon^\psi_{2}<\varepsilon^\psi_{3}$.
Next, we redefine the fundamental (i.e.~elementary) fermions $\psi_i$ via unitary rotations, so as to put $\lambda^\psi_{ai}$ in ``triangular" form,
\ba\label{PC3}
\lambda^\psi_{ai}(m_*)&\equiv& g_*\epsilon_{ai}^\psi~,
\\\no
\epsilon_{ai}^\psi&\equiv& 
\left(
\begin{array}{ccc}
 \epsilon_1^\psi &  &  \\
  & \epsilon_2^\psi &  \\
  &  & \epsilon_3^\psi 
\end{array}
\right)
\left(
\begin{array}{ccc}
 1 & c^\psi_1 & c^\psi_2 \\
 0 & 1 & c^\psi_3 \\
 0 & 0 & 1 \\
\end{array}
\right)
=
\left(
\begin{array}{ccc}
 \epsilon_1^\psi & \epsilon_1^\psi c_1^\psi & \epsilon_1^\psi c_2^\psi \\
 0 & \epsilon_2^\psi & \epsilon_2^\psi c_3^\psi \\
 0 & 0 & \epsilon_3^\psi \\
\end{array}
\right)~,
\ea
where $g_*$ is the strong-sector low-energy coupling, $c^\psi_{1,2,3}$ are unknown complex numbers of order unity, and
\be\label{usual}
\epsilon_i^\psi\equiv \frac{\lambda^\psi_{ii}(m_*)}{g_*}~, \qquad
\epsilon^\psi_{1}<\epsilon^\psi_{2}<\epsilon^\psi_{3}~.
\ee
Note that the $\epsilon^\psi_i$ can be taken real and positive by choosing the phase of $\psi_i$.~\footnote{The generalization to $n_a$ operators is conveniently done by arranging the composite fermions in an $n_a$-dimensional vector $O^\psi_a=(O^\psi_{n_a},O^\psi_{n_a-1},\cdots,O^\psi_5,O^\psi_4 | O^\psi_1,O^\psi_2,O^\psi_3)$, with $\Delta^\psi_{a>3}> \Delta^\psi_1$, that implies $\varepsilon^\psi_{a>3}<\varepsilon^\psi_{1}$. To make $\lambda^\psi$ triangular one defines $\psi'_3\propto\lambda^\psi_{i3}{\psi_i}$ as the combination that couples to the most relevant operator $O_3$. The remaining two states $\psi_{1,2}'$ are orthogonal and can be rotated among each other so that only $\psi'_2$ couples to $O_2$. With this choice $\psi_{1}'$ couples only to $O_1$, $\psi_{2}'$ couples to $O_{1,2}$ and $\psi_3'$ couples to $O_{1,2,3}$. The phases in the diagonal elements can be removed via $U(1)$ rotations of $\psi_i'$. (LV acknowledges K.~Agashe for illustrating this useful basis.) With these conventions (\ref{PC3}) is generalized to
\ba\no
\epsilon_{ai}^\psi\equiv 
\left(
\begin{array}{ccc|ccc}
\epsilon_{n_a}^\psi &  &  &  &  &  \\
 & \cdots &  &  &  &  \\
  &  & \epsilon_4^\psi &  &  &  \\
  \hline
 &  &  & \epsilon_1^\psi &  &  \\
 &  &  &  & \epsilon_2^\psi &  \\
 &  &  &  &  & \epsilon_3^\psi \\
\end{array}
\right)
\left(
\begin{array}{ccc}
c^\psi & c^\psi & c^\psi\\
\cdots & \cdots & \cdots\\
c^\psi & c^\psi & c^\psi\\
1 & c^\psi & c^\psi \\
0 & 1 & c^\psi \\
0 & 0 & 1 \\
\end{array}
\right)~. 
\ea
}
Note that the parameters $\epsilon_i^\psi$ inherit a hierarchy from $\varepsilon_a^\psi$. The quantities $\epsilon_{i}^\psi$ measure the amount of ``compositeness" of the SM fermions at scales of order $m_*$. Without loss of generality, these parameters are real, positive and normalised to one in the limit of a fully composite fermion. It is important to keep in mind that the quantities 
that actually enter the EFT at the scale $m_*$ are a function of a non-diagonal matrix, $\epsilon_{ai}^\psi$ (not just powers of $\epsilon_i^\psi$!). The off-diagonal elements of $\epsilon_{ai}^\psi$ will play an important role when discussing CP violation (see section~\ref{CKMphase}) and non-anarchic models (see section~\ref{sec:supp}).

The hierarchy in $\varepsilon_a^\psi$ finally translates into a hierarchy in the SM Yukawa couplings. Indeed, at leading order in $\lambda^\psi$, the Yukawa couplings of the charged fermions read
 \ba\label{Yukawa}
 y^u_{ij}&=&{g_*}(\epsilon_{ai}^q)^*\epsilon_{bj}^u~c^u_{ab}~,\\\no
 y^d_{ij}&=&{g_*}(\epsilon_{ai}^q)^*\epsilon_{bj}^d~c^d_{ab}~,\\\no
 y^e_{ij}&=&{g_*}(\epsilon_{ai}^\ell)^*\epsilon_{bj}^e~c^e_{ab}~,\\\no
 &=&g_*[(\epsilon^\ell)^\dagger c^e (\epsilon^e)]_{ij}~,
 \ea
where $c^\psi_{ab}$ are model-dependent parameters of order unity arising from the strong dynamics and we used the same naive-dimensional-analysis (NDA) counting described in \eq{NDA}. 
Within the assumption that the order-one coefficients $c$ in (\ref{Yukawa}) are all comparable (see appendix~\ref{sec:ass}), the masses and mixing angles of the charged fermions are fully controlled by the compositeness parameters $\epsilon_i^{q,u,d,\ell,e}$. For example, 
\be\label{ye}
y^e_{ij} \sim 
g_* \left(
\begin{array}{ccc}
\epsilon^\ell_1\epsilon_1^e & \epsilon^\ell_1\epsilon_2^e & \epsilon^\ell_1\epsilon_3^e\\ 
\epsilon^\ell_2\epsilon_1^e & \epsilon^\ell_2\epsilon_2^e & \epsilon^\ell_2\epsilon_3^e\\ 
\epsilon^\ell_3\epsilon_1^e & \epsilon^\ell_3\epsilon_2^e & \epsilon^\ell_3\epsilon_3^e
\end{array}
\right)~.
\ee
Remarkably, the structure (\ref{ye}) and the analogous ones for $y^{u,d}$ are consistent with experimental data on charged fermions \cite{Huber:2000ie, Agashe:2004cp}.
The charged lepton masses are given by 
\be \label{mass-eps}
m_i^e \sim g_* \epsilon^\ell_i \epsilon^e_i \frac{v}{\sqrt{2}},
\ee
where $v=246$ GeV. 
It also follows that $(y^ey^{e\dagger})_{ij} \propto \epsilon_{i}^\ell \epsilon_j^\ell$, $(y^{e\dagger} y^e)_{ij} \propto \epsilon_{i}^e\epsilon_j^e$,
therefore  the unitary matrices that diagonalize them have the structure 
\be \label{rot-eps}
U^\ell_{ij}\sim{\rm min}\left( \frac{\epsilon^\ell_i}{\epsilon^\ell_j},\frac{\epsilon^\ell_j}{\epsilon^\ell_i} \right) 
=\left(\begin{array}{ccc}
1 & \epsilon_1^\ell/\epsilon_2^\ell & \epsilon_1^\ell/\epsilon_3^\ell \\
\epsilon_1^\ell/\epsilon_2^\ell & 1 & \epsilon_2^\ell/\epsilon_3^\ell \\
\epsilon_1^\ell/\epsilon_3^\ell & \epsilon_2^\ell/\epsilon_3^\ell & 1
\end{array}\right)~,
\ee
and analogously for $U^e_{ij}$, with $\epsilon^\ell_i$ replaced by $\epsilon^e_i$.

It is useful to identify which of the $\epsilon^\psi_i$ are independent parameters and what is their range of variation. In the quark sector, the parameters are constrained by the requirement to reproduce the quark masses and CKM mixing angles. As a consequence,
\be
\frac{\epsilon^q_1}{\epsilon^q_2} \simeq \lambda_C \,,\quad\quad\frac{\epsilon^q_2}{\epsilon^q_3} \simeq \lambda_C^2 \,,\quad\quad \frac{\epsilon^q_1}{\epsilon^q_3} \simeq \lambda_C^3 \,,  
\ee
where $\lambda_C\simeq 0.225$ is the Cabibbo angle, which works by virtue of the relation $\theta_{13}\simeq \theta_{12}\theta_{23}$ among the CKM angles. On the other hand, the parameters $\epsilon^{u,d}_i$ can be traded for the quark masses. One thus concludes that the only free parameter is the overall scale of the $\epsilon^q$, that we conventionally choose to be controlled by $\epsilon^q_3$. Its range of variation is
\ba
\epsilon^q_3 \in \left[ \frac{\sqrt{2}m_t}{g_*v},1\right].
\ea
In the lepton sector one can choose $\epsilon^\ell_i$ as free parameters. Imposing that both the left- and right-handed compositeness are bounded by $1$, one finds 
\be
\epsilon^\ell_i \in \left[ \frac{\sqrt{2}m^e_i}{g_*v},1\right],~i=1,2,3,\quad\quad 
\frac{\epsilon^\ell_i}{\epsilon^\ell_j} \in %\left[ \frac{\sqrt{2}m^e_i}{g_*v} , 1\right],~j> i ~.
\left[ \frac{m^e_i}{m^e_j} , 1\right],~j> i ~.
\label{range-ell}\ee
The lepton singlet parameters $\epsilon^e_i$ are not independent, because of \eq{mass-eps}. Indeed,  $\epsilon^e_i$ and $\epsilon^e_i/\epsilon^e_j$ vary in exactly the same ranges as in \eq{range-ell}, but decreasing as the lepton doublet parameters increase. The numerical values for the ratios $\epsilon^\psi_i/\epsilon^\psi_j$ are summarized in table \ref{input}.

%%%%%%%%%%%%%%%%%%%%%
\begin{table}\begin{center}
\begin{tabular}{|l|l|}
\hline
fermion masses (GeV) & $\epsilon^\psi_i/\epsilon^\psi_j$ \\
\hline
$m_e = 0.490 \times 10^{-3}$ & %$2.8 \times 10^{-6}/g_* \le \epsilon^{\ell,e}_1/\epsilon^{\ell,e}_2 \le 1$ \\
$m_e/m_\mu \le \epsilon^{\ell,e}_1/\epsilon^{\ell,e}_2 \le 1$ \\
$m_\mu = 0.103$ & %$2.8 \times 10^{-6}/g_* \le \epsilon^{\ell,e}_1/\epsilon^{\ell,e}_3 \le 1$ \\ 
$m_e/m_\tau \le \epsilon^{\ell,e}_1/\epsilon^{\ell,e}_3 \le 1$ \\ 
$m_\tau = 1.76$ & %$5.9 \times 10^{-4}/g_* \le \epsilon^{\ell,e}_2/\epsilon^{\ell,e}_3 \le 1$\\ 
$m_\mu/m_\tau \le \epsilon^{\ell,e}_2/\epsilon^{\ell,e}_3 \le 1$\\ 
\hline
$m_u =1.2 \times 10^{-3}$ & $\epsilon^q_1/\epsilon^q_2 = \lambda_C = 0.225$\\
$m_c=0.54 $ & $\epsilon^q_2/\epsilon^q_3 = \lambda_C^2= 0.051$\\
$m_t=148$ & $\epsilon^u_1/\epsilon^u_2 = 0.010$ \\
$m_d =2.4 \times 10^{-3}$ & $\epsilon^u_2/\epsilon^u_3 = 0.072$ \\
$m_s = 0.05$ & $\epsilon^d_1/\epsilon^d_2 = 0.21$\\
$m_b = 2.4$ & $\epsilon^d_2/\epsilon^d_3 = 0.41$\\
\hline
\end{tabular}\end{center}
\caption{\small In the left column we show the values of the running fermion masses at $\mu=$ 1 TeV~\cite{Xing:2011aa}. In the right column are the corresponding ratios of PC parameters, assuming for definiteness a strict equality $m^\psi_i = g_* \epsilon^{\psi_L}_i \epsilon^{\psi_R}_i v/\sqrt{2}$ for each SM fermion $\psi$, and  
taking $\lambda_C=0.225$ for the Cabibbo angle.  Of course these ratios are sensitive to variations in the unknown order-one parameters $c$'s.}
\label{input}
\end{table}
%%%%%%%%%%%%%%%%%%%%%

\subsection{CKM phase and CP-invariant strong sectors \label{CKMphase}}
%%%%%%%%%%%%%%%%%%%%%%%%%%%%%%%%%%%%%%

From table \ref{input} we learn that in practice not all $\epsilon^\psi_i/\epsilon^\psi_j$ are hierarchical. This is important since it opens the possibility that CP violation, and in particular the CKM phase, 
comes entirely from the mixing $\lambda$ between elementary and strong sector. 
This will play a crucial role when discussing the bounds from the electron Electric Dipole Moment (EDM).

Because the diagonal elements in (\ref{PC3}) can always be taken to be {\emph{real}}, any complex phase in $\lambda^\psi$ enters a physical observable with a suppression proportional to the hierarchy $\epsilon_i^\psi/\epsilon_j^\psi$. 
In particular, if the strong sector coefficients $c_{ab}^\psi$ are real, then the Yukawa coupling matrices such as (\ref{ye}) and  the associated mixing matrices such as (\ref{rot-eps})  are real, up to ${\cal O}(\epsilon_i^\psi/\epsilon_{j>i}^\psi)$ corrections. The most remarkable consequence is that the Jarlskog invariant in the quark sector schematically scales as (here $\psi=q$ or $u$ or $d$)
\ba\label{J}
J\sim\lambda_C^6\left[{\rm arg}(c_{ab})+{\rm max}\left(\frac{\epsilon_i}{\epsilon_{j>i}}\right){\rm arg}({\lambda}_{ai})+{\cal O}\left(\frac{\epsilon^2_i}{\epsilon^2_{j>i}}\right)\right].
\ea
We thus see that if $\epsilon^\psi_{1}\ll\epsilon^\psi_{2}\ll\epsilon^\psi_{3}$ then a sizable CKM phase in the quark sector can be accommodated only if the strong sector itself has CP-violating parameters, i.e.~if $c^{u,d}_{ab}$ are complex. Equivalently, when the CFT respects CP the CKM phase is potentially suppressed by powers of $\epsilon^\psi_i/\epsilon^\psi_{j>i}$, which may not be phenomenologically acceptable unless some non-generic cancellation takes place. In practice, however, not all $\epsilon^\psi_i$'s are hierarchical, see table \ref{input}. Specifically, $\epsilon^d_2/\epsilon_3^d\simeq0.4$ is large enough to reproduce the observed value $J\sim10^{-5}$ even if the $c_{ab}$ in (\ref{Yukawa}) are all real. This statement, supported by the analytical expression (\ref{J}), has also been confirmed numerically.

In practice, what we find is that the assumption that CP is a good symmetry in the strong sector, which forces all composite couplings such as $c^\psi_{ab}$ to be real, is phenomenologically viable.~\footnote{{While we agree with the reasoning in \cite{Redi:2011zi} that the CKM phase is suppressed for hierarchical $\epsilon_i^\psi$, we do not reach the same conclusion in practice, precisely because phenomenologically not all the mixings are hierarchical.}}
The hypothesis of CP conservation in the strong sector has important consequences on the size of other CP-violating observables, associated to higher-dimensional operators. This is true especially in the lepton sector. The consequences for neutrino CP-violating phases and lepton EDMs will be discussed in the following sections. In the quark sector, however, the assumption of CP invariance of the CFT does not alleviate the stringent bounds on the new physics scale $m_*$, from the neutron EDM, because phases in the down sector are not small ({an efficient mechanism to achieve this suppression is mentioned in section~\ref{sec:u3imp}}).

%%%%%%%%%%%%%%%%%%%%%%%%%%%%%%%%%%%%%%
\section{Neutrino masses} \label{NeutrinoSection}
%%%%%%%%%%%%%%%%%%%%%%%%%%%%%%%%%%%%%%

One of the main goals of this paper is to assess the pattern PC generates within the neutrino sector and compare it with experiments. The most striking difference between the charged and neutral sectors of the SM is that neutrino masses and mixing angles reveal no large flavor hierarchies. The 
Pontecorvo-Maki-Nakagawa-Sakata (PMNS) matrix 
\ba
U_{\rm PMNS}=U^{\ell \dagger} U^\nu
\ea
is a measure of the misalignment between the rotations $U^\nu$ and $U^\ell$ needed to diagonalize the neutrino and charged lepton mass matrices, respectively. Experimentally, $U_{\rm PMNS}$ is found to have all entries of the same order. 
The structure of $U^\ell$ is provided in \eq{rot-eps}, whereas $U^\nu$ depends on the explicit form of $m_\nu$. 
To reproduce data it is then necessary to establish how the neutrinos get a mass. 

In generic scenarios the couplings to the strong sector allow the operator 
\ba\label{Weinberg}
\frac{m^\nu_{ij}}{v^2}\overline{\ell^c_i}\ell_j HH 
\ea
at the scale $m_*$, with $m^\nu\sim (g_*\epsilon^\ell v)^2/m_*$. Once the parameters $\epsilon^\ell$ are chosen so as to reproduce the charged lepton masses via (\ref{Yukawa}), the neutrino mass turns out to be unacceptably large if $m_*\ll10^{15}$ GeV. 
This indicates that viable
scenarios should approximately preserve a total SM lepton number $U(1)_{\ell+e}$,
where we adopt the shorthand notation $\ell =\ell_L$ and $e =e_R$. 

Technically speaking, the presence of a total lepton number implies that the CFT respects a global $U(1)_c$, such that the diagonal combination 
$U(1)_{L}$ of $U(1)_c\times U(1)_{\ell+e}$ is left intact by the mixings (\ref{PC1}). 
If this can be achieved, the mixings can be assigned spurion quantum numbers under $SU(3)_\ell \times U(1)_\ell \times SU(3)_e \times U(1)_e\times U(1)_c$,
\ba\label{spur1}
\lambda^\ell\sim(\bar 3_{-1},1_0,+1)~,\\\no
\lambda^e\sim(1_0,\bar 3_{-1},+1)~,
\ea
while $\lambda^{q,u,d}$ are all singlets. This amounts to assign a zero lepton number to the Higgs doublet. The choice (\ref{spur1}) is not only the simplest but also unique, up to unphysical redefinitions.~\footnote{
To show this, first note that there is an ambiguity in the definition of $U(1)_c$, and in general of $U(1)_\ell\times U(1)_e$. By assumption the strong sector has in fact a global $U(1)_Y\times U(1)_c$ symmetry, so the $U(1)_c$ charges are always defined up to a global $U(1)_Y$ transformation. Consider now the general assignment $c[O^\ell]=c^\ell,~c[O^e]=c^e$ and ask: which conditions should we impose on $c^\ell,c^e$ in order to ensure the existence of $U(1)_{L}$? Once the conventional assignments under $U(1)_\ell\times U(1)_e$ are assumed, the total lepton number exists if and only if one can find a generator $c'=\alpha Y+\beta c$, for some numbers $\alpha,\beta$, so that $c'[O^\ell]=c'[O^e]=+1$. It is easy to see that $\alpha,\beta$ exist unless $c^\ell-c^e/2=0$: the total lepton number exists if $c^\ell-c^e/2\neq0$. Then it is always possible to redefine $U(1)_c$ so that (\ref{spur1}) holds.}

We now turn to the generation of a Majorana mass for neutrinos. The case of a Dirac mass will be discussed along similar lines in appendix \ref{dirac}.
In Majorana neutrino models the $U(1)_{L}$ symmetry must be slightly broken by some CFT perturbation $\Delta{\cal L}$. The latter has coupling $\tilde\lambda$ at some high scale $\LambdaL$ and in general involves CFT as well as elementary degrees of freedom $\psi,\psi'\sim\ell,e,q,u,d$, 
\ba
\label{perturb}
\Delta{\cal L} = \tilde\lambda O~,~\tilde\lambda\psi O~,~\tilde\lambda\psi\psi' O~,~\cdots
\ea
where the dots refer to operators with additional $\psi$'s.
For definiteness, we will consider a single type of spurion $\tilde\lambda$ at a time, generalizations being relatively straightforward.

At scales $m_*$, the neutrino masses are described via the operator (\ref{Weinberg}), with $m^\nu$ proportional to powers of $\tilde \lambda$ renormalized at $m_*$,
\ba
\label{scalingtilde}
\tilde \lambda(m_*) \simeq (m_*/\LambdaL)^{\Delta-4} \tilde \lambda,
\ea
where $\Delta$ is the scaling dimension of the $U(1)_{L}$-breaking CFT perturbation~(\ref{perturb}).
Clearly, operators with large $\Delta$ are ineffective in the physically interesting regime $m_*\ll\LambdaL$. To quantify this statement, if we take $m_*\sim10$ TeV and $\LambdaL \sim 10^{15}$~GeV, one finds that $m^\nu\gtrsim0.05$~eV requires, for $m_\nu \propto \tilde \lambda^n$, $\Delta \lesssim 4+1/n$.
For this reason we will restrict our analysis to deformations (\ref{perturb}) with at most two elementary fermions~\footnote{Concerning deformations with two elementary fields, note that operators of the form $\tilde\lambda\psi^\dagger\bar\sigma^\mu\psi O_\mu$ have $\Delta>6$ in all unitary CFTs~\cite{Mack:1975je}, thus their contribution to neutrino masses is strongly suppressed for $m_*\ll\LambdaL$. 
Besides, in natural, non-supersymmetric realizations of PC one should at most consider single insertions of $\tilde\lambda\psi\psi'O$, i.e.~$m_\nu \propto \tilde\lambda$. The reason is that $O$ here is a scalar, and to avoid the complete singlet $|O|^2$ be strongly relevant --- thus introducing a fine-tuning problem similar to the Higgs mass in the SM --- one should demand $\Delta_O \gtrsim 2$, therefore $\Delta = 3+\Delta_O\gtrsim5$.}, and comment on some interesting subtleties concerning operators with more elementary fields in appendix~\ref{sec:textures}.

The flavor structure of the neutrino mass matrix is determined by the spurion quantum numbers of the coupling $\tilde \lambda$ under $SU(3)_\ell \times U(1)_\ell \times SU(3)_e \times U(1)_e\times U(1)_c$, as well as by the fact that the neutrino mass spurion~(\ref{Weinberg}) transforms as
\ba\label{spurnu}
m^\nu\sim(\bar 6_{-2}, 1_{0}, 0)~.
\ea
Under our (well-motivated) assumption that only operators in the classes $\tilde\lambda O, \tilde\lambda\psi O, \tilde\lambda\psi\psi' O$ can contribute, 
it turns out that the final flavor structure characterizing $m^\nu$ is extremely constrained by PC, with only three possible neutrino mass textures,
\ba
{\rm 2M}:&&m^\nu_{ij} = \epsilon^\ell_{ai}\epsilon^\ell_{bj}\tilde\epsilon_{ab} ~\frac{(g_*v)^2}{m_*}\propto\left(
\begin{matrix}
\epsilon^\ell_1\epsilon^\ell_1 & \epsilon^\ell_1\epsilon^\ell_2 & \epsilon^\ell_1\epsilon^\ell_3\\
\epsilon^\ell_1\epsilon^\ell_2 & \epsilon^\ell_2\epsilon^\ell_2 & \epsilon^\ell_2\epsilon^\ell_3\\
\epsilon^\ell_1\epsilon^\ell_3 & \epsilon^\ell_2\epsilon^\ell_3 & \epsilon^\ell_{3}\epsilon^\ell_3
\end{matrix}\right) \, , \no \\ \no
{\rm 1M}:&&m^\nu_{ij}= \left[\epsilon^\ell_{ai}\tilde\epsilon_{aj}+ \epsilon^\ell_{aj}\tilde\epsilon_{ai} \right]~\frac{(g_*v)^2}{m_*}\propto\left(
\begin{matrix}
\epsilon^\ell_1 & \epsilon^\ell_2 & \epsilon^\ell_3\\
\epsilon^\ell_2 & \epsilon^\ell_2 & \epsilon^\ell_3\\
\epsilon^\ell_3 & \epsilon^\ell_3 & \epsilon^\ell_{3,2}
\end{matrix}\right) \, ,\\
{\rm 0M}:&&m^\nu_{ij}= \tilde\epsilon_{ij}~\frac{(g_*v)^2}{m_*}\propto\left(
\begin{matrix}
1 & 1 & 1\\
1 & 1 & 1\\
1 & 1 & 1
\end{matrix}\right) \, ,
\label{Mclass}\ea
where the dependence on $\tilde \lambda$ is encoded in the $\tilde \epsilon$ parameters, the explicit relation between the two being model-dependent.
The nontrivial aspect of (\ref{Mclass}) is that the assumption of UV-anarchy forces the $\tilde\epsilon$ matrices to be anarchic, in both their SM and CFT flavor indices.
More precisely, $\tilde\epsilon_{ab}$ and $\tilde\epsilon_{ij}$ are symmetric matrices with all entries of the same order. {In class 1M the
matrix $\tilde\epsilon_{a i}$ is anarchical in the index $i$, however there are two possibilities for the index $a$. When $\tilde\epsilon_{a i}$  are of the same order
for each $a$, one finds $m^\nu_{33}\propto \epsilon^\ell_3$. On the other hand, when the third row ($a=3$) is suppressed by a factor $\epsilon^\ell_2/\epsilon^\ell_3$
(this can naturally happen because of gauge invariance, see section \ref{sec:Mmodels}),
then one finds $m^\nu_{33}\propto \epsilon^\ell_2$.}
We emphasize that, given the structure of $U^\ell$ in (\ref{rot-eps}), the parametric dependence of (\ref{Mclass}) on the 
parameters $\epsilon^\ell_i$ is unchanged when rotating to the field basis in which the charged lepton mass matrix is diagonal.

{The powers of $\epsilon^\ell$ in (\ref{Mclass}) are a simple consequence of the SM lepton charges of the spurions, see (\ref{spurnu}) and (\ref{spur1}).} The claim that the $\tilde\epsilon$ are anarchic, irrespective of the $U(1)_{L}$-violating deformation considered, is proven in appendix~\ref{sec:textures}, where the connection with $\tilde \lambda$ is explicitly worked out. The overall size of $\tilde\epsilon$ should be chosen to match the observed neutrino mass scale.
As anticipated above, the origin of such a small scale is model-dependent, and it will be further discussed in~\ref{sec:Mmodels}.

The beauty of our flavor story is that, thanks to (\ref{Mclass}), neutrino data may provide useful information on the parameters $\epsilon^\ell$. 
Specifically, in models of class 2M one finds $U_\nu\sim U_\ell$, as given in (\ref{rot-eps}), so the observed large neutrino mixing immediately tells us that $\epsilon^\ell_i\sim\epsilon^\ell_j$ is necessary. It follows that the charged-lepton mass hierarchy must be mostly encoded in the singlet spurions, $\epsilon^{e}_i$. More precisely, if we assume order-one coefficients really close to one, in class 2M neutrino oscillation data are found to imply
\cite{Frigerio:2002rd,Frigerio:2002fb}
\ba
\frac{\epsilon^\ell_2}{\epsilon^\ell_3} \sim 1 ~,\quad 0.2 \lesssim \frac{\epsilon^\ell_1}{\epsilon^\ell_{2,3}} \le 1 ~~~~~~~~~~~~({\rm for~class~2M})~.
\label{ratiosNEW}
\ea
This is a very strong restriction on the full range for $\epsilon^\ell_i$, that is allowed by charged lepton masses alone (see table~\ref{input}). 
A normal ordering of the neutrino spectrum is preferred when $\epsilon_1^\ell$ is a few times smaller than the others, while for all the $\epsilon^\ell_i$ close to each other
both quasi-degeneracy and inverted ordering can be realized, depending on the order-one coefficients.
Note that the smallness of $\Delta m^2_{12}/\Delta m^2_{23} \simeq 0.03$ requires some amount of tuning among order one parameters.
Besides, 2-3 mixing close to maximal does not require an exact degeneracy of $\epsilon_2^\ell$ and $\epsilon_3^\ell$, since when these two parameters are close, 
a large 2-3 mixing is also present from the charged lepton Yukawa.

In the scenario 1M the structure of $U_\nu$ has a different dependence on $\epsilon^\ell_i$ with respect to $U_\ell$. Irrespectively of the ambiguity in the 33 entry of $m^\nu$, see (\ref{Mclass}), when  $\epsilon^\ell_i$ are hierarchical the mass matrix has two eigenvalues of the same order, and one hierarchically smaller, 
and it predicts a mixing matrix with angles of order unity. Superficially, this looks like a viable option. Unfortunately, a careful analysis reveals that one diagonal entry of $U_{\rm PMNS}$ is always small if $\epsilon^\ell_{1,2}\ll\epsilon_3^\ell$, so data cannot be reproduced in this regime. 
A realistic model needs \cite{Frigerio:2002rd,Frigerio:2002fb}, 
\ba\label{nohp}
\epsilon^\ell_1 \lesssim \epsilon^\ell_2 \sim \epsilon^\ell_3~~~~~~~~~~~~~({\rm for~class~1M}).
\ea
A very small $\epsilon^\ell_1$ is allowed by the smallness of $m_e/m_{\mu,\tau}$, and it is compatible with normal hierarchy of the neutrino mass spectrum,
as long as the $12$ and $13$ entries of $m^\nu$ are smaller by a factor of a few with respect to the $22$, $23$ and $33$ entries.
It is interesting that, in this scenario, the main physical effect of a small $\epsilon^\ell_1$ is a suppression of neutrinoless $2\beta$ decay, 
that is controlled by the $11$-entry of $m^\nu$. 
The quantitative implications of a vanishing $m^\nu_{11}$ are discussed e.g.~in \cite{Merle:2006du}.

Finally, scenario 0M is automatically compatible with the observed large neutrino mixing angles, because $\tilde\epsilon_{ij}$ is anarchic. This scenario implies
\ba\label{nop}
\epsilon^\ell_i~~~{\rm unconstrained}~~~~~~~~~~~~~({\rm for~class~0M}),
\ea
since neutrino masses and mixing are independent from $\epsilon^\ell$. This shows that PC can be naturally compatible with anarchy in the neutrino sector. However, it also implies no testable correlation 
with flavor-violating processes in the charged lepton sector.

A comment is in order on the leptonic CP-violating phases. In the basis where $\epsilon^\ell_{ai}$ is given by (\ref{PC3}), the lepton-number violating spurions $\tilde\epsilon_{ab}$, $\tilde\epsilon_{ai}$, $\tilde\epsilon_{ij}$
are generic complex matrices, and order-one phases are expected. If the strong sector preserves $CP$, in some cases $\tilde\epsilon$ can be taken real. For example, when $\tilde\epsilon_{ab}\sim c_{ab} \tilde\lambda$  
(or when $\tilde\epsilon_{ai}\sim c_a \tilde\lambda_i$),  the CP-invariance of the 
strong sector implies that $c_{ab}$ ($c_a$) are real, and the overall phase of $\tilde\lambda$ ($\tilde\lambda_i$) can be rotated away. 
In these special cases (see section \ref{sec:Mmodels} for explicit realisations) one can check that CP-violating effects in scenarios 2M and 1M are suppressed by the ratios $\epsilon^\ell_i/\epsilon^\ell_{j>i}$, similarly to the case of the CKM phase discussed in section \ref{CKMphase}.
However, since neutrino mixing requires $\epsilon^\ell_2\sim\epsilon^\ell_3$, at least one complex coefficient from $\epsilon^\ell_{ai}$ induces order-one CP-violating phases in $m^\nu$.
We conclude that, barring cancellations, the anarchic PC scenario implies a large CP-violating phase in neutrino oscillations, 
that is slightly favoured in current global fits \cite{globalfits}, as well as large Majorana-type phases, that in principle can be probed in lepton-number violating processes
such as neutrinoless $2\beta$ decay.

%%%%%%%%%%%%%%%%%%%%%%%%%%%%%%%%%%%%%%
\subsection{{Lepton-number breaking sources}}
\label{sec:Mmodels}
%%%%%%%%%%%%%%%%%%%%%%%%%%%%%%%%%%%%%%

Let us end this analysis with a more detailed discussion of the most interesting $U(1)_L$-violating operators:
\ba\label{list1}
\Delta{\cal L}&=&\tilde\lambda_{\tilde a}O_{\tilde a, c=2}~~~~~~~~~~~~~~~~~~~~~~~~~~~~~~~~~\in~{\rm class~2M,}\no\\\no
&&\tilde\lambda_{ij\tilde a}\ell^i\ell^j O^{(3)}_{\tilde a, c=0}~,~\tilde\lambda_{i\tilde a}\ell_iO_{\tilde a, c=0}~~~~~~~~~\in~{\rm class~0M,}\\
&&\tilde\lambda_{i\tilde a}\ell_iO_{\tilde a, c=1}~,~\tilde\lambda_{ij\tilde a}\ell^i\ell^j O^{(0)}_{\tilde a, c=0}~~~~~~~~~\in~{\rm class~1M},
\ea 
where we explicitly show the $U(1)_c$ charge of the composite operator. Two of these operators previously appeared in the PC literature. 

The first is $\Delta{\cal L}=\tilde\lambda_{\tilde a}O_{\tilde a, c=2}$, {that induces $\tilde\epsilon_{ab} g_* = c_{ab}\tilde\lambda_{\tilde a^*} (m_*/\LambdaL)^{\Delta_O-4}$, where ${\tilde a^*}$ labels the most relevant of the
operators $O_{\tilde a}$ with scaling dimension $\Delta_O$, and $c_{ab}={\cal O}(1)$.}
This deformation of the CFT may naturally emerge e.g.~from heavy singlet neutrinos, $O_{c=2}= (O^N)^2$, as discussed in~\cite{Agashe:2015izu}. Singlet neutrinos for Dirac neutrino masses are discussed in appendix \ref{dirac}. 

The second operator, $\tilde\lambda_{ij\tilde a}\ell^i\ell^j O^{(3)}_{\tilde a}$ with the superscript $^{(3)}$ indicating the composite operator is an $SU(2)_L$ triplet, was first proposed by Keren-Zur et al.~\cite{KerenZur:2012fr}. In this scenario $\tilde\epsilon_{ij} g_* \simeq \tilde\lambda_{ij\tilde a^*}(m_*/\LambdaL)^{\Delta_O-1}$, where $O^{(3)}_{\tilde a^*}$ is the most relevant of the scalar composites $O^{(3)}_{\tilde a}$ and $\Delta_O$ is its scaling dimension.

The operator $\tilde\lambda_{i\tilde a}\ell^i O_{\tilde a, c=0}$ gives 
$\tilde\epsilon_{ij}g_*^2 \simeq \tilde\lambda_{i\tilde a}\tilde\lambda_{j\tilde b}c_{\tilde a\tilde b}(m_*/\LambdaL)^{2(\Delta_O-5/2)}$ with $c_{\tilde a\tilde b} = \mathcal{O}(1)$. If a single operator $O_{\tilde a^*}$ dominates, the neutrino mass matrix would be rank one,
therefore a viable model must have at least two operators of comparable scaling dimension. This is an interesting alternative to the proposal of~\cite{KerenZur:2012fr}, because the scaling dimension of $\Delta{\cal L}$ can naturally be close to four, for $\Delta_O \approx 5/2$, 
which allows for an arbitrarily large $\LambdaL$. In contrast, in the case of $\tilde\lambda_{ij\tilde a}\ell^i\ell^j O^{(3)}_{\tilde a}$ a not too large $\LambdaL/m_*$ is necessary to generate a realistic $m^\nu$, given the naturalness bound $\Delta_O \gtrsim 2$. 

The two operators in class 1M are associated to the two inequivalent predictions shown in (\ref{Mclass}). The first, $\tilde\lambda_{i\tilde a}\ell^i O_{\tilde a, c=1}$, gives $\tilde\epsilon_{ai}g_* \simeq 
\tilde\lambda_{i\tilde a}c_{a\tilde a}(m_*/\LambdaL)^{\Delta_O-2.5}$ with $c_{a \tilde a} = \mathcal{O}(1)$. This implies that the entries $\tilde\epsilon_{ai}$ are all of the same order, resulting in $(m_\nu)_{3i} \propto \epsilon^\ell_3$ for all values of $i=1,2,3$. 
Note that, since $m^\nu$ requires symmetrization of the flavor indices, two nonvanishing masses are generated even if a single operator $O_{\tilde a^*}$ dominates (the sum of two rank-one matrices has generically rank two).

On the other hand, gauge invariance implies antisymmetry in the flavor indices $ij$ in $\tilde\lambda_{ij\tilde a}\ell^i\ell^j O^{(0)}_{\tilde a, c=0}$, where $O^{(0)}$ is an $SU(2)_L$ singlet. In this scenario one finds
$\tilde\epsilon_{ai} \simeq c_{a\tilde b c} \tilde\lambda_{ij\tilde b}\epsilon_{cj}^{\ell  *} \times (m_*/\LambdaL)^{\Delta_O-1} g_*/(16 \pi^2)$, with $c_{a \tilde b c} = \mathcal{O}(1)$,
where the sum over $j$ corresponds to integrate over a loop of the elementary fermion $\ell_j$. 
One can then verify that the antisymmetry of $\tilde\lambda$ implies $\tilde\epsilon_{a1,a2}\propto \epsilon^\ell_3$ while $\tilde\epsilon_{a3}\propto \epsilon^\ell_2$, 
thus in (\ref{Mclass}) one has $m^\nu_{33} \propto \epsilon^\ell_2$.
 This latter scenario may be interpreted as a generalization of the Zee model \cite{Zee:1980ai}. 
 
Note that, in general, there is a crucial difference between weakly-coupled radiative neutrino models, and composite scenarios where $m^\nu$ arises from loops
of elementary fermions $\psi$: the assumption of 
composite flavor violation (that is to say, anarchy in the indices $a,\tilde a, \dots$), that lies at the heart of PC, tends to screen the potential hierarchies induced by $\psi$-loops. The only remaining hierarchies are determined by symmetry considerations alone, 
independently of how many loops are required. For example, in the case of $O^{(0)}$ discussed above, the fact that $\tilde\lambda_{ij} \sim 3$ under $SU(3)_\ell$ leads to a hierarchy in the values of $\tilde\epsilon_{ai}$.
{Other radiative neutrino models may arise from $U(1)_L$-breaking sources that involve elementary fields $\psi\ne \ell$. In this case $\tilde\lambda$ is contracted with the appropriate $\epsilon^\psi$, corresponding to a $\psi$-loop, and one reduces again to one of the textures in (\ref{Mclass}).}

%%%%%%%%%%%%%%%%%%%%%%%%%%%%%%%%%%%%%%
\section{Charged-lepton flavor and CP violation}\label{CLFV}
%%%%%%%%%%%%%%%%%%%%%%%%%%%%%%%%%%%%%%

In this section we derive constraints on charged-lepton PC parameters from the experimental upper bounds on flavor and CP-violating observables. 
The most relevant observables are collected in table~\ref{boundsPDG}.~\footnote{The list does not include rare scattering like $\sigma(e^+e^-\to e^\pm \tau^\mp)$ 
and $\sigma(e^+e^-\to \mu^\pm \tau^\mp)$. These have been constrained at LEP and must be $10^{-6}$ times smaller than $\sigma(ee\to\mu\mu)$. We find that 
the resulting bounds are weaker than those derived below.}

%%%%%%%%%%%%%%%%%%%%%%%%%%%%%%%%%%%%%%
\begin{table}[p]
\begin{center}
\resizebox{.5\hsize}{!}{
\begin{tabular}{c|c} 
\hline
Observable & Upper bound on Br ($90\%$ CL)\\
\hline
$\mu\to e\gamma$ & ~~~~~~$4.2\times10^{-13}$~\cite{TheMEG:2016wtm}\\
$\mu^-\to e^+e^-e^-$ & $1.0\times10^{-12}$\\
$\mu^- Au \to e^- Au$ & $7.0\times 10^{-13}$\\
\hline
$\tau\to e\gamma$ & $3.3\times10^{-8}$\\
$\tau\to \mu\gamma$ & $4.4\times10^{-8}$\\
$\tau^-\to e^+e^-e^-$ & $2.7\times10^{-8}$\\
$\tau^-\to \mu^+\mu^-\mu^-$ & $2.1\times10^{-8}$\\
$\tau^-\to \mu^+\mu^-e^-$ & $2.7\times10^{-8}$\\
$\tau^-\to e^+e^-\mu^-$ & $1.8\times10^{-8}$\\
$\tau^-\to e^+\mu^-\mu^-$ & $1.7\times10^{-8}$\\
$\tau^-\to \mu^+e^-e^-$ & $1.5\times10^{-8}$\\
\hline
Observable & Upper bound\\
\hline
$|d_e|$ & $1.1\times10^{-29}~e\,{\rm cm}$ ($90\%$ CL)~\cite{Andreev:2018ayy}\\
$|d_\mu|$ & $1.9\times10^{-19}~e\,{\rm cm}$ ($95\%$ CL)~\cite{Bennett:2008dy}\\
$|d_\tau|$ & $\sim1\times10^{-17}~e\,{\rm cm}$ ($95\%$ CL)\\
\hline
$\Delta a_e$ & $-1.05(0.82)\times10^{-12}$\\
$\Delta a_\mu$ & 
$2.68(0.63)_{exp}(0.43)_{th}\times10^{-9}$\\
$\Delta a_\tau$ & $[-0.052,0.013]$ ($95\%$ CL)\\
\hline
\end{tabular}}
\end{center}
\caption{\small Current upper bounds on the most relevant lepton flavor-violating observables, as well as on the lepton electric and magnetic dipole moments. Bounds are all 
taken from the Particle Data Group~\cite{PDG}, unless stated otherwise. We defined $\Delta a=a_{\rm exp}-a_{\rm SM}$. A $\sim$ in the EDM of the $\tau$ lepton emphasizes this quantity suffers from large experimental uncertainties (see~\cite{PDG} for precise values). 
\label{boundsPDG}}
\end{table}
%%%%%%%%%%%%%%%%%%%%%
%
%%%%%%%%%%%%%%%%%%%%%
\begin{table}[p]
\begin{center}
\begin{tabular}{l|l} 
Effective operator  & Wilson coefficient  \\
\hline
$Q_{eW}^{ij}=
\left( \bar{\ell}_L^i \sigma^{\mu \nu}  e^j_{R} \right) \sigma^I H W^I_{\mu \nu}$ 
& $\frac{C^{eW}_{ij}}{\Lambda^2}=\frac{1}{16 \pi^2} \frac{g_{*}^3}{m_{*}^2} \epsilon^{\ell}_i \epsilon^{e}_j g \, c^{eW}_{ij}
=\frac{1}{16 \pi^2} \frac{g_{*}^2}{m^2_{*}} \frac{\epsilon^{\ell}_i}{\epsilon^{\ell}_j}\frac{\sqrt{2}m^e_j}{v} g \, c^{eW}_{ij}$ \\
$Q_{eB}^{ij}=
\left( \bar{\ell}^i_{L} \sigma^{\mu \nu} e^j_{R} \right)  H B_{\mu \nu}$ 
& $\frac{C^{eB}_{ij}}{\Lambda^2}=\frac{1}{16 \pi^2} \frac{g_{*}^3}{m_{*}^2} \epsilon^{\ell}_i \epsilon^{e}_j g' \, c^{eB}_{ij} 
=\frac{1}{16 \pi^2} \frac{g_{*}^2}{m^2_{*}} \frac{\epsilon^{\ell}_i}{\epsilon^{\ell}_j}\frac{\sqrt{2}m^e_j}{v} g' \, c^{eB}_{ij}$ \\
\hline
$Q_{eH}^{ij}
=\left(H^{\dagger} H\right) \left(\overline{\ell}^i_L  e^j_R H\right)$ 
& $\frac{C^{eH}_{ij}}{\Lambda^2}= \frac{g^3_*}{m_*^2} \epsilon^{\ell}_i \epsilon^{e}_j \, c^{eH}_{ij} 
=\frac{g_*^2}{m_*^2} \frac{\epsilon^{\ell}_i}{\epsilon^{\ell}_j} \frac{\sqrt{2}m^e_j}{v}\, c^{eH}_{ij}$ \\
\hline
$Q^{(1)ij}_{H \ell}
=\left( H^{\dagger} i \overset{\leftrightarrow}{D}_{\mu} H \right) \left(\overline{\ell}^i_L \gamma^{\mu}\ell^j_L\right)$ 
& $\frac{C^{H \ell(1)}_{ij}}{\Lambda^2}=\frac{g^2_*}{m_*^2} \epsilon^{\ell}_i \epsilon^{\ell}_j \, c^{H \ell(1)}_{ij}$ \\
$Q^{(3)ij}_{H \ell} 
=\left( H^{\dagger} \sigma^I i\overset{\leftrightarrow}{D}_{\mu}  H \right) \left(\overline{\ell}^i_L \sigma^I \gamma^{\mu}\ell^j_L\right)$ 
& $\frac{C^{H \ell(3)}_{ij}}{\Lambda^2}=\frac{g^2_*}{m_*^2} \epsilon^{\ell}_i \epsilon^{\ell}_j \, c^{H \ell(3)}_{ij}$ \\
$Q_{H e}^{ij}
=\left( H^{\dagger} i \overset{\leftrightarrow}{D}_{\mu}  H \right) \left(\overline{e}^i_R  \gamma^{\mu} e^j_R\right)$ 
& $\frac{C^{H e}_{ij}}{\Lambda^2}=\frac{g^2_*}{m_*^2} \epsilon^{e}_i \epsilon^{e}_j \, c^{H e}_{ij} 
=\frac{1}{m_*^2} \frac{2m^e_i m^e_j}{v^2} \frac{1}{\epsilon^{\ell}_i \epsilon^{\ell}_j} \, c^{H e}_{ij}$ \\
\hline
$Q_{\ell \ell}^{ijmn}
=\left(\overline{\ell}^i_L \gamma_{\mu}\ell^j_L\right) \left(\overline{\ell}^m_L \gamma^{\mu}\ell^n_L\right)$ 
& $\frac{C^{\ell \ell}_{ijmn}}{\Lambda^2}=\frac{g^2_*}{m_*^2} \epsilon^{\ell}_i \epsilon^{\ell}_j \epsilon^{\ell}_m \epsilon^{\ell}_n \, c^{\ell \ell}_{ijmn}$ \\
$Q_{\ell e}^{ijmn}
=\left(\overline{\ell}^i_L \gamma_{\mu}\ell^j_L\right) \left(\overline{e}^m_R \gamma^{\mu} e^n_R\right)$ 
& $\frac{C^{\ell e}_{ijmn}}{\Lambda^2}=\frac{g^2_*}{m_*^2} \epsilon^{\ell}_i \epsilon^{\ell}_j \epsilon^{e}_m \epsilon^{e}_n \, c^{\ell e}_{ijmn} 
=\frac{1}{m_*^2}\frac{2m^e_m m^e_n}{v^2} \frac{\epsilon^{\ell}_i \epsilon^{\ell}_j}{\epsilon^{\ell}_m \epsilon^{\ell}_n}\, c^{\ell e}_{ijmn}$ \\
$Q_{e e}^{ijmn}
=\left(\overline{e}^i_R \gamma_{\mu}e^j_R\right) \left(\overline{e}^m_R \gamma^{\mu} e^n_R\right)$ 
& $\frac{C^{e e}_{ijmn}}{\Lambda^2}=\frac{g^2_*}{m_*^2} \epsilon^{e}_i \epsilon^{e}_j \epsilon^{e}_m \epsilon^{e}_n \, c^{e e}_{ijmn}
=\frac{1}{g_*^2m_*^2} \frac{4m^e_i m^e_j m^e_m m^e_n}{v^4\epsilon^{\ell}_i \epsilon^{\ell}_j \epsilon^{\ell}_m \epsilon^{\ell}_n}\, c^{e e}_{ijmn}$ \\
\hline
\end{tabular}
\end{center}
\caption{\small Dimension-six operators involving leptons and no quarks, and the NDA estimate of their Wilson coefficients, in the anarchic PC scenario. In the last equalities we used \eq{mass-eps} to eliminate the parameters $\epsilon^e_i$ in favour of $\epsilon^\ell_i$.}
\label{ops}
\end{table}
%%%%%%%%%%%%%%%%%%%%%

We first translate the bounds into constraints on the coefficients of the leptonic dimension-six operators, listed in table \ref{ops}, as defined in the Warsaw basis \cite{Grzadkowski:2010es}. 
In subsequent sections we will confront these bounds first with an anarchic composite sector, characterised by the two parameters $g_*$ and $m_*$ and no flavor nor CP symmetries, 
and then discuss how to relax them via symmetries or dynamical separation of mass scales.

Our constraints are summarized in tables~\ref{Ferruccio5} and~\ref{Ferruccio6}. They are derived ignoring the (small) RG effects from the matching scale down to the experimental scale, and assuming that a single coefficient dominates the rate under consideration. See appendix~\ref{sec:bounds} for details. Our bounds agree, when overlap exists, with e.g.~\cite{Pruna:2014asa,Feruglio:2015gka,Feruglio:2015yua,Crivellin:2017rmk,Calibbi:2017uvl}. The strongest bounds on the dipole operators $Q_{eW,eB}$ come from $\mu\to e\gamma$ 
and the electric dipole moment (EDM) of the electron, $d_e$. On the other hand, $Q_{He,H\ell}$ are dominantly constrained by their tree-level contribution to the exotic lepton decays in table \ref{boundsPDG}. 
Finally, $Q_{\ell e,\ell\ell,ee,eH}$ contribute at tree-level to the latter processes, and at loop-level to $l_i\to l_j\gamma$. Loop-induced contributions to 
the dipole transitions are not negligible under our hypothesis of single-coupling dominance  (see also the discussion in appendix~\ref{sec:bounds}). 

%%%%%%%%%%%%%%%%%%%%%
\begin{table}[p]
\begin{center}
\resizebox{0.75\hsize}{!}{
\begin{tabular}{c|c|c} 
\rule{0pt}{1.2em}%
 & Upper bound on $|C|$ for $\Lambda=1$ TeV & Observable \\
%\hline
\hline
\hline
$C^{e\gamma}_{12,21}$ & $2.1\times10^{-10}$ & $\mu\to e\gamma$\\
$C^{e\gamma}_{13,31}$ & $2.4\times10^{-6}$ & $\tau\to e\gamma$\\
$C^{e\gamma}_{23,32}$ & $2.7\times10^{-6}$ & $\tau\to \mu\gamma$\\
${\rm Im}\, C^{e\gamma}_{11},~ {\rm Re}\, C^{e\gamma}_{11}$ & $ 4.8 \times 10^{-13},~ 2.4\times10^{-6}$ & $d_e, \Delta a_e$\\
${\rm Im}\, C^{e\gamma}_{22},~ {\rm Re}\, C^{e\gamma}_{22}$ & $ 8.4 \times10^{-3},~ 1.8\times10^{-5}$ & $d_\mu, \Delta a_\mu$\\
${\rm Im}\, C^{e\gamma}_{33},~ {\rm Re}\, C^{e\gamma}_{33}$ & $ 4.4\times10^{-1},~ 3.2 $ & $d_\tau, \Delta a_\tau$\\
\hline
$C^{eH}_{12,21}$ & $3.5\times10^{-5}$ & $\mu\to e\gamma$~(2-loop)\\
$C^{eH}_{13,31}$ & $3.0\times10^{-1}$ & $\tau\to e\gamma$~(1- and 2-loop)\\
$C^{eH}_{23,32}$ & $3.4\times10^{-1}$ & $\tau\to \mu\gamma$~(1- and 2-loop)\\
${\rm Im}\, C^{eH}_{11},~ {\rm Re}\, C^{eH}_{11}$ & $ 8.2\times10^{-8},~ 8.4\times10^{-2}$ & $d_e, \Delta a_e$~(2-loop)\\
\hline
$C^{He}_{12}$ & $4.9 (39)\times10^{-6}$ & $\mu\, Au \to e\, Au \ (\mu\to eee)$\\
$C^{He}_{13}$ & $1.5 (1.8)\times10^{-2}$ & $\tau\to eee \ (\tau\to e\mu^+\mu^-)$\\ 
$C^{He}_{23}$ & $1.3 (1.5) \times10^{-2}$ & $\tau\to \mu\mu\mu\ (\tau\to \mu e^+e^-)$\\
$C^{H\ell(1,3)}_{12}$ & $4.9 (37)\times10^{-6}$ & $\mu\, Au \to e\, Au \ (\mu\to eee)$\\
$C^{H\ell(1,3)}_{13}$ & $1.4 (1.8)\times10^{-2}$ & $\tau\to eee (\tau\to e\mu^+\mu^-)$\\
$C^{H\ell(1,3)}_{23}$ & $1.3 (1.5)\times10^{-2}$ & $\tau\to \mu\mu\mu \ (\tau\to \mu e^+e^-)$
\end{tabular}}
\end{center}
\caption{\small Most relevant constraints on the Wilson coefficients of two-lepton operators. 
These values (when there is overlap) agree pretty well with \cite{Feruglio:2015yua}, except for the 1/2-loop effects (see section \ref{QeH}), because we use the latest experimental constraints on radiative decays.
\label{Ferruccio5}}
\end{table}
%%%%%%%%%%%%%%%%%%%%%

%%%%%%%%%%%%%%%%%%%%%
\begin{table}[p]
\begin{center}
\resizebox{.75\hsize}{!}{
\begin{tabular}{c|c|c} 
\rule{0pt}{1.2em}%
 & Upper bound on $|C|$ for $\Lambda=1$ TeV & Observable \\
%\hline
\hline
\hline
$C^{\ell\ell,ee}_{2111}$ & $2.3\times10^{-5}$ & $\mu\to eee$\\
$C^{\ell\ell,ee}_{3111}$ & $9.1\times10^{-3}$ & $\tau\to eee$\\
$C^{\ell\ell,ee}_{3222}$ & $8.0\times10^{-3}$ & $\tau\to \mu\mu\mu$\\
$C^{\ell\ell,ee}_{2321}$ & $7.2\times10^{-3}$ & $\tau^-\to e^+\mu^-\mu^-$\\
$C^{\ell\ell,ee}_{1312}$ & $6.8\times10^{-3}$ & $\tau^-\to \mu^+e^-e^-$\\
$C^{\ell\ell}_{1322,1223},C^{ee}_{1322}$ & $1.3\times10^{-2}$ & $\tau\to e\mu^+\mu^-$\\
$C^{\ell\ell}_{2311,2113},C^{ee}_{2311}$ & $1.0\times10^{-2}$ & $\tau\to \mu e^+e^-$\\
$C^{\ell\ell}_{2212,3312,3213},C^{ee}_{2212,3312}$ & $\sim6\times10^{-2}$ & $\mu\to e\gamma$ (2-loop)\\
\hline
$C^{\ell e}_{2111,1121}$ & $3.3\times10^{-5}$ & $\mu\to eee$\\
{$C^{\ell e}_{1311,1113}$} & $1.3\times10^{-2}$ & $\tau\to eee$\\
{$C^{\ell e}_{2322,2223}$} & $1.1\times10^{-2}$ & $\tau\to \mu\mu\mu$\\ 
$C^{\ell e}_{2321,2123}$ & $1.0\times10^{-2}$ & $\tau^-\to e^+\mu^-\mu^-$\\
$C^{\ell e}_{1312,1213}$ & $9.6\times10^{-3}$ & $\tau^-\to \mu^+e^-e^-$\\
{$C^{\ell e}_{1322,2213,2312,1223}$} & $1.3\times10^{-2}$ & $\tau\to e\mu^+\mu^-$\\\
{$C^{\ell e}_{2311,1123,1321,2113}$} & $1.0\times10^{-2}$ &  $\tau\to \mu e^+e^-$\\
$C^{\ell e}_{1332,2331}$ & $1.1\times10^{-5}$ & $\mu\to e\gamma$ (1-loop)\\
$C^{\ell e}_{1222,2221}$ & $1.7\times10^{-4}$ & $\mu\to e\gamma$ (1-loop)\\
$C^{\ell e}_{1333,3331}$ & $1.2\times10^{-1}$ & $\tau\to e\gamma$ (1-loop)\\
$C^{\ell e}_{2333,3332}$ & $1.4\times10^{-1}$ & $\tau\to \mu\gamma$ (1-loop)\\
${\rm Im}\, C^{\ell e}_{1331},~ {\rm Re}\, C^{\ell e}_{1331}$ & $ 2.5\times10^{-8}, 2.6\times10^{-2}$ & $d_e, \Delta a_e$ (1-loop)\\ 
${\rm Im}\, C^{\ell e}_{1221},~ {\rm Re}\, C^{\ell e}_{1221}$ & $ 4.2\times10^{-7}, 4.3\times10^{-1}$ & $d_e, \Delta a_e$ (1-loop)%\\ 
%${\rm Im}\, C^{\ell e}_{1111}$ & $ 7.0\times10^{-4}$ & $d_e$ (1-loop)\\
\end{tabular}}
\end{center}
\caption{\small 
Most relevant constraints on the Wilson coefficients of four-lepton operators. 
We do not show constraints that allow Wilson coefficients of order unity or larger.
\label{Ferruccio6}}
\end{table}
%%%%%%%%%%%%%%%%%%%%%

%%%%%%%%%%%%%%%%%%%%%%%%%%%%%%%%%%%%%%
\subsection{Constraints on the anarchic scenario}
%%%%%%%%%%%%%%%%%%%%%%%%%%%%%%%%%%%%%%

Let us translate the constraints of tables~\ref{Ferruccio5} and~\ref{Ferruccio6} into bounds on the PC scenario,
assuming anarchical flavor structure in the composite sector.

PC allows us to determine the Wilson coefficients of the dimension-six operators, up to unknown and model-dependent numbers $c$ expected to be of order unity. The structure of the Wilson coefficients can be easily extracted from \eq{NDA}. To illustrate this we briefly discuss explicitly the case of the dipole operator, defined in the Warsaw basis as 
\begin{equation}
Q_{eW}^{ij} \equiv \left( \bar{\ell}^i_{L} \sigma^{\mu \nu} e^j_{R} \right) \sigma^I H W^I_{\mu \nu} ~.
\end{equation}
From~(\ref{NDA}) one immediately reads~\cite{KerenZur:2012fr}
\begin{equation}\label{sim}
\frac{C^{eW}_{ij}}{\Lambda^2}  = \frac{m^4_{*}}{g^2_{*}} \frac{g^2_{*}}{16 \pi^2} \frac{g_{*} \epsilon^{\ell}_i}{m^{3/2}_{*}} \frac{g_{*} \epsilon^{e}_j}{m^{3/2}_{*}} \frac{g_{*}}{m_*} \frac{g}{m^2_{*}} c^{eW}_{ij} .
\end{equation}
Here $c^{eW}_{ij}$ is an unknown matrix, that in anarchic models is taken to be complex of order one and no particular structure.
To make contact with the experimental bounds from $\ell_i\to\ell_j\gamma$ it is useful to introduce the combinations of dipole operators that correspond to the mass eigenstates for the gauge bosons,
\begin{eqnarray}
Q_{e \gamma }^{ij} & \equiv & \cos \theta_w Q_{e B }^{ij} - \sin \theta_w Q_{e W}^{ij} ~,\\
Q_{e Z}^{ij} & \equiv & \sin \theta_w Q_{e B }^{ij} + \cos \theta_w Q_{e W}^{ij} ~,
\end{eqnarray}
with the dominant constraints coming from the photon observables.

The coefficients for all leptonic operators can be estimated similarly to what done in (\ref{sim}), and they are displayed in table \ref{ops} again up to unknown anarchic matrices with complex numbers of order unity. {Some technical aspects of this procedure are analyzed in appendix~\ref{compApp}. }

We now translate the model-independent constraints of tables \ref{Ferruccio5} and \ref{Ferruccio6} into bounds on the PC parameters introduced in section~\ref{sec:PC}. 
We report in table \ref{bounds2eps} the most stringent bounds on the products of two mixing parameters, $\epsilon^{\ell}_i$ and/or $\epsilon^{e}_i$.
In fact,  the dipole ($Q^{ij}_{e\gamma}$) and vector ($Q^{ij}_{H \ell}$ and $Q^{ij}_{H e}$) operators are the most constraining for PC, and they determine all the bounds  in table \ref{bounds2eps}.
An extended list of bounds in the anarchic PC scenario is displayed in appendix~\ref{compApp}, table~\ref{boundsall}.

For the ease of the discussion we considered two interesting phenomenological limits:
\begin{itemize}
\item[(i)] {\it Left-right symmetry}: $\epsilon^{\ell}_i \sim \epsilon^{e}_i$.

This limit minimizes the bounds coming from the flavor-violating dipole operators. 
Indeed, the contribution to the decay rate scales as $\Gamma( \mu \to e \gamma) \sim \left( |\epsilon^{\ell}_2 \epsilon^e_1|^2 +  |\epsilon^{e}_2 \epsilon^{\ell}_1|^2 \right)$ \cite{KerenZur:2012fr}
and, taking into account \eq{mass-eps}, it gets minimized when $\epsilon^{\ell}_i \sim \epsilon^{e}_i$. For the numerical analysis the equality $\epsilon^{\ell}_i = \epsilon^{e}_i$ has been imposed.
\item[(ii)] {\it Left anarchy}: $\epsilon^{\ell}_i \sim \epsilon^{\ell}_j $.

We have seen in section~\ref{NeutrinoSection} that, when the neutrino masses are linked to the compositeness of the lepton doublets (classes 2M and 1M), 
then large mixing angles imply $\epsilon^\ell_{1,2,3}$ of the same order.
For the numerical analysis the equality $\epsilon^{\ell}_i = \epsilon^{\ell}$ has been imposed for $i=1,2,3$.
Notice that the bounds depend on the unique parameter $\epsilon^{\ell}$, and the hierarchies in the charged lepton masses correspond to the degrees of compositeness of the 
lepton singlets, $\epsilon^{e}_i /  \epsilon^{e}_j = m_i / m_j$.
\end{itemize}

A couple of remarks are in order. Firstly, we quote bounds in terms of the coefficients $c$ which, in anarchic PC scenarios, are expected to be order one numbers. Hence, the PC structure provides enough flavor suppression when in table \ref{bounds2eps} $c$ is bounded by a number larger than one. On the other hand, when the bounds are much smaller than one, 
one concludes that the PC predictions fail to provide the required flavor suppression to be compatible with the experimental observables. 
Our second comment concerns the reference value for the scale of the strongly coupled dynamics, here taken to be $m_{*}= 10$ TeV. This value has been chosen as reference for two reasons: it provides quite enough suppression in the quark sector (see for example \cite{KerenZur:2012fr}), and it is also large enough to be consistent with the non-observation of composite states in direct searches at the LHC. As all the operators under consideration have dimension six, the constraints scale as $m_*^{-2}$. 

%%%%%%%%%%%%%%%%%%%%%
\begin{table}[t]
\begin{center}
\resizebox{0.9\hsize}{!}{$
\begin{tabular}{L|L|L|L}
\hline
\textrm{Structure} & \textrm{Bound} & \epsilon^{\ell}_i \sim \epsilon^{e}_i &  \epsilon^{\ell}_i \sim \epsilon^{\ell}_j \\
\hline
%Dipole12%
\epsilon^{\ell}_1 \epsilon^{e}_2 \frac{g_*^3}{m_*^2}  & 
c \times \left(\frac{g_*}{4 \pi}\right)^2 \frac{\epsilon^{\ell}_1}{\epsilon^{\ell}_2} < 1.1 \times 10^{-4} & 
c \times \left(\frac{g_*}{4 \pi}\right)^2 < 1.6 \times 10^{-3} &
c \times \left(\frac{g_*}{4 \pi}\right)^2 < 1.1 \times 10^{-4}
\\
%Dipole21%
\epsilon^{\ell}_2 \epsilon^{e}_1 \frac{g_*^3}{m_*^2}  & 
c \times \left(\frac{g_*}{4 \pi}\right)^2 \frac{\epsilon^{\ell}_2}{\epsilon^{\ell}_1} < 2.4 \times 10^{-2} & 
c \times \left(\frac{g_*}{4 \pi}\right)^2 < 1.6 \times 10^{-3} &
c \times \left(\frac{g_*}{4 \pi}\right)^2 < 2.4 \times 10^{-2}
\\
%Vector-12-Left%
\epsilon^{\ell}_1 \epsilon^{\ell}_2 \frac{g_*^2}{m_*^2}  &
c \times \left(\frac{g_*}{4 \pi}\right)^2 \epsilon^{\ell}_1 \epsilon^{\ell}_2 < 3.1 \times 10^{-6} & 
c \times \left(\frac{g_*}{4 \pi}\right) < 1.0 &
c \times \left(\frac{g_*}{4 \pi}\right)^2 (\epsilon^{\ell})^2<  3.1 \times 10^{-6}
\\
%Vector-12-Right%
\epsilon^{e}_1 \epsilon^{e}_2 \frac{g_*^2}{m_*^2}  &
c \times \frac{1}{ \epsilon^{\ell}_1 \epsilon^{\ell}_2} < 2.9 \times 10^5 & 
c \times \left(\frac{g_*}{4 \pi}\right) < 1.0 &
c \times  \frac{1}{ (\epsilon^{\ell})^2}<  2.9 \times 10^5
\\
\hline
%Dipole13%
\epsilon^{\ell}_1 \epsilon^{e}_3 \frac{g_*^3}{m_*^2}  & 
c \times \left(\frac{g_*}{4 \pi}\right)^2 \frac{\epsilon^{\ell}_1}{\epsilon^{\ell}_3} < 7.8 \times 10^{-2} & 
c \times \left(\frac{g_*}{4 \pi}\right)^2 <  4.6  &
c \times \left(\frac{g_*}{4 \pi}\right)^2 < 7.8 \times 10^{-2}
\\
%Dipole31%
\epsilon^{\ell}_3 \epsilon^{e}_1 \frac{g_*^3}{m_*^2}  & 
c \times \left(\frac{g_*}{4 \pi}\right)^2 \frac{\epsilon^{\ell}_3}{\epsilon^{\ell}_1} < 2.7 \times 10^{2} & 
c \times \left(\frac{g_*}{4 \pi}\right)^2 < 4.6 &
c \times \left(\frac{g_*}{4 \pi}\right)^2 < 2.7 \times 10^{2} 
\\
%Vector-13-Left%
\epsilon^{\ell}_1 \epsilon^{\ell}_3 \frac{g_*^2}{m_*^2}  &
c \times \left(\frac{g_*}{4 \pi}\right)^2 \epsilon^{\ell}_1 \epsilon^{\ell}_3 < 8.9 \times 10^{-3} & 
c \times \left(\frac{g_*}{4 \pi}\right) < 6.6 \times 10^2  &
c \times \left(\frac{g_*}{4 \pi}\right)^2 (\epsilon^{\ell})^2< 8.9 \times 10^{-3}
\\
%Vector-13-Right%
\epsilon^{e}_1 \epsilon^{e}_3 \frac{g_*^2}{m_*^2} &
c \times \frac{1}{ \epsilon^{\ell}_1 \epsilon^{\ell}_3} < 5.3 \times 10^7 & 
c \times  \left(\frac{g_*}{4 \pi}\right) < 7.1 \times 10^2 &
c \times  \frac{1}{ (\epsilon^{\ell})^2}< 5.3 \times 10^{7}
\\
\hline
%Dipole23%
\epsilon^{\ell}_2 \epsilon^{e}_3 \frac{g_*^3}{m_*^2}  & 
c \times \left(\frac{g_*}{4 \pi}\right)^2 \frac{\epsilon^{\ell}_2}{\epsilon^{\ell}_3} < 8.7 \times 10^{-2} & 
c \times \left(\frac{g_*}{4 \pi}\right)^2 < 3.6 \times 10^{-1} &
c \times \left(\frac{g_*}{4 \pi}\right)^2 < 8.7 \times 10^{-2}
\\
%Dipole32%
\epsilon^{\ell}_3 \epsilon^{e}_2 \frac{g_*^3}{m_*^2}  & 
c \times \left(\frac{g_*}{4 \pi}\right)^2 \frac{\epsilon^{\ell}_3}{\epsilon^{\ell}_2} < 1.5 & 
c \times \left(\frac{g_*}{4 \pi}\right)^2 < 3.6 \times 10^{-1} &
c \times \left(\frac{g_*}{4 \pi}\right)^2 < 1.5
\\
%Vector-23-Left%
\epsilon^{\ell}_2 \epsilon^{\ell}_3 \frac{g_*^2}{m_*^2} &
c \times \left(\frac{g_*}{4 \pi}\right)^2 \epsilon^{\ell}_2 \epsilon^{\ell}_3 < 8.2 \times 10^{-3} & 
c \times \left(\frac{g_*}{4 \pi}\right) < 42  &
c \times \left(\frac{g_*}{4 \pi}\right)^2 (\epsilon^{\ell})^2<  8.2 \times 10^{-3}
\\
%Vector-23-Right%
\epsilon^{e}_2 \epsilon^{e}_3 \frac{g_*^2}{m_*^2}  &
c \times \frac{1}{ \epsilon^{\ell}_2 \epsilon^{\ell}_3} < 2.2 \times 10^{5} & 
c \times \left(\frac{g_*}{4 \pi}\right) < 42 &
c \times  \frac{1}{ (\epsilon^{\ell})^2}< 2.2 \times 10^5
\\
\hline
\epsilon^{\ell}_1 \epsilon^{e}_1  \frac{g_*^3}{m_*^2}  &
\textrm{Im} (c) \times \left(\frac{g_*}{4 \pi}\right)^2 < 5.7 \times 10^{-5} & 
\textrm{Im} (c)  \times \left(\frac{g_*}{4 \pi}\right)^2 < 5.7 \times 10^{-5} &
\textrm{Im} (c)  \times \left(\frac{g_*}{4 \pi}\right)^2 < 5.7 \times 10^{-5}
\\
\epsilon^{\ell}_1 \epsilon^{e}_1 \frac{g_*^3}{m_*^2}  &
\textrm{Re} (c) \times \left(\frac{g_*}{4 \pi}\right)^2 < 2.6 \times 10^{2} & 
\textrm{Re} (c)  \times \left(\frac{g_*}{4 \pi}\right)^2 < 2.6 \times 10^{2}&
\textrm{Re} (c)  \times \left(\frac{g_*}{4 \pi}\right)^2 < 2.6 \times 10^{2}
\\
\hline
\epsilon^{\ell}_2 \epsilon^{e}_2  \frac{g_*^3}{m_*^2}  &
\textrm{Im} (c) \times \left(\frac{g_*}{4 \pi}\right)^2 < 4.6 \times 10^{3} & 
\textrm{Im} (c)  \times \left(\frac{g_*}{4 \pi}\right)^2 < 4.6 \times 10^{3} &
\textrm{Im} (c)  \times \left(\frac{g_*}{4 \pi}\right)^2 < 4.6 \times 10^{3}
\\
\epsilon^{\ell}_2 \epsilon^{e}_2  \frac{g_*^3}{m_*^2}  &
\textrm{Re} (c) \times \left(\frac{g_*}{4 \pi}\right)^2 < 9.4 & 
\textrm{Re} (c)  \times \left(\frac{g_*}{4 \pi}\right)^2 < 9.4 &
\textrm{Re} (c)  \times \left(\frac{g_*}{4 \pi}\right)^2 < 9.4 
\\
\hline
\epsilon^{\ell}_3 \epsilon^{e}_3  \frac{g_*^3}{m_*^2}  &
\textrm{Im} (c) \times \left(\frac{g_*}{4 \pi}\right)^2 < 1.4 \times 10^{4} & 
\textrm{Im} (c)  \times \left(\frac{g_*}{4 \pi}\right)^2 < 1.4 \times 10^{4} &
\textrm{Im} (c)  \times \left(\frac{g_*}{4 \pi}\right)^2 < 1.4 \times 10^{4}
\\
\epsilon^{\ell}_3 \epsilon^{e}_3  \frac{g_*^3}{m_*^2}  &
\textrm{Re} (c) \times \left(\frac{g_*}{4 \pi}\right)^2 < 1.0 \times 10^{2} & 
\textrm{Re} (c)  \times \left(\frac{g_*}{4 \pi}\right)^2 < 1.0 \times 10^{2} &
\textrm{Re} (c)  \times \left(\frac{g_*}{4 \pi}\right)^2 < 1.0 \times 10^{2}
\\
\hline
\end{tabular}
$}
\end{center}
\caption{\small Bounds on the coefficients of the anarchic scenario (i.e. the matrices $c$ are anarchic, complex, order one) {for $m_*=10$ TeV}. In the first column we show the combination of parameters that is constrained, and in the second we used \eq{mass-eps} to remove redundant mixing parameters. 
The flavor and operator indices of the coefficients $c$ are understood.
In the last two columns, the bounds are specialized to two phenomenologically relevant limits: left-right symmetry and left anarchy.
\label{bounds2eps}}
\end{table}
%%%%%%%%%%%%%%%%%%%%%

An inspection of the results of table \ref{bounds2eps} shows that it is quite easy to find regions of parameters space that survive all the constraints coming from observables that involve the third family of leptons. The true obstacle is represented by transitions between the second and first generations ($\mu \to e$). In the left-right symmetry limit, the optimal case, it is in principle possible to pass the bound with $g_{*} = 1$, in which case one gets $c^{e \gamma}_{12,21} < 0.25$. Of course the bound can be satisfied by having a large composite scale, for example 
$c^{e \gamma}_{12,21}<1$ when $g_{*} = 4 \pi$ and $m_{*} =250$ TeV. In both cases the price to pay is in terms of an unnatural electroweak scale; a crude estimate of the fine-tuning is provided by $\xi \sim g^2_{*} v^2/m^2_{*} \lesssim 0.015 \%$. 

An even more severe problem of the anarchic scenario is represented by the experimental bound on the EDM of the electron. This observable, despite being flavor diagonal and suppressed by the small mass of the electron, is sensitive to very tiny CP violating effects. In terms of the fundamental parameters of our model we get
\begin{equation}
{\rm Im} (c^{e \gamma}_{11}) \left( \frac{g_{*}}{4 \pi} \right)^2 \left( \frac{\textrm{10 TeV}}{m_{*}} \right)^2 < 5.7 \times 10^{-5}~.
\end{equation}
Without invoking any CP protection we expect Im$(c)\sim 1$ and the bound gets saturated when $m_{*}=$~1300 TeV and $g_{*} = 4 \pi$, or in the more ``weakly'' coupled scenario when $g_{*} =1$ and $m_{*}=$~110 TeV. These values clearly imply a large tuning of the electroweak scale.

%%%%%%%%%%%%%%%%%%%%%%%%%%%%%%%%%%%%%%
\section{Scenarios with suppressed flavor and CP violation}
\label{sec:supp}
%%%%%%%%%%%%%%%%%%%%%%%%%%%%%%%%%%%%%%

We have shown that, if one uses the NDA estimates for PC defined by  \eq{NDA}, for an anarchic composite sector with generic complex coefficients $c={\cal O}(1)$,
the flavor and CP violating observables, in particular $\mu\to e\gamma$ and the electron EDM, push the new physics scale $m_*$ well beyond the 10 TeV frontier. On the other hand, naturalness arguments suggest the new physics scale should be not far above the TeV. Thus, the severe experimental constraints may be interpreted as an indication that the composite sector cannot be anarchic. In this section we will pursue two main alternative avenues.

One possibility we will consider is that the composite sector carries some accidental symmetries, possibly broken by the spurion couplings $\lambda$ in (\ref{PC1}). This possibility sounds reasonable because strongly-coupled systems generically possess global symmetries, as in the case of QCD. However, while a large flavor symmetry appears to be more adequate to suppress flavor-violation, such an option is typically at odds with partial compositeness. Indeed, if the strong sector enjoys a large non-abelian symmetry that forces all $O_a$ to have approximately the same scaling dimension, then (\ref{PC1}) would not be able to generate the desired hierarchies. We are therefore led to consider scenarios with a $U(1)^3$ symmetry: this is the largest symmetry compatible with a dynamical generation of flavor. As we will argue in  section \ref{sec:u1cube}, this option can simultaneously suppress flavor violation and allow $m_*\sim10$ TeV.~\footnote{Since the strongest experimental constraints on flavor violation involve the electron, one could actually restrict the requirement to $U(1)_e\times CP$. It is rather straightforward to adapt our results to this more minimal case. Note also that a flavor symmetry $SU(3)_{comp}$ has been considered in \cite{Redi:2013pga}. Assuming fully-composite $\ell_i$ or $e_i$ that picture naturally reproduces the ``minimal flavor violation assumption". This successfully suppresses $\mu\to e$ transitions but unfortunately not the electron EDM. In addition, as stressed above, the mass hierarchy is not explained.} $CP$ can then be added to suppress EDMs.

An alternative possibility to make $m_*\sim10$ TeV compatible with data is to postulate the strong dynamics has more, flavor-dependent mass gaps rather than a single, flavor-universal $m_*$. An interesting direction is to associate a different $m_{a*}$ to each of the composite operators ${\cal O}_a$ of \eq{PC1}. This way any flavor-violating process involving the composite index $a$ is controlled by $m_{a*}$ and may thus be suppressed by taking $m_{a*}>m_*$ (see section~\ref{sec:multi}). The possibility of a large compositeness scale for the fermionic operators common to the three lepton families ($m_{1*}=m_{2*}=m_{3*}$) has been suggested in \cite{Vecchi:2012fv}. This is enough to suppress the dipole operators, that govern the most dangerous observables. 
{However, to be more general we will allow different compositeness scales for each family, along the lines of \cite{Panico:2016ull}. }

\subsection{$U(1)^3\times CP$ symmetry} \label{sec:u1cube}
%%%%%%%%%%%%%%%%%%%%%%%%%%%%%%%%%%%%%%

Let us now show how a $U(1)^3$ symmetry in the strong sector suppresses $\mu\to e$ and, when combined with a $CP$ symmetry (still in the strong sector only), also $d_e$, while preserving the generation of the fermion mass hierarchy. As far as we know, a study of this scenario has not been presented in the literature.

To start we postulate the strong sector has a 
\ba\label{sym3}
U(1)_{c_1}\times U(1)_{c_2}\times U(1)_{c_3}
\ea
family symmetry, with the operators $O^{\ell,e}_a$ having the {\emph{same}} charges under $U(1)_{c_a}$ for each $a=1,2,3$. In other words, the mixings in (\ref{PC1}) have spurious charges
\ba\label{ass3}
\lambda_{a = 1 \, i}^{\ell,e} \sim (+1, 0, 0) \, , \quad \lambda_{a = 2 \, i}^{\ell,e} \sim (0, +1, 0) \, , \quad  \lambda_{a = 3 \, i}^{\ell,e} \sim (0, 0, +1) \, ,
\ea
under (\ref{sym3}). This is assumed to ensure that our model generates Yukawa couplings for the charged leptons.

A combination of the $U(1)$ associated with each fundamental lepton flavor and (\ref{sym3}) may be respected by the mixings in (\ref{PC1}), in which case one obtains a framework with an exact $U(1)_e\times U(1)_\mu\times U(1)_\tau$, satisfied by the strong sector as well as the spurions $\lambda$.
{However, within our view that all global symmetries are accidental, such a possibility appears very unlikely unless we gauge part or all of the anomaly-free SM lepton symmetries. If we decide to follow this path, then there is no lepton flavor-violation whatsoever and all the coefficients in table \ref{ops} reduce to combinations of Kronecker deltas.} Additional sources of flavor-violation need to be introduced in order to reproduce the neutrino mixing pattern, but this effect can be naturally small and presents no serious problem (see end of section~\ref{U13nu} for a more detailed discussion). In fact, only the EDMs can set a non-trivial constraint on this exact $U(1)^3$ scenario. Nevertheless, even the latter may be suppressed if the $U(1)^3$ symmetry is combined with $CP$. This extreme picture can thus be easily made consistent with data with $m_*$ of order a few TeV. Its main drawback is perhaps the lack of distinctive signatures and correlations among observables. For this reason we will not discuss it further until section~\ref{anoB}.

In the following we will instead allow the symmetry (\ref{sym3}) to be broken by the mixings $\lambda$ with the SM fermions (\ref{PC1}). This option is more convincing theoretically, and certainly more interesting phenomenologically. We will now show that scenarios with a $U(1)^3$ symmetry in the strong sector unambiguously predict, up to numbers expected to be of order unity, the flavor structure of the operators $Q_{eW,eB,eH}$ in terms of the charged lepton masses. The Wilson coefficients of the remaining operators, $Q_{He,H\ell,\ell\ell,ee,\ell e}$, instead crucially depend on the ratios $\epsilon_i^\psi/\epsilon_j^\psi$ and are thus more model-dependent.

To see this it is useful to choose a convenient field basis. Let us work under the assumption that $O_a^{\ell,e}$ have the same $U(1)^3$ charges, see (\ref{ass3}), but make no a priori assumption on their scaling dimensions. In other words we do not impose any constraint on the ratios $\epsilon_i^\psi/\epsilon^\psi_j$ at this stage. Yet, we rotate the fundamental fermions to put the mixings in the triangular form~(\ref{PC3}). Note that this step does not rely on the assumption $\epsilon^\psi_1 < \epsilon^\psi_2 < \epsilon^\psi_3$: one can always choose a basis for the $\psi_i$'s in which the mixings take the form~(\ref{PC3}) whatever the relative size of the $\epsilon_i^\psi$'s is. Once this is done we have no more freedom to rotate the fields $\psi$. However, we are still free to order the composite index $a$, or in other words the ratios $\epsilon_i^\psi/\epsilon^\psi_j$ of (\ref{PC3}), as we wish. A convenient ordering for the composite fermions is eventually identified inspecting the charged lepton Yukawa couplings. These are formally the same as in (\ref{Yukawa}), $y^e_{ij}= g_* \epsilon^{\ell *}_{ai} \epsilon^e_{bj} c^e_{ab}$, though by $U(1)^3$ symmetry the coefficients of order unity must satisfy $c^e_{ab}=\delta_{ab}c_a$. Thus we get:
\be\label{yukFlip}
y^e = g_* \left[\epsilon^\ell_1\epsilon_1^e {\cal O}\left(
\begin{array}{ccc}
1 & 1 & 1 \\ 
1 & 1 & 1 \\ 
1 & 1 & 1
\end{array}\right)
+  \epsilon^\ell_2\epsilon_2^e {\cal O}\left(
\begin{array}{ccc}
0 & 0 & 0 \\ 
0 & 1 & 1 \\ 
0 & 1 & 1
\end{array}\right)
+ \epsilon^\ell_3\epsilon_3^e {\cal O}\left(
\begin{array}{ccc}
0 & 0 & 0 \\ 
0 & 0 & 0 \\ 
0 & 0 & 1
\end{array}
\right)\right],
\ee
which is a sum of three rank-one matrices because $c^e$ is a diagonal matrix. This expression gives us all the information necessary to single out the phenomenologically viable $U(1)^3$ scenarios. First, an inspection of (\ref{yukFlip}) teaches us that a fermion mass hierarchy can only be obtained, barring unnatural cancellations, if there is a hierarchy among $\epsilon_i^\ell\epsilon_i^e/\epsilon_j^\ell\epsilon_j^e$. Second, there is a unique choice of ordering of the composite fermions $O_{1,2,3}$ that makes sure the mixing angles between gauge and mass basis are small. Such an ordering is useful because with small angles an understanding of the pattern of flavor violation beyond the SM may be achieved via a perturbative expansion. From (\ref{yukFlip}) one sees that this requirement is realized when 
\ba\label{123}
\epsilon^\ell_1\epsilon^e_1 \ll \epsilon^\ell_2\epsilon^e_2\ll \epsilon^\ell_3\epsilon^e_3~. 
\ea
To summarize, we first found a field basis in which the Yukawa matrix acquires the form (\ref{yukFlip}), without loss of generality; from that expression we then learnt that in all $U(1)^3$ models that can generate a realistic pattern of charged lepton masses we can always label the composite operators  according to (\ref{123}). Each independent ordering of $O^{\ell}_a$ and $O^{e}_a$, which select different ratios $\epsilon_i^\psi/\epsilon_j^\psi$ while still preserving (\ref{123}), constitutes a different variant of the $U(1)^3$ scenario.

Having identified a convenient flavor basis, we can now study the phenomenological implications of $U(1)^3$. Thanks to (\ref{123}) the expression for the Yukawa matrix simplifies
\be\label{yuksym3}
y^e\sim {g_*}\left(
\begin{array}{ccc}
\epsilon^\ell_1\epsilon_1^e & \epsilon^\ell_1\epsilon_1^e & \epsilon^\ell_1\epsilon_1^e \\ 
\epsilon^\ell_1\epsilon_1^e & \epsilon^\ell_2\epsilon_2^e & \epsilon^\ell_2\epsilon_2^e \\ 
\epsilon^\ell_1\epsilon_1^e & \epsilon^\ell_2\epsilon_2^e & \epsilon^\ell_3\epsilon_3^e
\end{array}
\right)
\sim \frac{\sqrt{2}m_3}{v}
\left(
\begin{array}{ccc}
\frac{m_1}{m_3} & \frac{m_1}{m_3} & \frac{m_1}{m_3}\\ 
\frac{m_1}{m_3} & \frac{m_2}{m_3} & \frac{m_2}{m_3}\\ 
\frac{m_1}{m_3} & \frac{m_2}{m_3} & 1
\end{array}
\right).
\ee
Moreover, the matrices $U^\ell,U^e$ that diagonalize it have the form
\ba\label{uniLR}
U^{\ell,e}_{ij} \sim
\left(
\begin{array}{ccc}
1 & \frac{\epsilon^\ell_1}{\epsilon^\ell_2}\frac{\epsilon^e_1}{\epsilon^e_2} & \frac{\epsilon^\ell_1}{\epsilon^\ell_3}\frac{\epsilon^e_1}{\epsilon^e_3} \\ 
\frac{\epsilon^\ell_1}{\epsilon^\ell_2}\frac{\epsilon^e_1}{\epsilon^e_2} & 1 & \frac{\epsilon^\ell_2}{\epsilon^\ell_3}\frac{\epsilon^e_2}{\epsilon^e_3}\\ 
\frac{\epsilon^\ell_1}{\epsilon^\ell_3}\frac{\epsilon^e_1}{\epsilon^e_3} & \frac{\epsilon^\ell_2}{\epsilon^\ell_3}\frac{\epsilon^e_2}{\epsilon^e_3} & 1
\end{array}
\right)_{ij}
\sim
\left(
\begin{array}{ccc}
1 & \frac{m_1}{m_2} & \frac{m_1}{m_3}\\ 
\frac{m_1}{m_2} & 1 & \frac{m_2}{m_3}\\ 
\frac{m_1}{m_3} & \frac{m_2}{m_3} & 1
\end{array}
\right)_{ij}={\rm min}\left(\frac{m_i}{m_j},\frac{m_j}{m_i}\right)
\label{Us}\ea
and the eigenvalues are $m_e=m_1,~m_\mu=m_2,~m_\tau=m_3$. As promised, the ordering defined by (\ref{123}) implies $|U^{\ell, e}_{ij}|\ll1$ for $i\neq j$ irrespective of the ratios $\epsilon_i^\psi/\epsilon_j^\psi$ for $\psi=\ell,e$ separately. 
The Yukawa matrix structure (\ref{yuksym3}) is to be contrasted with the one for anarchic models, see (\ref{ye}). Importantly, the off-diagonal elements here depend only on the lepton masses $m_k$, rather than on $\epsilon^{\ell,e}_k$ separately as it was in the anarchic case (\ref{rot-eps}). The smaller mixing angles will have important implications for charged-lepton flavor and CP-violating observables, as we will see in the next subsection. 

Before proceeding it is important to emphasize that the $U(1)^3$ symmetry is here assumed to be satisfied by the ``lepton sector" of the strong dynamics, specifically by $O^{\ell,e}_a$. The mixing structure (\ref{uniLR}) may be extended to the quark sector only provided one is willing to accept moderate hierarchies (of the order of {one power of} the Cabibbo angle) among the order-one parameters {of the mixing matrix $\epsilon^{d}$ in~(\ref{Yukawa}). In that case the CKM matrix, being determined by the rotation in the down sector, can be reproduced~\cite{Panico:2016ull}}.

\subsubsection{Dipole operators} \label{sec:u3imp}
%%%%%%%%%%%%%%%%%%%%%%%%%%%%%%%%%%%%%%

Let us now consider the implications of $U(1)^3$ on the Wilson coefficients of the dim-6 operators of table~\ref{ops}. Using the field basis identified in the previous subsection, the flavor structure of these operators can be written schematically as:
\be\begin{array}{ll}
Q^{ij}_{eW,eB,eH}: & \epsilon^{\ell*}_{ai} \epsilon^e_{aj} c_a~,\\
Q^{ij}_{H\ell,He}: & \epsilon^{\ell*}_{ai}\epsilon^{\ell}_{aj}c_a~,~~~ \epsilon^{e*}_{ai}\epsilon^{e}_{aj}c_a~,\\
Q_{\ell e}^{ijmn}: & {\epsilon^{\ell*}_{ai}\epsilon^{\ell}_{cj}\epsilon^{e*}_{bm}\epsilon^{e}_{dn} (\delta_{ac}\delta_{bd}c_{ab}+\delta_{ad}\delta_{bc}c'_{ab}) ~,}\\
Q_{\ell\ell}^{ijmn}: & \epsilon^{\ell*}_{ai}\epsilon^{\ell}_{cj}\epsilon^{\ell*}_{bm}\epsilon^{\ell}_{dn} (\delta_{ac}\delta_{bd}c_{ab}+\delta_{ad}\delta_{bc}c'_{ab}) ~,\\
Q_{ee}^{ijmn}: & {\epsilon^{\ell*}_{ai}\epsilon^{\ell}_{aj}\epsilon^{\ell*}_{bm}\epsilon^{\ell}_{bn} c_{ab}~,~~~ 
\epsilon^{e*}_{ai}\epsilon^{e}_{aj}\epsilon^{e*}_{bm}\epsilon^{e}_{bn} c_{ab}~,}
\end{array}
\label{stru}
\ee
where for the last line we took into account the Fierz identity $Q_{ee}^{ijmn}=Q_{ee}^{inmj}$.

To go to the mass basis and identify the coefficients to be constrained by tables~\ref{Ferruccio5} and \ref{Ferruccio6}, one needs to contract with the rotation matrices (\ref{uniLR}). With the aid of \eq{PC3} one obtains
\be\label{impl}
\epsilon^{\psi}\to\epsilon^{\psi}U^\psi = {\cal O}\left(\begin{array}{ccc}
\epsilon^\psi_1 & \epsilon^\psi_1 & \epsilon^\psi_1 \\
{\frac{m_1}{m_2}} \epsilon^\psi_2 & \epsilon^\psi_2 & \epsilon^\psi_2 \\
{\frac{m_1}{m_3}} \epsilon^\psi_3 & 
{\frac{m_2}{m_3}} \epsilon^\psi_3 & \epsilon^\psi_3 
\end{array}\right)~,\quad \psi=\ell,e.
\ee
We will first focus on the $\mu\to e\gamma$ transition and the electron EDM, that set by far the most remarkable constraints on new physics coupled to leptons.

The operators of interest are associated, together with $Q_{eH}$, to the first class of flavor structures in (\ref{stru}). Being the same as the Yukawa, this structure is not modified significantly by the rotation to the mass basis. In particular, their Wilson coefficients in the mass basis continue to be functions dominantly of ratios of masses, irrespective of what the ratios $\epsilon_i^\psi/\epsilon_j^\psi$ are,
\be\label{questa}
(\epsilon^\ell U^{\ell})^\dag c (\epsilon^{e} U^{e}) = 
\frac{\sqrt{2}m_3}{g_*v}
{\cal O}\left(\begin{array}{ccc}
\frac{m_1}{m_3} & \frac{m_1}{m_3} & \frac{m_1}{m_3} \\
\frac{m_1}{m_3} & \frac{m_2}{m_3} & \frac{m_2}{m_3} \\
\frac{m_1}{m_3} & \frac{m_2}{m_3} & 1 
\end{array}\right)~.
\ee
As a result the phenomenology is common to all realistic $U(1)^3$ variants. For example, let us start with $\mu\to e\gamma$, for which the $U(1)^3$ symmetry implies
\be\label{U13suppr}
\frac{C^{e\gamma}_{12,21}}{\Lambda^2}\simeq \frac{g_*^2}{16\pi^2} \frac{e}{m_*^2} \frac{\sqrt{2}m_e}{v} \hat c^{e\gamma}_{12,21}~,
\ee
with $\hat c$ a matrix of order unity (we included a hat to distinguish it from the corresponding matrix in the anarchic scenario). This is to be compared with the best case option ${C^{e\gamma}_{12,21}}/{\Lambda^2}\simeq ({g_*^2}/{16\pi^2}) ({e}/{m_*^2}) ({\sqrt{2m_em_\mu}}/{v}) c^{e\gamma}_{12,21}$ of the anarchic scenario, found when $\epsilon^\ell_i\sim\epsilon^e_i$ in table \ref{ops}. Eq.~(\ref{U13suppr}) shows that $U(1)^3$ can weaken the constraint by a factor $\sqrt{m_e/m_\mu}\simeq1/15$ relative to the anarchic case. The current $90\%$ CL bound from MEG, ${\rm BR}(\mu\to e\gamma)<5.7\times10^{-13}$, implies $|\hat c^{e\gamma}_{12,21}|<0.02 (0.1)$ for $m_*=10 (20)$ TeV and $g_*=4\pi$. We may allow $\hat c^{e\gamma}_{12,21}$ to be order unity if $m_*/g_*\gtrsim 5$ TeV.

The electron EDM $d_e$ can be significantly suppressed by combining the $U(1)^3$ symmetry with the assumption that the entire strong dynamics is symmetric under $CP$. Note that this is a viable option since we have shown in section \ref{CKMphase} that a realistic value for the CKM phase can arise from complex entries in the spurions $\lambda$. In the CP-symmetric scenarios all strong-sector order-one parameters such as $c^e_{ab}$ are real. To see the implications of these ingredients, observe that $d_e$ is controlled by the imaginary part of the $ee$-entry of the dipole operator, 
\ba
d_e=2 \frac{{\rm Im} (U^{\ell\dag}C^{e\gamma}U^e)_{ee}}{\Lambda^2} \frac{v}{\sqrt{2}} = \frac{g_*^2}{8\pi^2} \frac{e}{m_*^2} 
{\rm Im}(U^{\ell\dag}\epsilon^{\ell\dag}c^{e\gamma}\epsilon^e U^e)_{ee} \frac{g_*v}{\sqrt{2}} ~.
\ea
The $U(1)^3$ symmetry by itself ($c^{e\gamma}_{ab}=c^{e\gamma}_a\delta_{ab}$) in this case does not improve the situation compared to the anarchic case. On the other hand, when we force $c^{e\gamma}$ to be real by $CP$ the imaginary part controlling $d_e$ gets suppressed by the small mixing angles of the $U(1)^3$ model. Specifically, the CP-odd phases must be found in $(\epsilon^{\ell,e}U^{\ell,e})_{ae}$ with $a\neq1$ because, as we saw in (\ref{PC3}), the diagonal elements of $\epsilon_{ai}$ can be made real via rotations of the SM fields. Generalizing (\ref{questa}) to include the CP-odd phases arising from the mixings $\lambda$ one can show that ${\rm Im}(U^{\ell\dag}\epsilon^{\ell\dag}c^{e\gamma}\epsilon^e U^e)_{ee}\sim \epsilon_1^\ell\epsilon_1^em_e/m_\mu$, which in turn gives
\ba\label{de}
d_e\simeq \frac{g_*^2}{8\pi^2} \frac{e m_e}{m_*^2} \frac{m_e}{m_\mu}\hat c^{e\gamma}_{11}~.
\ea
{This is a significant $m_e/m_\mu$ improvement compared to the anarchic, CP-violating scenario (see table~\ref{ops}). There is also an improvement with respect to the anarchic scenario with CP invariance, for which we find 
${\rm Im}(U^{\ell\dag}\epsilon^{\ell\dag}c^{e\gamma}\epsilon^e U^e)_{ee} \sim 
\epsilon_1^\ell\epsilon_1^e \sqrt{m_\mu/m_\tau}$.}
Numerically, for $m_*=10$ TeV and $g_*=4\pi$ one verifies that (\ref{de}) is below the current bound of table \ref{boundsPDG} provided $|{\rm Im}\,\hat c^{e\gamma}_{11}|\lesssim 0.012$. 
In other words, combining $U(1)^3$ with CP invariance makes our model compatible with data for $m_*/g_*\gtrsim 7.3$ TeV.

{If one further assumes that the $U(1)^3$ in the composite lepton sector carries over to the quark sector, then the neutron EDM gets suppressed by a factor $\sim m_d/m_s\sim\lambda_C^2$ compared to the anarchic scenario. This could be enough to make the model with $g_*\sim4\pi, m_*\sim10$ TeV consistent with {quark} data.} 

\subsubsection{$Q_{He,H\ell,\ell\ell,ee,\ell e}$} \label{sec:boh}
%%%%%%%%%%%%%%%%%%%%%%%%%%%%%%%%%%%%%%

As opposed to what we saw for the dipoles and $Q_{eH}$, the rotation matrices $U^{\ell,e}$ may modify significantly the flavor structures in the other lines of (\ref{stru}). The key difference is that for the bilinears 
$(\epsilon^{\psi\dagger} c \, \epsilon^\psi)$ (this applies to both $\psi=\ell,e$), the flavor structure in the mass basis depends not only on the charged lepton masses, but also on the {\emph{relative size} of the parameters $\epsilon_{i}$}. Because (\ref{123}) can be achieved with different choices of ratios $\epsilon_i^\psi/\epsilon_j^\psi$, one thus finds that realistic models with a $U(1)^3$ symmetry can have several, qualitatively different phenomenological predictions for the Wilson coefficients of $Q_{He,H\ell,\ell\ell,ee,\ell e}$.
Explicitly we obtain
\ba\label{LLRR}
&& \!\!\!\!\!\!\!\!\!\!\!\! (\epsilon U)^\dag c (\epsilon U) \sim \\
&& \left(\begin{matrix}
(\epsilon_1)^2 + \frac{m^2_1}{m^2_2}(\epsilon_2)^2 +  \frac{m^2_1}{m^2_3} (\epsilon_3)^2 & 
(\epsilon_1)^2 + \frac{m_1}{m_2}(\epsilon_2)^2 +  \frac{m_1m_2}{m^2_3} (\epsilon_3)^2 & 
(\epsilon_1)^2 + \frac{m_1}{m_2}(\epsilon_2)^2 +  \frac{m_1}{m_3} (\epsilon_3)^2 \\
\dots & 
(\epsilon_1)^2 + (\epsilon_2)^2 +  \frac{m^2_2}{m^2_3} (\epsilon_3)^2 & 
(\epsilon_1)^2 + (\epsilon_2)^2 +  \frac{m_2}{m_3} (\epsilon_3)^2 \\
\dots & \dots & 
(\epsilon_1)^2 + (\epsilon_2)^2 +  (\epsilon_3)^2 
\end{matrix}\right),\no
\ea
where we spared ourselves from writing all entries because the matrix is manifestly hermitian. Curiously, these coefficients are ordered in size, i.e.~the elements $ij$ of (\ref{LLRR}) satisfy $11\lesssim12\lesssim13\lesssim22\lesssim23\lesssim33$, independently of the $\epsilon_i$'s. This is important because it implies that all $U(1)^3$-invariant realizations satisfy the following general properties: first, transitions involving heavier leptons are always faster; second, there exists a lower bound on flavor violation. The most relevant constraints on the off-diagonal entries come from $C^{He,H\ell}_{ij}$. For $m_*=10$ TeV and $g_*=4\pi$, the bounds shown in table \ref{Ferruccio5} roughly translate into an upper bound $\sim 10^{-6}$ ($\sim 10^{-2}$) on the $e\mu$ ($e\tau$ and $\mu\tau$) entry of (\ref{LLRR}).

The off-diagonal coefficients in (\ref{LLRR}) control the magnitude of flavor-violation, but their size strongly depends on $\epsilon^\psi_{i}/\epsilon^\psi_j$. Here we mention a few paradigmatic examples:
\begin{itemize}
\item[(i)] {\it Left-right symmetry}: $\epsilon^{\ell}_i \sim \epsilon^{e}_i$

Together with (\ref{123}), this implies automatically a normal ordering of the indices, i.e.~$\epsilon_1<\epsilon_2<\epsilon_3$ for both $\ell,e$.
The Wilson coefficients for the various operators, expressed in the mass basis, are compactly collected in table \ref{Wilson3}. For each pair $ij$ of lepton indices the coefficients are suppressed by a factor ${\rm min}(\sqrt{m_i/m_j},\sqrt{m_j/m_i})$ compared to the anarchic case. Explicitly, in the $U(1)^3$-symmetric scenario the coefficients $c_{ij}, c_{ijmn}$, that are of order unity in table \ref{ops}, become
\ba
\label{LRde}
&&c_{ij}\sim{\rm min}\left(\sqrt{\frac{m_i}{m_j}},\sqrt{\frac{m_j}{m_i}}\right)\hat c_{ij},\\\no
&&c_{ijmn}\sim{\rm min}\left(\sqrt{\frac{m_i}{m_j}},\sqrt{\frac{m_j}{m_i}}\right){\rm min}\left(\sqrt{\frac{m_m}{m_n}},\sqrt{\frac{m_n}{m_m}}\right)\hat c_{ijmn}+{\rm permut.},
\ea
with $\hat c_{ij}, \hat c_{ijmn}$ coefficients of order unity emerging from the $U(1)^3$-symmetric dynamics. All flavor-violating channels are therefore suppressed by $U(1)^3$ compared to the anarchic scenario. Overall in this scenario the dominant bound comes from the dipole operators that we discussed above, from which we get $m_*/g_*\gtrsim 7$ TeV.
\item[(ii)] {\it Left anarchy}: $\epsilon^{\ell}_i \sim \epsilon^{\ell}_j $

As in the previous case, when combined with (\ref{123}), this option implies $\epsilon^e_1<\epsilon^e_2<\epsilon^e_3$. In this case however the flavor-violating coefficients with structure $\epsilon^{\ell*}_{ai} \epsilon^e_{aj}$ similar to the Yukawa (see \eq{stru}) are suppressed by a factor ${\rm min}(m_i/m_j,1)$, while all the others remain (always at leading order in the mass ratios) parametrically the same as in the anarchic case,
\ba
\label{Lde}
&&c^{eW,eB,eH}_{ij}\sim{\rm min}\left({\frac{m_i}{m_j}},1\right)\hat c^{eW,eB,eH}_{ij},\\\no
&&c^{H\ell,He}_{ij}\sim\hat c^{H\ell,He}_{ij},\\\no
&&c^{\ell\ell,ee,\ell e}_{ijmn}\sim\hat c^{\ell\ell,ee,\ell e}_{ijmn}.\no
\ea
More explicit expressions for the Wilson coefficients are presented in table \ref{Wilson3}. Note that the relative suppression of the dipole operator in (\ref{Lde}) is different from that in (\ref{LRde}) because the predictions of the anarchic scenario are different in the left-right symmetric and the left-anarchic cases. In reality, as shown above, the $U(1)^3$ prediction is actually the same in the two cases, see table \ref{Wilson3}. In summary, for this scenario $\mu\to e\gamma$ is still important, but also the other bounds are relevant. Fixing $g_*\simeq1$ and $m_*\simeq10$ TeV, so that the former rate is safely consistent with data, the other bounds in table \ref{bounds2eps} imply
\ba\label{46}
\epsilon_{\ell} \lesssim 2.2 \cdot 10^{-2}~.
\ea
This reduces the range $\sqrt{2}m_\tau/(g_*v)\simeq 1.0\cdot 10^{-2}<\epsilon^\ell<1$, that follows from requiring PC to correctly reproduce the tau lepton mass.
\item[(iii)] {\it Example of ``flipped" scenario}

Flipped scenarios are those in which the $\epsilon_i^\psi$'s do not satisfy the usual ordering, but are still consistent with the defining property of our field basis, namely (\ref{123}). These are typical of the $U(1)^3$ models and cannot arise in the anarchic case. Many options are available and discussing all of them is beyond the scope of this paper. We prefer to illustrate some of the possible phenomenological implications by focusing on an explicit example. We consider $\epsilon^\ell_1 \sim \epsilon^e_1$, $\epsilon^\ell_2 \sim \epsilon^e_2$, and $\epsilon^\ell_3 \sim (m_\mu/m_\tau)^2 \epsilon^e_3$, that correspond to $\epsilon^\ell_1/\epsilon^\ell_2\sim \sqrt{m_e/m_\mu}<1$ and the flipped hierarchy $\epsilon^\ell_2/\epsilon^\ell_3\sim \sqrt{m_\tau/m_\mu}>1$. To assess the viability of this model recall that the constraints on dipole operators are the same as in section~\ref{sec:u3imp} and are therefore unaffected by the flipping $\epsilon_1^\ell<\epsilon_3^\ell<\epsilon_2^\ell$. The next to relevant constraints come from  
$\mu\, Au \to e\, Au \ (\mu\to eee)$
or $\tau\to\ell_i\ell_j\ell_k$. In the present flipped scenario the $e^\dagger e$ Wilson coefficients are much bigger than the $\ell^\dagger\ell$ ones,
and very close to 
the experimental bounds (again for $m_*=10$ TeV and $g_*=4\pi$). 
Importantly, it may well be that the dominant experimental signature comes from $Q_{He}$ rather than the dipole operators as in the anarchic scenarios.
\end{itemize}
The above three examples reveal that the signatures of the $U(1)^3$ scenario associated to the operators $Q_{He,H\ell,\ell\ell,ee,\ell e}$ do depend significantly on the relative size of the $\epsilon^{\psi}_{i}$'s for $\psi = \ell, e$. This model-dependence only affects these operators because, as we saw in section~\ref{sec:u3imp}, our analysis of the dipoles apply to all $U(1)^3$-symmetric models, irrespective of the numerical values of the ratios $\epsilon^{\psi}_{i}/\epsilon^\psi_j$.

%%%%%%%%%%%%%%%%%%%%%
\begin{table}[tb]
\begin{center}
\begin{tabular}{l|l|l} 
$U(1)^3$  &  $\epsilon_i^\ell\sim\epsilon_i^e$ & $\epsilon_i^\ell\sim\epsilon^\ell_j$ \\
\hline
$\frac{C^{eW}_{ij}}{\Lambda^2}$ 
& $\frac{g_*^2}{16 \pi^2} \frac{g_*}{m_*^2}  \frac{\sqrt{2}{\rm min}(m_i,m_j)}{g_*v}  g\, \hat c^{eW}_{ij}$ 
&  $\frac{g_*^2}{16 \pi^2} \frac{g_*}{m_*^2}  \frac{\sqrt{2}{\rm min}(m_i,m_j)}{g_*v} g\,
\hat c^{eW}_{ij}$ \\
$\frac{C^{eB}_{ij}}{\Lambda^2}$ 
& $\frac{g_*^2}{16 \pi^2} \frac{g_*}{m_*^2}  \frac{\sqrt{2}{\rm min}(m_i,m_j)}{g_*v} g'\,
\hat c^{eB}_{ij}$
&  $\frac{g_*^2}{16 \pi^2} \frac{g_*}{m_*^2}  \frac{\sqrt{2}{\rm min}(m_i,m_j)}{g_*v} g'\,
\hat c^{eB}_{ij}$ \\
$\frac{C^{eH}_{ij}}{\Lambda^2}$ 
& $\frac{g^3_*}{m_*^2}  \frac{\sqrt{2}{\rm min}(m_i,m_j)}{g_*v}   \hat c^{eH}_{ij}$ 
& $\frac{g^3_*}{m_*^2}  \frac{\sqrt{2}{\rm min}(m_i,m_j)}{g_*v}   \hat c^{eH}_{ij}$ \\ 
$\frac{C^{H \ell(1,3)}_{ij}}{\Lambda^2}$ 
& $\frac{g^2_*}{m_*^2} \frac{\sqrt{2} {\rm min}(m_i,m_j)}{g_*v}  \hat c^{H \ell(1,3)}_{ij}$ 
& $\frac{g^2_*}{m_*^2} (\epsilon^\ell)^2 \hat c^{H \ell(1,3)}_{ij}$ \\
$\frac{C^{H e}_{ij}}{\Lambda^2}$ 
& $\frac{g^2_*}{m_*^2} \frac{\sqrt{2} {\rm min}(m_i,m_j)}{g_*v}  \hat c^{H e}_{ij}$ 
& $\frac{g^2_*}{m_*^2} \frac{2m_i m_j}{g^2_*v^2}\frac{1}{(\epsilon^\ell)^2}  \hat c^{H e}_{ij}$  \\
$\frac{C^{\ell e}_{ijmn}}{\Lambda^2}$ 
& $\frac{g^2_*}{m_*^2} 
\left[ \begin{array}{c} \frac{2 {\rm min}(m_i,m_j) {\rm min}(m_m,m_n) }{g^2_*v^2} \hat c^{\ell e}_{ijmn}  \\
+\frac{2 {\rm min}(m_i,m_n) {\rm min}(m_m,m_j)}{g^2_*v^2}  \hat c'^{\ell e}_{ijmn} \end{array}\right]$
& $\frac{g^2_*}{m_*^2} 
\left[ \begin{array}{c} \frac{2 m_m m_n }{g^2_*v^2}\hat c^{\ell e}_{ijmn}  \\ +
\frac{2 {\rm min}(m_i,m_n) {\rm min}(m_m,m_j) }{g^2_*v^2} \hat c'^{\ell e}_{ijmn} \end{array}\right]$ \\
$\frac{C^{\ell \ell}_{ijmn}}{\Lambda^2}$ 
& $\frac{g^2_*}{m_*^2} 
\left[ \begin{array}{c} \frac{2 {\rm min}(m_i,m_j) {\rm min}(m_m,m_n) }{g^2_*v^2} \hat c^{\ell \ell}_{ijmn}  \\
+\frac{2 {\rm min}(m_i,m_n) {\rm min}(m_m,m_j)}{g^2_*v^2}  \hat c'^{\ell \ell}_{ijmn} \end{array}\right]$
& $\frac{g^2_*}{m_*^2} (\epsilon^\ell)^4 \hat c^{\ell \ell}_{ijmn}$\\
$\frac{C^{e e}_{ijmn}}{\Lambda^2}$ & $\frac{g^2_*}{m_*^2} \frac{2 {\rm min}(m_i,m_j) {\rm min}(m_m,m_n) }{g^2_*v^2}  \hat c^{e e}_{ijmn}$ 
& $\frac{g^2_*}{m_*^2} \frac{4m_im_jm_m m_n  }{g^4_*v^4}\frac{1}{(\epsilon^\ell)^4}   \hat c^{e e}_{ijmn}$ \\
\hline
\end{tabular}
\end{center}
\caption{\small Wilson coefficients of the dim-6 operators in the mass basis, assuming the strong dynamics has a $U(1)^3$ symmetry broken by the mixings (\ref{PC1}) (to be compared to the anarchic case of table~\ref{ops}). Here we show the result for two different assumptions on the
values of $\epsilon^{\ell,e}_i$. As demonstrated in the text the predictions for $C_{eW,eB,eH}$ are however general and hold for any value of $\epsilon_i^\psi$.}
\label{Wilson3}
\end{table}
%%%%%%%%%%%%%%%%%%%%%

\subsubsection{Neutrino masses} \label{U13nu}
%%%%%%%%%%%%%%%%%%%%%%%%%%%%%%%%%%%%%%

In section~\ref{NeutrinoSection} the operators $O$ in (\ref{perturb}) responsible for generating the neutrino masses had specific charges under the composite lepton number $U(1)_c$. Here the strong sector operators have well-defined quantum numbers under $U(1)^3$. As a consequence, the mixings in (\ref{PC1}) have definite $U(1)^3$-spurious charges, that we chose without loss of generality as in (\ref{ass3}), whereas the lepton-breaking couplings are now $\tilde \lambda \sim (c_1,c_2,c_3)$ under the full $U(1)_{c_1}\times U(1)_{c_2}\times U(1)_{c_3}$ symmetry. The total lepton number of $\tilde \lambda$ is generally non-vanishing, $L = c_1 + c_2 + c_3 + \ell\ne 0$.

There is a major departure from our treatment in section~\ref{NeutrinoSection}. When a $U(1)^3$ is assumed the $\tilde \epsilon$ matrices for the classes 2M and 1M in (\ref{Mclass}) are no longer expected to be anarchic, since their entries have different $U(1)_{c_1}\times U(1)_{c_2}\times U(1)_{c_3}$ quantum numbers:
\ba\label{Mclasscharges}
&&{\rm 2M}:~\tilde \epsilon_{ab} \sim \left(
\begin{matrix}
(-2,0,0) & (-1,-1,0) & (-1,0,-1) \\
(-1,-1,0)  & (0,-2,0) & (0,-1,-1) \\
(-1,0,-1) & (0,-1,-1) & (0,0,-2)
\end{matrix}\right) \,,~~~ \\\no
&&{\rm 1M}:~ \tilde\epsilon_{ai} \sim \left(
\begin{matrix}
(-1,0,0) & (-1,0,0) & (-1,0,0) \\
(0,-1,0) & (0,-1,0) & (0,-1,0) \\
(0,0,-1) & (0,0,-1) & (0,0,-1) 
\end{matrix}\right) \, ,~~~ \\\no
&&{\rm 0M}:~ \tilde\epsilon_{ij} \sim (0,0,0) \, .
\ea
The phenomenology is completely determined by these charge assignments.

Because we do not actually know the UV dynamics that gives rise to (\ref{perturb}), we do not know which operators and hence which of the representations in (\ref{Mclasscharges}) will be present in the EFT below $\Lambda_{\rm UV}$. We therefore take an agnostic perspective and discuss a few possibilities. In particular we will focus on the case in which a single representation $\tilde\lambda$ in (\ref{Mclasscharges}) dominates. There are several reasons for this. The first is minimality; interestingly, we will find a few cases in which a single charge is enough to generate a realistic neutrino texture. The second is more technical: if several lepton-breaking operators with different $U(1)^3$ quantum numbers were to have non-vanishing coefficients $\tilde\lambda$ in the UV, then their effect at low energies would be dependent on their anomalous dimensions. But these latter are generically different because by assumption the operators have different quantum numbers. Therefore, barring non-generic coincidences, one of the CFT deformations will typically be more relevant than the others and determine the neutrino mass texture. One recovers the results of the anarchic scenarios discussed in section \ref{NeutrinoSection} when several lepton-breaking operators break $U(1)^3$ maximally and the corresponding scaling dimensions are similar.

{Scenarios belonging to 0M in (\ref{Mclass}) reveal no new features compared to the anarchic case and will not be discussed. We instead analyze a few examples within classes 2M and 1M. We will demonstrate that in a few interesting cases a single spurion, $\tilde \epsilon\ne 0$, is sufficient to yield a realistic neutrino mass matrix. 
} 

As a first concrete case let us consider the $U(1)_L$-violating operator $\tilde\lambda_{\tilde a}O_{\tilde a, (c_1 = 1, c_2 = 1, c_3 = 0)}$, which breaks lepton number by two units. Such a deformation gives rise to a neutrino mass matrix which 
belongs to class 2M, since two insertions of the mixings $\epsilon^\ell$ are required. According to (\ref{Mclasscharges}), only $\tilde \epsilon_{12} = \tilde \epsilon_{21}$ is non-zero. Thus, recalling the triangular form of the mixings, \eq{PC3}, we find the neutrino mass matrix reads
\be
m^\nu \propto \epsilon^\ell_1\epsilon^\ell_2 \left(
\begin{matrix}
0 & 1 & 1 \\
1 & 1 & 1 \\
1 & 1 & 1
\end{matrix}\right) \, .
\label{newmnu}\ee
This texture for $m^\nu$ is completely uncorrelated with the values of $\epsilon^\ell_i/\epsilon^\ell_j$. Its entries are all automatically of the same order except for $m_{11}^\nu$ which, 
after rotation to the field basis in which the charged lepton mass matrix is diagonal, is suppressed by $m_e/m_\mu$. {Note that because the matrix $\tilde\epsilon$ is rank two, so is $m^\nu$.} Therefore one eigenvalue is vanishing and strict normal hierarchy of the neutrino spectrum is predicted, together with a strong suppression
of neutrinoless $2\beta$-decay. We remark that the assumption of single-spurion dominance implies that the order one numbers in \eq{newmnu} depend exclusively on the three complex, order-one coefficients of $\epsilon^\ell_{ai}$, 
shown in \eq{PC3}, with no dependence on strong-sector order-one coefficients: this makes single-spurion scenarios particularly predictive, {in particular the values of CP-violating phases are correlated with neutrino masses and mixing angles.}
One can check that, within the class 2M, this is the only viable possibility to fit neutrino data with only a single non-zero entry in the spurion matrix $\tilde\epsilon_{ab}$.

Moving to class 1M, the very same rank-two texture as in \eq{newmnu} happens to be realized by a non-zero spurion $\tilde\epsilon_{a=2\, i} \sim (0,-1,0)$.
In this case the three complex, order-one coefficients that fully determine the matrix texture are the one in $\epsilon^\ell_{23}$ and the two independent ones in $\tilde\epsilon_{2i}$.
A second viable 1M scenario is provided by $\tilde\epsilon_{a=1\, i} \sim (-1,0,0)$, that induces a neutrino mass texture with {\it all} entries of the same order. Also in this scenario $m_\nu$ has rank two, and the vanishing eigenvalue may correspond to either normal or inverted hierarchy of the neutrino spectrum.
Inverted hierarchy is preferred, because in this case all the entries of $m_\nu$ can have the same size within a factor of two,
while normal hierarchy requires entries different by a factor of five or so \cite{Frigerio:2002rd,Frigerio:2002fb}.\\

We end this section observing that, as neutrino mixing breaks explicitly $U(1)^3$, one may wonder whether the analysis of charged lepton flavor violation in section \ref{sec:u3imp}, that relied on the existence of {such (approximate)}
symmetry, still holds. To quantify how accurate the $U(1)^3$ symmetry needed to be, recall that main achievement was a significant suppression of flavor and CP violation in charged leptons, by a factor as small as $m_e/m_\mu\sim1/200$ in the case of $d_e$, see \eq{de}.
Thus, in order to be consistent with data when $m_*\sim10$ TeV, it is sufficient that the $U(1)^3$ symmetry of the strong sector be respected at the percent to permille level. 
Now, the naive expectation is that in a concrete model of neutrino masses a typical source of $U(1)^3$ breaking is controlled by powers of $m^\nu/m_*$, and therefore it can be neglected, very much like in the SM. 
This expectation holds generically in all neutrino mass models we consider in section~\ref{NeutrinoSection}, with the notable exception of scenarios based on the lepton-breaking perturbation $\tilde\lambda_{ij\tilde a}\ell_i\ell_j O_{\tilde a}$. The peculiarity of this model is that $O_{\tilde a}$ is necessarily a scalar 
of dimension not far from 2, so that $|O_{\tilde a}|^2$ is nearly marginal. From this follows that the theory might contain an additional, nearly marginal source of $U(1)^3$ breaking controlled by 
\ba
\tilde c_{{\tilde a}{\tilde b}}O_{\tilde a}O^*_{\tilde b}~, 
\ea 
where $\tilde c_{ab}\gtrsim\tilde\lambda_{ij\tilde a}\tilde\lambda_{ij\tilde b}^*/{16\pi^2}$ (of course no $U(1)^3$ breaking would be present if $O_{\tilde a}$ is a singlet). In none of the other models of section~\ref{NeutrinoSection} there {\emph{necessarily}} exists a nearly marginal operator within the strong sector that could potentially give rise to a sizable violation of $U(1)^3$ from loops of elementary leptons. The conservative condition $|\tilde c_{ab}|\lesssim0.1-1\%$, identified above, translates into $|\tilde\lambda_{ij\tilde a}|\lesssim0.1-1$. As a result, the
$U(1)^3$ symmetry invoked in our study of the charged-lepton flavor and CP violation is accurate in virtually all realistic scenarios for neutrino masses.

\subsection{Multiple flavor scales}\label{sec:multi}
%%%%%%%%%%%%%%%%%%%%%%%%%%%%%%%%%%%%%%

In this section we discuss the possibility that the strong sector gives rise to flavor-dependent scales, $m_{a*}^{\psi}$, for the various composite operators $O_a^\psi$ in (\ref{PC1}), see \cite{Vecchi:2012fv}\cite{Panico:2016ull}. For simplicity we will consider $m_{a*}^\ell=m_{a*}^e$, a scale which we simply denote by $m_{a*}$, though a similar analysis can be extended to the general case {(see also the discussion at the end of sections~\ref{sec:neutrMulti} and \ref{anoB})}. We take 
\ba\label{assDyn}
m_{a*} > m_*~. 
\ea
This implies that below $m_{a*}$ the operator $O_a$ disappears from the CFT (or, as we will loosely denote in the following,  it ``decouples''), and we have to match two different descriptions, the one at scales $\mu>m_{a*}$ with the one at scales $\mu<m_{a*}$.~\footnote{We implicitly assume that both theories are approximate CFTs.}

The correct degrees of freedom to describe physics at scales above $m_*$ are the fundamental fermions $\psi$ and the operators $O$ of the strong sector. Recall in particular that, if the strong sector is to provide a natural explanation of the electroweak hierarchy or, in other words, to be free of unnatural tunings, then at scales well above $m_*$ (those relevant to our matching procedure) there should be no weakly-coupled scalar, and in particular no weakly-coupled Higgs field. Yet the UV description could certainly contain a Higgs operator $O_H$ with the same quantum numbers of the Higgs doublet but, being not weakly-coupled, having scaling dimension $d_H>1$. Such a picture does not suffer from a hierarchy problem if $d_H\gtrsim2$, so that $|O_H|^2$ is irrelevant. At scales above $m_*$ we should therefore describe the strong dynamics in terms of CFT operators $O$; it is only at energies of order $m_*$ that the weakly-coupled Higgs doublet, and possibly other resonances of the strong dynamics, emerge. At this last stage all operators with the appropriate quantum numbers can interpolate the Higgs, for instance $O_H \to g_* m_*^{d_H -1} H$.

We are now ready to present our matching procedure. Schematically:
\be\label{decO}
\delta{\cal L}_{\mu>m_{a*}}=\sum_b \lambda^\psi_{bi}\overline{O}^\psi_b{\psi_i} \quad \stackrel{m_{a*}}{\longrightarrow} \quad 
\delta{\cal L}_{\mu<m_{a*}}=\sum_{b \neq a} \lambda^\psi_{bi}\overline{O}^\psi_b{\psi_i}  + \Delta \mathcal{L}_a.
\ee
Here $\Delta \mathcal{L}_a$ contains all terms allowed by symmetries written in terms of the CFT operators $O$ that remain in the EFT, including $O_{b\neq a}$, $O_H$, and possibly others (to be discussed below). A naive dimensional analysis estimate gives:
\be
\label{matchO}
\Delta \mathcal{L}_a = 
\frac{(m_{a*})^4}{g_*^2} \hat{\mathcal{L}}^a_0  \left(\frac{O}{(m_{a*})^{d_{O}}},\epsilon^\psi_{bi}\frac{g_*\psi_i}{(m_{a*})^{3/2}},\frac{D_\mu}{m_{a*}}\right) + {\cal O}\left(\frac{g_*^2}{16\pi^2}\right) \, .
\ee
All parameters, including $\epsilon^\psi$, are evaluated at the scale $m_{a*}$, and we have assumed a universal, $a$-independent $g_*$ coupling for simplicity. The effect of the $O$'s is weighted by the corresponding scaling dimension $d_O$. One proceeds similarly for all mass thresholds $m_{a*}$. 

Without loss of generality it is convenient to label the composite operators in such a way that $O_1^\psi$ decouples before $O_2^\psi$, that decouples before $O_3^\psi$, that is 
\ba\label{assDyn1}
m_{3*}^\psi \ll m_{2*}^\psi \ll m_{1*}^\psi~, 
\ea
irrespective of their anomalous dimensions.
With this convention, and keeping the same triangular form of (\ref{PC3}), one finds that operators containing $\psi_i$ will only be generated at scales $m_{a*}^{\psi}$ with $a \leqslant i$. Furthermore, given a large separation among the $m_{a*}$, at low energies the most relevant operators containing $\psi_i$ will be those generated at $m_{a*}^{\psi}$ with $a = i$.

As a first concrete application of (\ref{decO}) and (\ref{matchO}) we identify the form of the charged lepton Yukawa couplings that arise within this scenario. First, observe that below $m_{a*}$ the combination of fundamental fermions mixed with $O_a$, i.e.~$\lambda_{ai}^\psi \psi_i$, no longer couples \emph{linearly} to the remaining CFT. Its dominant interaction to the strong sector at lower scales is assumed to be mediated by %$c_{ab}(\lambda_{ai}^\psi)^*\lambda_{bi}^{\psi'}\bar\psi O_H\psi'$
$c_{ab}(\lambda_{ai}^\psi)^*\lambda_{bj}^{\psi'}\bar\psi_i O_H\psi'_j$, with $O_H$ a composite operator with the Higgs quantum numbers and $\psi'$ another fundamental fermion with appropriate charges. The Yukawa operators $\bar\psi O_H\psi'$ have scaling dimension $d_H + 3$, causing a suppression of order $(\mu/m_{a*})^{d_H-1}$ of the Yukawa couplings evaluated at a lower scale $\mu$.

Technically speaking, the appearance of %$c_{ab}(\lambda_{ai}^\psi)^*\lambda_{bi}^{\psi'}\bar\psi O_H\psi'$
$c_{ab}(\lambda_{ai}^\psi)^*\lambda_{bj}^{\psi'}\bar\psi_i O_H\psi'_j$ within the EFT (\ref{matchO}) relies on the existence of a CFT 3-point function involving $O_a^{\psi}, O_b^{\psi'}, O_H$. However, in generic scenarios such a correlator with $a\neq b$ would be hard to reconcile with (\ref{assDyn1}): if $O_1$ has unsuppressed interactions with $O_3$ it is not a priori clear why the hypothesis $m_{3*}\ll m_{1*}$ is radiatively stable. To ensure these hierarchies are stable we will assume that the strong sector enjoys a $U(1)^3$ symmetry distinguishing the three composite flavors $a=1,2,3$. We will see that this operational assumption will lead to phenomenological implications analogous to those of section~\ref{sec:u1cube}. The main modification is in the order one coefficients.

We now have all the ingredients necessary to determine the charged lepton Yukawa couplings. These take the same form as in (\ref{Yukawa}), $y^e_{ij}= g_* \epsilon^{\ell *}_{ai} \epsilon^e_{bj} c^e_{ab}$, the presence of flavored scales being encoded in a non-trivial structure of the $U(1)^3$-invariant coefficients,
\ba\label{cdyn}
c^e_{ab} \!\!\! &=& \!\!\! \left( \frac{m_*}{m_{1*}} \right)^{d_H-1} {\cal O}\left(\!\!
\begin{array}{ccc}
1 & 0 & 0 \\ 
0 & 1 & 0 \\ 
0 & 0 & 1
\end{array}
\!\!\right)
+  \left( \frac{m_*}{m_{2*}} \right)^{d_H-1} {\cal O}\left(\!\!
\begin{array}{ccc}
0 & 0 & 0 \\ 
0 & 1 & 0 \\ 
0 & 0 & 1
\end{array} \!\!\right)
+ \left(\frac{m_*}{m_{3*}} \right)^{d_H-1} {\cal O}\left(\!\! 
\begin{array}{ccc}
0 & 0 & 0 \\ 
0 & 0 & 0 \\ 
0 & 0 & 1
\end{array}
\!\! \right) \no \\
\!\!\! &\simeq& \!\!\!
c_a \left( \frac{m_*}{m_{a*}} \right)^{d_H-1} \delta_{ab}~,
\ea
where the second line follows from (\ref{assDyn1}) and $c_a$ are complex numbers of order unity. The Yukawa matrix is well approximated by
\be\label{yedyn}
y^e \sim \frac{\sqrt{2}m_3}{v}
\left(
\begin{array}{ccc}
\frac{m_1}{m_3} & \frac{m_1}{m_3} & \frac{m_1}{m_3}\\ 
\frac{m_1}{m_3} & \frac{m_2}{m_3} & \frac{m_2}{m_3}\\ 
\frac{m_1}{m_3} & \frac{m_2}{m_3} & 1
\end{array}
\right)~,
\ee
where
\be\label{mdyn}
m_i \sim g_* \epsilon_i^\ell \epsilon_i^e \left( \frac{m_*}{m_{i*}} \right)^{d_H-1} \frac{v}{\sqrt{2}}~,
\ee
and we assumed the ordering $m_1\ll m_2\ll m_3$ (similarly to (\ref{123})) so the terms that have been neglected in (\ref{yedyn}) are automatically of order $m_i/m_{j>i}$.
The structure (\ref{yedyn}) is the same as in the scenario with no hierarchical scales and $U(1)^3$ symmetry.~\footnote{In the absence of a $U(1)^3$ symmetry the strong-sector coefficients would read
\ba\label{cdyngen}\no
c^e_{ab} \!\!\! &=& \!\!\! \left( \frac{m_*}{m_{1*}} \right)^{d_H-1} {\cal O}\left(\!\!
\begin{array}{ccc}
1 & 1 & 1 \\ 
1 & 1 & 1 \\ 
1 & 1 & 1
\end{array}
\!\!\right)
+  \left( \frac{m_*}{m_{2*}} \right)^{d_H-1} {\cal O}\left(\!\!
\begin{array}{ccc}
0 & 0 & 0 \\ 
0 & 1 & 1 \\ 
0 & 1 & 1
\end{array} \!\!\right)
+ \left(\frac{m_*}{m_{3*}} \right)^{d_H-1} {\cal O}\left(\!\! 
\begin{array}{ccc}
0 & 0 & 0 \\ 
0 & 0 & 0 \\ 
0 & 0 & 1
\end{array}
\!\! \right) \no \\
\!\!\! &\simeq& \!\!\!\no
 \left( \frac{m_*}{m_{\mathrm{min\mathrm{}}(a,b)*}} \right)^{d_H-1} c_{ab}~.
\ea
In this case the Yukawa matrix of charged leptons reads the same as \eq{yedyn}, as long as the $\epsilon^{\ell,e}_i$ do not exhibit large hierarchies, i.e.~$\epsilon^{\ell,e}_1 \sim \epsilon^{\ell,e}_2 \sim \epsilon^{\ell,e}_3$, which is equivalent to the statement that the hierarchies in the lepton masses are completely controlled by the hierarchies in $m_{a*}$. In fact, as long as this last statement holds, the flavor structure of this scenario is effectively that corresponding to a $U(1)^3$-symmetric multi-scale scenario of~(\ref{cdyn}), the phenomenological consequences being the same.}

The hierarchies in the masses of charged leptons can be entirely reproduced by the hierarchies in $m_{a*}$. In other words, in this multi-scale scenario there is no need for hierarchical $\epsilon_i$. For instance, taking $\epsilon^{\ell,e}_i = 1$ for all flavors, $g_* = 4 \pi$, $m_* = 10$ TeV, and considering for definiteness a value $d_H = 2$, a realistic pattern of lepton masses can be achieved with
\be\label{dynscales}
m_{1*} \sim \, 5 \times 10^{10} \, \mathrm{GeV}, \quad m_{2*} \sim \, 2 \times 10^{8} \, \mathrm{GeV}, \quad m_{3*} \sim \, 1 \times 10^{7} \, \mathrm{GeV}~.
\ee
For $\epsilon^{\ell}_i \epsilon^{e}_i < 1$ and/or $d_H > 2$, the scales $m_{a*}$ at which the Yukawa operators are generated are necessarily lower than in (\ref{dynscales}).

\subsubsection{Flavor and CP violation}
\label{sec:multi-impl}
%%%%%%%%%%%%%%%%%%%%%%%%%%%%%%%%%%%%%%

The virtue of the scenario with flavored scales is that it leads to a natural suppression of flavor and CP-violating transitions.
The operators responsible for such processes are generated at scales $m_{a*}$ significantly above $m_*$. This gives rise to suppressing factors $(m_*/m_{a*})^\alpha$ with respect to both the anarchic and $U(1)^3$ scenarios discussed in the previous sections. 
Obviously, since the Yukawa operators are also generated at high scales, they are suppressed as well. Nevertheless, the important question is what is the relative suppression of the latter compared to the former.

Let us then see how this works for the operators controlling the most important deviations from the SM. First, the dipole operators $Q_{eW, eB}$ are here built in terms of $\psi$'s and $O_H$, rather than the Higgs doublet, and have dimension $d_H+5$. 
For the remaining operators a few more comments are necessary. $Q_{eH}$ at high scales can be obtained by replacing $(H^\dagger H) H \to O_H$, but in so doing one would get back the proto-Yukawa $\bar \ell O_H e$.~\footnote{An equivalent way to see this is to observe that $O_H$ interpolates both $H$ and $(H^\dagger H)H$ at scales of order $m_*$.} In that case the flavor structure of the Wilson coefficient would be aligned with the SM Yukawa and not mediate new effects. A more interesting option is to postulate there is another scalar operator $O_H'$ with the quantum numbers of the Higgs, and that it is this that governs $Q_{eH}$. Then the scaling dimension of $Q_{eH}$ above $m_*$ would be $d_H'+3$. A minimal option is to take $O_H'=O_HO_H^\dagger O_H$ so that $d_H'\sim3d_H$. We will consider this possibility in the following and declare $Q_{eH}$ has dimension $3(d_H+1)$. Regarding $Q_{H\ell, He}$ we note that a vector CFT operator is needed. By unitarity this has dimension $\geq3$, so $Q_{H\ell, He}$ must have dimension $\geq6$. To be conservative we will assume the scaling dimension is minimal. Finally, $Q_{\ell \ell, \ell e, ee}$ are made up of 4 fundamental fermions and always have dimension 6, irrespective of the CFT dynamics.

The overall suppression of a given operator will be determined by its scaling dimension, while the flavor structure of such a suppression will depend on the scale at which the leading contribution is generated. Given the $U(1)^3$ symmetry assumed to be present in this scenario (see the discussion above (\ref{cdyn})), we can make use of the results presented in section~\ref{sec:u3imp} with the novelty that now the order one coefficients in (\ref{stru}) are hierarchical:
\be\begin{array}{ll}
Q^{ij}_{eW,eB}: & c_a \sim ( m_*/m_{a*} )^{d_H+1}~,\\
Q^{ij}_{eH}: & c_a \sim ( m_*/m_{a*})^{3d_H-1}~,\\
Q^{ij}_{H\ell,He}: & c_a \sim ( m_*/m_{a*})^{2}~,\\
Q_{\ell e, \ell\ell, ee}^{ijmn}: & c_{ab} \sim ( m_*/m_{\mathrm{min}(a,b)*})^{2}~.
\end{array}
\label{strudyn}
\ee
The Wilson coefficients of the dim-6 operators in Table (\ref{ops}), in the gauge basis, are obtained by dressing the strong-dynamics coefficients (\ref{strudyn}) with the mixings $\lambda^\psi$ in (\ref{PC1}), renormalized at the appropriate flavor scale. Then one can verify that the relative suppression of these operators compared to the anarchic scenario with one scale $m_*$ is of order $(m_*/m_{a*})^2$ for $Q^{ij}_{eW,eB}$, $(m_*/m_{a*})^{2d_H}$ for $Q^{ij}_{eH}$, $(m_*/m_{a*})^{3-d_H}$ for $Q^{ij}_{H\ell,He}$ and finally $(m_*/m_{a*})^{4-2d_H}$ for $Q^{ij}_{\ell e,\ell\ell,ee}$ \cite{Vecchi:2012fv}. 
In the limit $d_H \to 1$ of a fundamental Higgs all operators are effectively dimension 6 and one completely decouples the flavor and CP problems as in the SM. In a composite Higgs picture, as long as $d_H < 3$, all flavor-violating operators other than the 4-fermion ones will be relatively suppressed at low energies compared to the standard anarchic scenario. $Q_{\ell e,\ell\ell,ee}$ on the other hand are suppressed only if $d_H<2$. For definiteness our benchmark point will be $d_H=2$, that is also close to the minimal dimension compatible with a solution of the hierarchy problem. 

To identify the largest contribution to a given dim-6 operator in the mass basis two factors are important, first that going to the mass basis introduces mixing angles of order $m_i/m_{j>i}$, following (\ref{uniLR}), and second that the relative size of the operators generated at different scales is controlled by $m_{j*}/m_{i*} \sim (m_i/m_j)^{{1}/({d_H-1})}$ and by the ratios of $\epsilon_i^\psi$, according to (\ref{mdyn}). 
Rather than showing the general expressions for the Wilson coefficients we decide to focus on the interesting case $\epsilon_{i}^{\ell,e}\sim\epsilon_{j}^{\ell,e}$ (and $d_H = 2$) in which the lepton mass hierarchies are entirely given by $m_{*}/m_{i*}$. {In this case the coefficients would be the same even if we did not assume the $U(1)^3$ symmetry.}

Explicit expression for all the Wilson coefficients in the mass basis are shown in table~\ref{Wilson4} for this particular case. We find that the leading contribution to $Q_{eW, eB}$ and $Q_{eH}$ is generated at the lowest scale $m_{3*}$, implying the suppressions parametrized by the factors of $m_3/g_*v$. {The reason for this can be traced back to the power of $m_*/m_{a*}$ in~(\ref{strudyn}) being larger than two. On the contrary, the other operators are suppressed by $(m_*/m_{a*})^2$ and the leading contribution to their Wilson coefficients depends on their flavor.} The inverse proportionality of the coefficients on $\epsilon^{\ell, e}$ is just a consequence of the fact that the smaller the degree of compositeness, the lower the flavor scales $m_{i*}$ need to be to reproduce the lepton masses. 

%%%%%%%%%%%%%%%%%%%%%%%%%%%%%%%%%%%%%%%%%%%
\begin{table}[tb]
\begin{center}
\begin{tabular}{l|l} 
$m_{a*}$  &  $\epsilon_{i}^{\ell,e}\sim\epsilon_{j}^{\ell,e} \, , \, d_H = 2$ \\
\hline
$\frac{C^{eW}_{ij}}{\Lambda^2}$ 
& $\frac{g_*^2}{16 \pi^2} \frac{g_*}{m_*^2} \frac{1}{(\epsilon^\ell \epsilon^e)^2}  \frac{2 \sqrt{2} \, m_i m_j m_3}{g_*^3 v^3}  g\, \hat c^{eW}_{ij}$ \\
$\frac{C^{eB}_{ij}}{\Lambda^2}$ 
& $\frac{g_*^2}{16 \pi^2} \frac{g_*}{m_*^2} \frac{1}{(\epsilon^\ell \epsilon^e)^2}  \frac{2 \sqrt{2} \, m_i m_j m_3}{g_*^3 v^3} g'\, \hat c^{eB}_{ij}$ \\
$\frac{C^{eH}_{ij}}{\Lambda^2}$ 
& $\frac{g^3_*}{m_*^2}  \frac{1}{(\epsilon^\ell \epsilon^e)^4}  \frac{4 \sqrt{2} \, m_i m_j m_3^3}{g_*^5 v^5}   \hat c^{eH}_{ij}$ \\
$\frac{C^{H \ell(1,3)}_{ij}}{\Lambda^2}$ 
& $\frac{g^2_*}{m_*^2} \frac{1}{(\epsilon^e)^2} \frac{2 \, m_i m_j}{g_*^2v^2}  \hat c^{H \ell(1,3)}_{ij}$ \\
$\frac{C^{H e}_{ij}}{\Lambda^2}$ 
& $\frac{g^2_*}{m_*^2} \frac{1}{(\epsilon^\ell)^2} \frac{2 \, m_i m_j}{g_*^2v^2}  \hat c^{H e}_{ij}$ \\
$\frac{C^{\ell e}_{ijmn}}{\Lambda^2}$ 
& $\frac{g^2_*}{m_*^2} \frac{2 \, {\rm min}(m_i,m_j,m_m,m_n) {\rm min}'(m_i,m_j,m_m,m_n) }{g^2_*v^2} \hat c^{\ell e}_{ijmn}$ \\
$\frac{C^{\ell \ell}_{ijmn}}{\Lambda^2} $ & $\frac{g^2_*}{m_*^2} \left(\frac{\epsilon^\ell}{\epsilon^e}\right)^2 \frac{2 \, {\rm min}(m_i,m_j,m_m,m_n) {\rm min}'(m_i,m_j,m_m,m_n) }{g^2_*v^2} \hat c^{\ell \ell}_{ijmn}$ \\
$\frac{C^{e e}_{ijmn}}{\Lambda^2}$ & $\frac{g^2_*}{m_*^2} \left(\frac{\epsilon^e}{\epsilon^\ell}\right)^2 \frac{2 \, {\rm min}(m_i,m_j,m_m,m_n) {\rm min}'(m_i,m_j,m_m,m_n) }{g^2_*v^2}  \hat c^{e e}_{ijmn}$ \\
\hline
\end{tabular}
\end{center}
\caption{\small Here we present the Wilson coefficients of the dim-6 operators in the mass basis, assuming the strong dynamics gives rise to flavor dependent dynamical scales $m_{a*}$. The results shown are for $\epsilon_{i}^{\ell,e}\sim\epsilon_{j}^{\ell,e}$ and $d_H = 2$, therefore they hold whether a $U(1)^3$ symmetry is assumed or not. We defined ${\rm min}(\{\})$ as usual whereas ${\rm min}'(\{\})$ as the operation of identifying the {\emph{next to smallest}} element, e.g.~{${\rm min}(m_e,m_\tau,m_\mu,m_\mu)=m_e$ and ${\rm min}'(m_e,m_\tau,m_\mu,m_\mu)=m_\mu$}.}
\label{Wilson4}
\end{table}
%%%%%%%%%%%%%%%%%%%%%%%%%%%%%%%%%%%%%%%%%%%

The comparison between the present $U(1)^3$-symmetric multi-scale scenario and the one with a single scale, see table~\ref{Wilson3}, shows that, for $d_H = 2$ and $\epsilon^{\ell,e} = 1$, the presence of dynamical scales is always more effective in suppressing flavor and CP-violating transitions (except for some 4-lepton operators whose coefficients are equal, up to order one factors). Let us discuss specifically the predictions for $\mu \to e \gamma$ and the electron EDM, which lead to the most stringent constraints in generic scenarios. For the former the dependence on $g_*$ cancels out and we get
\be
\label{mutoedyn}
\frac{C^{e\gamma}_{12,21}}{\Lambda^2}\simeq \frac{1}{16\pi^2} \frac{e}{m_*^2} \frac{2 \sqrt{2}m_e m_\mu m_\tau}{v^3} \frac{\hat c^{e\gamma}_{12,21}}{(\epsilon^\ell \epsilon^e)^2}.
\ee
For $m_* = 10$~TeV, $\epsilon^\ell \epsilon^e = 1$, $d_H=2$, and a natural $\hat c^{e\gamma}_{12,21} = 1$, the result is more than six orders of magnitude below the current experimental bound. This is a factor $m_\mu m_\tau/g_*^2 v^2$ smaller than in the $U(1)^3$ single-scale scenario, where the prediction for $\mu \to e \gamma$ was barely compatible with data.
The constraint associated with (\ref{mutoedyn}) can also be read as a lower bound on $\epsilon^\ell \epsilon^e$ and thus on $m_{3*}$, the scale where the leading contribution to $Q_{e \gamma}$ is generated, resulting in $m_{3*} \gtrsim 15 (g_*/4\pi)$~TeV, just slightly above $m_*$ for maximal $g_*$.
The bound on the electron EDM is easily satisfied as well due to the extra suppression, of order $m_\mu m_\tau/g_*^2 v^2$ once again, with respect to the $U(1)^3 \times CP$ scenario,
\ba\label{deDYN}
d_e \simeq \frac{1}{8\pi^2} \frac{e}{m_*^2} \frac{2 \sqrt{2} m_e^2 m_\tau}{v^3} \frac{\hat c^{e\gamma}_{11}}{(\epsilon^\ell \epsilon^e)^2}.
\ea
Note there is no need to invoke CP-invariance of the strong sector here. The very presence of multiple scales is enough.

To conclude this section, we emphasize that in the present multi-scale scenario the electron EDM receives, on top of the tree-level effects discussed here, an additional loop-level contribution generated at the scale $m_*$ \cite{Panico:2016ull}. This can be seen to arise from {one}-loop diagrams involving flavor-independent CP-violating operators suppressed by $m_*$ (in particular affecting the Higgs coupling to photons). These turn out to give the largest contribution to $d_e$ and keeping them under control requires $m_* \gtrsim$~10 TeV \cite{Panico:2017vlk}.

\subsubsection{Neutrino masses}
\label{sec:neutrMulti}
%%%%%%%%%%%%%%%%%%%%%%%%%%%%%%%%%%%%%%

The structure of the Majorana mass matrix (\ref{Weinberg}) depends, as in the scenarios discussed above, on the properties of the lepton-breaking perturbation $\tilde \lambda$. In this regard, note that we should still assume the strong sector has at least an approximately conserved lepton number $U(1)_c$, since otherwise neutrino masses would be unacceptably large. To see this explicitly, note that the Weinberg operator (\ref{Weinberg}) would receive a minimum contribution from a term $\ell \ell O_H O_H$, generated along with the proto-Yukawa of the tau at the scale $m_{3*}^\ell$,
\ba\no
m^\nu\sim(\epsilon^\ell_3)^2\frac{(g_*v)^2}{m_*}\left(\frac{m_*}{m_{3*}^\ell}\right)^{2d_H-1}
\gtrsim  \frac{2m_\tau^2}{m^\ell_{3*}} \gtrsim
\frac{2m_\tau^2}{m_*} \left(\!\frac{\sqrt{2}m_\tau}{g_*v}\!\right)^{\frac{1}{d_H-1}}
\gtrsim 500~{\rm eV}~,
\ea
where in the first and second inequalities we made use of (\ref{mdyn}) and $\epsilon_3^{\ell,e} \le 1$, while in the last inequality we used $d_H \geq 2$, $g_*=4\pi$ and $m_*=10$ TeV, that correspond to
$m_{3*}^\ell \sim 10^7$~GeV, see (\ref{dynscales}).
Such a neutrino mass scale is at least three orders of magnitude too large. 

Having stablished the necessity of a composite lepton number,
the parametric dependence of the neutrino mass matrix is still provided by three classes 0M, 1M, 2M as in (\ref{Mclass}), though now the strong sector coefficients are not anarchic. Here we find
\ba\label{Mclassdyn}
{\rm 2M}:&&m^\nu_{ij} = \epsilon^\ell_{ai}\epsilon^\ell_{bj}\tilde\epsilon_{ab} ~\frac{(g_*v)^2}{m_*} \, , \,\,\, \tilde\epsilon_{ab} = \tilde \epsilon c_{ab}^\nu \, , \quad 
c_{ab}^\nu \simeq \delta_{ab}c_{a} \left( \frac{m_*}{m^\ell_{a*}} \right)^{d_T -1} \, ,  \\\no
{\rm 1M}:&&m^\nu_{ij}= \left[\epsilon^\ell_{ai}\tilde\epsilon_{aj}+ \epsilon^\ell_{aj}\tilde\epsilon_{ai} \right]~\frac{(g_*v)^2}{m_*}\, , \,\,\, 
\tilde \epsilon_{ai} = \tilde \epsilon_{i} c_{a i}^\nu \, , \quad c_{ai}^\nu \simeq c_{ai} \left( \frac{m_*}{m_{a*}^\ell} \right)^{d_T -1} \, ,\\\no
{\rm 0M}:&&m^\nu_{ij}= \tilde\epsilon_{ij} \,\frac{(g_*v)^2}{m_*} ,
\ea
where the last equation in each class shows the expectation for the $U(1)^3$-invariant strong sector coefficients, with 
$c_{a},c_{ai}$ complex numbers of order unity.~\footnote{If we relax the assumption of $U(1)^3$ the only difference with respect to (\ref{Mclassdyn}) is in class 2M, where we now have $c_{ab}^\nu \simeq c_{ab} ( m_*/m^\ell_{\mathrm{min}(a,b)*})^{d_T -1} $.}
Here $d_T$ is the scaling dimension of a triplet scalar operator, $O_T$, which at $m_*$ interpolates to $HH$, that is $O_T \to g_*^2 m_*^{d_T -2} HH$. At high energy scales, $m_{a*}^\ell$, the operator $\ell \ell O_T$ is generated, finally interpolating the Weinberg operator (\ref{Weinberg}) at $m_*$.
The minimal example would be $O_T = O_H O_H$, in which case $d_T \simeq 2 d_H$. Other possibilities are well-motivated as well, for instance that the CFT contains $O_T$ as an independent scalar operator, as long as (in analogy with $O_H$) the dimension of the singlet operator $|O_T|^2$ is $\gtrsim 4$.

For class 0M, the $U(1)_L$-breaking deformation $\tilde \epsilon_{ij}$ is associated with couplings of $\ell_i$ to the strong sector that are independent  from the multiple flavor scales, therefore the neutrino mass texture is not affected compared to scenarios with a single mass scale.

Because of $U(1)^3$, neutrino masses in classes 1M and 2M depend on which lepton-breaking perturbation (\ref{Mclasscharges}) is turned on. The viable neutrino textures turn out to be the same as those discussed in section \ref{U13nu}: the spurion $\tilde \epsilon \sim (-1, -1, 0)$ in class 2M, and either $\tilde \epsilon \sim (-1, 0, 0)$ or $\tilde \epsilon \sim (0,-1, 0)$ in class 1M can fit neutrino data. The only difference in the multi-scale scenario is in the neutrino mass scale.
The 2M neutrino mass matrix in \eq{newmnu} is here further suppressed by a factor $(m_*/m^\ell_{1*})^{d_T-1} \sim (m_1/\epsilon_1^\ell \epsilon_1^e)^\Delta$, where $\Delta = \frac{d_T-1}{d_H-1}$ and we used the relation (\ref{mdyn}) with $m^\ell_{a*} = m^e_{a*}$. Similarly, the overall size of the neutrino matrix in the two viable models in class 1M is determined by the fact that now $\tilde \epsilon_{a=1\, i} \sim (m_*/m^\ell_{1*})^{d_T-1}$ and $\tilde \epsilon_{a=2\, i} \sim (m_*/m^\ell_{2*})^{d_T-1} \sim (m_2/\epsilon_2^\ell \epsilon_2^e)^\Delta$.

If one drops the assumption of $U(1)^3$ symmetry, there are important consequences for the neutrino mass matrix in classes 1M and 2M. Let us focus our discussion on the former, the conclusions for the latter being qualitatively the same. 
In class 1M the structure of $m_\nu$ is determined by the fact that below the scale $m_{a*}$, where the operator $O_a^{\ell}$ decouples, the strong sector remains linearly coupled to the $\ell$'s via the lepton-breaking perturbation. Recalling that in the absence of $U(1)^3$ the $\tilde \epsilon$ matrix is anarchic, the resulting neutrino mass matrix is
\be
\label{mnudyn1M}
m^\nu \propto  \left(\!\!
\begin{array}{ccc}
m_1^{\Delta}  & m_2^{\Delta} & m_3^{\Delta} \\ 
m_2^{\Delta} & m_2^{\Delta} & m_3^{\Delta} \\ 
m_3^{\Delta} & m_3^{\Delta} & m_3^{\Delta}
\end{array}
\!\!\right)~,
\ee
where we chose $\epsilon_i^{\ell,e} \sim \epsilon_j^{\ell,e}$, such that the hierarchy of flavor scales completely determines the charged lepton masses, according to (\ref{mdyn}).
The texture (\ref{mnudyn1M}) exhibits a hierarchical structure, at odds with the observed neutrino mass anarchy. Similar hierarchies are obtained in class 2M, as well as
in models with $U(1)^3$-symmetric strong sector, but several comparable $U(1)_L$-violating spurions $\tilde \epsilon$ at the relevant scales,
see the discussion below (\ref{Mclasscharges}).

Finally, let us comment on one interesting multi-scale scenario where the PMNS matrix is reproduced, still in classes 1M and 2M and without $U(1)^3$ symmetry, relaxing the assumption that both operators $O^\ell_a$ and $O^e_a$ decouple at the same scale. {This case is the multi-scale analog of scenarios with $\epsilon^\ell_i\sim\epsilon^\ell_j$, where the charged lepton masses are determined by the $\epsilon^e_i$'s.} We assume all the $O^\ell_a$ (or at least $O^\ell_{2,3}$ if class 1M) decouple at a single scale $m^\ell_* < m_{a*}^e$, that is below any of the scales where $O^e_a$ decouple. The resulting neutrino mass matrix is, for $\epsilon^\ell_i \sim \epsilon^\ell_j$, completely anarchic
\be
\label{mnudyndef}
m^\nu {\propto}\, \tilde \epsilon \left( \frac{m_*}{m_*^\ell} \right)^{d_T-1} {\cal O}\left(\!\!
\begin{array}{ccc}
\zeta & 1 & 1 \\ 
1 & 1 & 1 \\ 
1 & 1 & 1
\end{array}
\!\!\right)~,
\ee
its overall size being determined by $\tilde \epsilon$ and $m^\ell_*$. {These latter quantities are now uncorrelated with the charged lepton masses, that are controlled by $m_{a*}^e$}. In (\ref{mnudyndef}) we introduced the parameter $\zeta$ to differentiate between class 1M with $m^\ell_{1*} \gg m^\ell_{2*,3*} \equiv m^\ell_{*}$, where $\zeta = m_e/m_\mu$, and class 2M or class 1M with a flavor-universal $m^\ell_{*}$, for which $\zeta = 1$.

Remarkably, in the present scenario the results of section~\ref{sec:multi-impl} concerning operators involving right-handed leptons, most importantly the Yukawa couplings and dipole operators, are not affected.
However, operators with only lepton doublets, i.e.~$Q_{H\ell}$ and $Q_{\ell \ell}$, could be significantly enhanced if $m^\ell_*$ is low and $\epsilon^\ell \sim 1$, since their Wilson coefficients are predicted to be $C^{H\ell, \ell \ell}/\Lambda^2 \sim (g_*/m^\ell_*)^2$. Constraints from $\mu \to e$ conversion in nuclei and $\mu \to eee$ then require
\ba
m^\ell_{*} \gtrsim \left\{
\begin{matrix}
\!\!\! 5.7 \times 10^3 \, \mathrm{TeV} \left( \frac{g_* \epsilon^\ell}{4 \pi} \right) && (Q_{H\ell}^{(1),(3)}) \\
2.6 \times 10^3 \, \mathrm{TeV} \left( \frac{g_* \epsilon^\ell \epsilon^\ell}{4 \pi} \right) &&(Q_{\ell\ell})
\end{matrix}\right. ~,
\label{ellbounds}
\ea
while, if in class 1M $m^\ell_{1*} \gg m^\ell_*$, the bounds are relaxed by a factor $\sqrt{m_e/m_\mu}\simeq1/15$.
We should also point out that, if the strong sector enjoys a custodial $SU(2)_L \times SU(2)_R$ symmetry, as required in composite Higgs models, then a $P_{LR}$ parity \cite{Agashe:2006at} can enforce a suppression of the non-standard (flavor-changing) couplings of the $Z$ boson to fermions.
If the composite operators $O^\ell$ coupled to the lepton doublets are such that $P_{LR}$ is respected,
then the coefficients of the operators $Q_{H\ell}^{(1)}$ and $Q_{H\ell}^{(3)}$ satisfy $C^{H\ell(1)} = - C^{H\ell(3)}$ at tree level. Then, \eq{match1} implies that the first bound in (\ref{ellbounds}) is significantly relaxed,
leaving the bound from $Q_{\ell \ell}$ as the only relevant constraint on $m_*^\ell$.
We will invoke this scenario in section \ref{anoB}, to address the anomaly in the $b\to sl^+l^-$ transitions.

%%%%%%%%%%%%%%%%%%%%%%%%%%%%%%%%%%%%%%
\section{Anomalies in semi-leptonic $B$ decays} \label{anoB}
%%%%%%%%%%%%%%%%%%%%%%%%%%%%%%%%%%%%%%

In recent years, a series of experimental results have been showing a coherent pattern of deviations from the SM predictions in semi-leptonic decays of $B$-mesons. These ``flavor anomalies" can be grouped in two sets of observables: deviations in semi-leptonic decays in flavor changing neutral current (FCNC)  \cite{BsphimumuLHCb1,BK*mumuLHCb1,Aaij:2014pli,RKexpt,BsphimumuLHCb2,BK*mumuLHCb2,BK*mumuLHCb2,BK*mumuBelle,RK*expt} and deviations in semi-leptonic decays in flavor changing charged current (FCCC) \cite{RD_BaBar,RD_Belle,RD_LHCb} . In both cases the significance of the departure from the SM surpasses the 4$\sigma$ level. 
Assuming (optimistically) that these effects are coming from new physics, what data is suggesting is a departure from lepton flavor universality (LFU), a key feature of the gauge interactions of the SM. The structure of the violation of LFU hinted by data can be summarised as follows:
\begin{enumerate}
\item an enhanced rate of the decays involving the $\tau$ lepton in $b \to c \tau \nu$ compared to the same transitions involving muons or electrons, whose rates agree with the SM predictions;
\item a destructive interference of the new physics with the SM in processes involving the muon in $b \to s \mu^+ \mu^-$, without significant deviation in similar processes involving electrons;
\item absence of evidence of lepton flavor violating effects (LFV);
\item a good description of data in both the FCCC and FCNC processes is obtained invoking new physics in left currents only (both for quarks and for leptons), in particular at the scale of the bottom mass we can make use of the effective Lagrangian 
\begin{equation}
\label{LeffRDRK}
\mathcal{L}_{\rm eff} \supset 
- \frac{1}{\Lambda^2_{CC}}\, \overline{c}_L \gamma^{\mu} b_L \overline{\tau}_L \gamma_{\mu} \nu_L 
+ \frac{1}{\Lambda^2_{NC}} \overline{s}_L  \gamma^{\mu} b_L \overline{\mu}_L \gamma_{\mu} \mu_L 
+ \textrm{h.c.} \, ,   
\end{equation}
with a best fit value for the charged and neutral current given by $\Lambda_{CC}= 2.4 \textrm{ TeV} $ and $\Lambda_{NC}= 31 \textrm{ TeV}$ respectively, see for example \cite{DiLuzio:2017chi}.  
\end{enumerate}
We observe that the new physics effects required to explain the FCCC anomalies have to be very large, as hinted by the low scale $\Lambda_{CC}=2.4$ TeV. This is in part due to fact that the anomalies are observed in decay channels where the SM contributes at tree level. Any explanation of the charged current anomalies beyond the SM has to face a series of stringent constraints coming from other flavor observables like the decay rate of $B \to K^{(*)} \nu \nu$, meson mixing observables, LFV decays of the $\tau$ lepton, possible  modification of the $W$ and $Z$ couplings to the third families of quarks and leptons and also direct resonance searches at the LHC. The explanation of this class of anomalies in PC looks disfavoured and a complete assessment of the viability of potential explanations requires a non-trivial analysis that goes beyond the purpose of this work. We refer to \cite{Barbieri:2017tuq,Marzocca:2018wcf,Azatov:2018knx} for recent discussions on this topic.

We will focus then on the flavor anomalies in $b \to s \ell \ell$ transitions. The higher value $\Lambda_{NC}= 31$ TeV compared to the previous case makes the interpretation in terms of new physics much more feasible (in absolute terms the ratio of amplitudes scales like $\Lambda^2_{NC} / \Lambda^2_{CC} = \mathcal{O} (100)$). The analysis of the semi-leptonic operators, such as those in \eq{LeffRDRK}, inevitably requires a discussion of the quark sector. It is well known that in the context of PC (see e.g.~\cite{KerenZur:2012fr,Bellazzini:2014yua,Panico:2015jxa}) the anarchic scenario cannot remain natural while evading all the bounds coming from indirect searches (in particular those coming from the EDM of the neutron and from mixing observables in the $K$ and charm systems).~\footnote{Interesting attempts to explain the anomalies in the context of compositeness include \cite{Gripaios:2014tna,Niehoff:2015bfa,Megias:2016bde,Cline:2017aed,Carmona:2017fsn}.}
On the other hand the focus of our work is on the lepton sector so as a first step we want to understand the potential physical implications of the anomalies relying as little as possible on the structure of the quark sector. After having identified what are the viable options from a purely leptonic point view, we will sketch at the end of the section a possible realization that is phenomenologically viable also in the quark sector.

The needed operator $\overline{s}_L \gamma^{\mu} b_L \overline{\mu}_L \gamma_{\mu} \mu_L$
matches at the scale $m_{*}$ to $SU(2)_L \times U(1)_Y$ gauge invariant operators of the following form
\begin{eqnarray}
\label{PCsemi-leptonic}
 \mathcal{L}_{eff} \supset  \sum _{ij l k}  \frac{g^2_*}{m_*^2} \left[ c^{ijkl}_{1} (\overline{q}^i \gamma^{\mu} q^j) ( \overline{\ell}^k \gamma_{\mu} \ell^l) + c^{ijkl}_{3}  ( \overline{q}^i \gamma^{\mu} \sigma^a q^j) (\overline{\ell}^k \gamma_{\mu} \sigma^a \ell^l) \right]~.
\end{eqnarray}
The flavor anomaly in $b\to s\mu^+\mu^-$ is reproduced when 
\begin{equation}
\label{PCfitRD}
 \left( \frac{g_*}{4 \pi} \right)^2 \left( \frac{10 \textrm{ TeV}}{m_*} \right)^{2} \left( \frac{c^{2322}_1 + c_3^{2322}}{0.11} \right)^2 \simeq \left( \frac{31 \textrm{ TeV}}{\Lambda_{NC}} \right)^2~,
\end{equation}
where for instance in the anarchic PC case we have $c_i^{2322} \sim \epsilon^q_2\epsilon^q_3 \epsilon^{\ell}_2 \epsilon^{\ell}_2$.

Our strategy in this section is simple: we fix the new physics contribution in the $b \to s \mu^+ \mu^-$ as in (\ref{PCfitRD}) and then derive bounds on the new physics parameters in the lepton sector using processes with the same down-to-strange flavor transitions $(b \to s)$ but different lepton flavor combinations.~\footnote{The assessment of the constraints from other transitions, associated with other semi-leptonic operators, or with purely leptonic or quark operators, requires a comprehensive analysis within a complete flavor scenario, which is beyond the scope of this simple work (see however the proposal at the end of section~\ref{sec:pcanom}).}

As a first step we introduce a normalisation of the new physics effect required to reproduce the best fit value of the anomalies, comparing its strength with those induced by the SM in the same left-currents vector operator
\begin{equation}
k \equiv \frac{C^{\textrm{BSM}}_{b_L \ell_L}}{C^{\textrm{SM}}_{b_L \ell_L}} = -0.15~,
\end{equation}
where the values and the definitions of the Wilson coefficients $C_{b_L \ell_L}$ are taken from \cite{DAmico:2017mtc}. This means that the new physics gives a destructive interference of about $15 \%$ of the SM amplitude.

Operators in \eq{PCsemi-leptonic} induce flavor transitions with different leptons in the final states both with charged particles $b \to s l_i^+ l_j^-$ and with neutrinos $b \to s \overline{\nu}_i  \nu_j$.~\footnote{We denote with $l_i$ the charged lepton contained in the lepton doublet $\ell_i$.} To overcome the model dependence coming from the quark sector we define the following dimensionless ratios of new physics amplitudes induced by the operators in \eq{PCsemi-leptonic}~\footnote{The new physics amplitudes might receive extra contributions from other operators such as $H^{\dagger} \overset{\leftrightarrow}{D}_{\mu} H \, \overline{\ell}^i \gamma^{\mu} \ell^j$ or $H^{\dagger} \overset{\leftrightarrow}{D}_{\mu} H \, \overline{q}^i \gamma^{\mu} q^j$, involving the Higgs current. Let us note that the contribution of the first only arises at loop level or in combination with the second, thus subleading. Also, the second operator, associated with $Z$-mediated quark flavor transitions, does not contribute to the ratios $R_{K^{(*)}}$ at leading order in the new physics scale, being LFU. We will neglect the effects of such type of operators in our analysis, which in practice amounts at neglecting non-generic cancellations between non-standard contributions.
The bounds presented in this section can then be considered as conservative when applied only to 4-fermion semi-leptonic operators.}
\begin{equation}
\label{Xell}
X^{\ell}_{ij} \equiv \frac{\mathcal{A}^\ell_{b \to s l^+_i l^-_j}}{\mathcal{A}^\ell_{b \to s \mu \mu}} \sim \frac{\mathcal{A}^\ell_{b \to s \overline{\nu}_i \nu_j}}{\mathcal{A}^\ell_{b \to s \overline{\nu}_{\mu} \nu_{\mu}}}~.
\end{equation}
All the chiralities of the particles involved in the above expression are left-handed and the last equality has to be understood up to order-one numbers. Correlations between different isospin elements of the lepton doublets depend on the ratio of the triplet and singlet Wilson coefficients in (\ref{PCsemi-leptonic}), but again in the spirit of our analysis effects in charged leptons and neutrino channels are equal up to order 1 numbers. For similar reasons $X^{\ell}_{ij} \sim X^{\ell}_{ji}$. The most relevant bounds on the $X^{\ell}_{ij}$ are coming from three distinct classes of processes:

\begin{enumerate}

\item \textit{Neutrino channels} ($b \to s \overline{\nu}_i \nu_j$)

Due to the absence of experimental information on the flavor of the neutrinos in the final state, this channel can be used to bound all the entries of the $X^{\ell}$ matrix.
The phenomenological input is given by \cite{Lees:2013kla,Lutz:2013ftz}
\begin{equation}
\frac{\mathcal{B} (B \to K^{(*)} \overline{\nu} \nu)}{\mathcal{B}(B \to K^{(*)} \overline{\nu} \nu)_{\textrm{SM}} } < 4.8~,
\end{equation}
which can be translated into
\begin{equation}
\frac{1}{3} \left(3 + 2 k \sum_{i=1}^3 X^{\ell}_{ii} + k^2 \sum^3_{i,j=1} (X^{\ell}_{ij})^2\right) < 4.8 \quad \to \quad
\begin{cases}
-17 < X^{\ell}_{ee}, X^{\ell}_{\tau \tau} < 30 \\
 |X^{\ell}_{e\mu}|, |X^{\ell}_{e \tau}|, |X^{\ell}_{\mu \tau}| < 16
\end{cases}.
\end{equation} 

\item \textit{Flavor conserving charged lepton channels} ($b \to s l^+_i l^-_i$) 

More stringent bounds on $X^{\ell}_{ee}$ can be derived using inclusive $b \to s$ transitions as well as requiring not too large effects in the anomalous observable $R_K$. This leads to 
\begin{eqnarray}
\frac{\mathcal{B} (B^+ \to X_s e^+ e^-)_{[1,6]}}{\mathcal{B} (B^+ \to X_s e^+ e^-)^{\rm SM}_{[1,6]}} \approx (1+k X^{\ell}_{ee})^2  &\to& -2.1 <X^{\ell}_{ee} < 2.0~, \\
 R_K = \frac{\mathcal{B} (B^+ \to K^+ \mu^+ \mu^-)_{[1,6]}}{\mathcal{B} (B^+ \to K^+ e^+ e^-)_{[1,6]}} \approx \frac{(1+k)^2}{(1+k X^{\ell}_{ee})^2}  &\to& -0.6 < X^{\ell}_{ee} < 0.8~,
 \end{eqnarray}
where the subscript $[1,6]$ denotes the $q^2$ region (in GeV$^2$) where the differential branching ratios have been integrated, with $q^2$ the invariant mass of the di-lepton system.
  
Concerning $X_{\tau \tau}$, a bound can be derived from the upper limit on $\mathcal{B} (B^+ \to K^+ \tau^+ \tau^-)$
\begin{eqnarray}
\frac{\mathcal{B} (B^+ \to K^+ \tau^+ \tau^-)}{\mathcal{B} (B^+ \to K^+ \mu^+ \mu^-)} \approx \frac{(1+k X^{\ell}_{\tau \tau})^2}{(1+k)^2}  &\to& |X^{\ell}_{\tau \tau}| < 400~,
\end{eqnarray}
which is much less constraining than that derived from $\mathcal{B} (B \to K^{(*)} \overline{\nu} \nu)$. Notice that in the approximate formula above we also considered the $\tau$ to be massless.

The phenomenological inputs used are summarized in the table below:
\begin{equation}
\begin{array}{c|c}
\textrm{Input} & \textrm{Reference} \\
\hline
\hline
R_K = 0.745^{+0.090}_{0.074} \pm 0.036 & \cite{RKexpt} \\
\mathcal{B} (B^+ \to X_s e^+ e^-)_{[1,6]} = (1.73 \pm 0.12) \times 10^{-6} & \cite{Lees:2013nxa} \\
\mathcal{B} (B^+ \to X_s e^+ e^-)^{\rm SM}_{[1,6]}  = (1.93 \pm 0.55) \times 10^{-6} & \cite{Hurth:2014vma} \\
\mathcal{B}(B^+ \to K^+ \mu^+ \mu^-) = (4.29 \pm 0.22) \cdot 10^{-7} & \cite{Aaij:2014pli}\\
\mathcal{B}(B^+ \to K^+ \tau^+ \tau^-) < 2.25 \cdot 10^{-3} & \cite{Patrignani:2016xqp} \\
\end{array}
\end{equation}

\item \textit{Flavor violating charged lepton channels} ($b \to s l^+_i l^-_j$) 

The most relevant constraints are coming from the $B^+ \to K^+ l^+_i l^-_j$ decays. Once normalised to the muon mode, their expression simplifies to
\begin{equation}
\frac{\mathcal{B} (B^+ \to K^+ \l_i^+ \l_j^-)}{\mathcal{B} (B^+ \to K^+ \mu^+ \mu^-)} = \frac{k^2 }{(1+k)^2} |X^{\ell}_{ij}|^2~.
\end{equation}

The experimental inputs with the associated bounds in our parametrization are reported in the following table
\begin{equation}
\begin{array}{c|c}
\textrm{Experimental bound} & X^{\ell} \textrm{ bound} \\
\hline
\hline
\mathcal{B}(B^+ \to K^+ e^{+} \mu^{-}) < 9.1 \times 10^{-8} \, \cite{Patrignani:2016xqp} & |X^{\ell}_{e \mu}| < 2.5 \\
\mathcal{B}(B^+ \to K^+ e^{-} \mu^{+}) < 1.3 \times 10^{-7} \, \cite{Patrignani:2016xqp} & |X^{\ell}_{\mu e}| < 3.0 \\
\mathcal{B}(B^+ \to K^+ e^{+} \tau^{-}) < 4.3 \times 10^{-5} \, \cite{Patrignani:2016xqp} & |X^{\ell}_{e \tau}| < 55 \\
\mathcal{B}(B^+ \to K^+ e^{-} \tau^{+}) < 1.5 \times 10^{-5} \, \cite{Patrignani:2016xqp} & |X^{\ell}_{\tau e}| <  32 \\
\mathcal{B}(B^+ \to K^+ \mu^{+} \tau^{-}) < 4.5 \times 10^{-5} \, \cite{Patrignani:2016xqp} & |X^{\ell}_{\mu \tau}| < 56 \\
\mathcal{B}(B^+ \to K^+ \mu^{-} \tau^{+}) < 2.8 \times 10^{-5} \, \cite{Patrignani:2016xqp} & |X^{\ell}_{\tau \mu}| < 44
\end{array}
\end{equation}
\end{enumerate}
In PC, operators with lepton singlets are also \textit{generically} induced with a strength correlated with those of the lepton doublets, via their Yukawa couplings.~\footnote{A breaking of this correlation could arise, for instance, in the presence of pseudo-Goldstone bosons with specific gauge quantum numbers, see for example \cite{Gripaios:2014tna}. See also the discussion at the end of section~\ref{sec:pcanom}.} 
For this reason we report also the bounds on these operators. As before, we can define the matrix $X^e$ (induced by $\overline{q} \gamma^{\mu} q \overline{e} \gamma_{\mu} e$ operators) normalized to the effect required by the anomalies:
\begin{equation}
\label{Xell}
X^{e}_{ij} \equiv 
\frac{\mathcal{A}^e_{b \to s l^+_{i} l^-_{j}}}{\mathcal{A}^\ell_{b \to s \mu \mu}~.
}
\end{equation}
In this case we miss the channels with neutrinos, the bounds from lepton flavor conserving transitions gets softened because of the very small interference with the SM, while LFV bounds on $X^{e}$ are the same as for $X^{\ell}$.

Collecting the most stringent constraints we have
\begin{equation}
\label{Xbounds}
|X^{\ell}| \leq
\left(
\begin{array}{ccc}
0.8 & 2.5 & 55 \\
3.0 & 1 & 56 \\
 32 & 44 & 30
\end{array}
\right) \,,
\qquad
|X^e| \leq
\left(
\begin{array}{ccc}
3.5 & 2.5 & 55 \\
3.0 & 1 & 56 \\
32 & 44 & 400
\end{array}
\right)~.
\end{equation}

%%%%%%%%%%%%%%%%%%%%%%%%%%%%%%%%%%%%%%
\subsection{Predictions from partial compositeness}\label{sec:pcanom}
%%%%%%%%%%%%%%%%%%%%%%%%%%%%%%%%%%%%%%

With these inputs we now make the connection with the PC framework discussed in the previous sections.
The strong constraints from the electron EDM and $\mu \to e \gamma$ require to go beyond the anarchic scenario in the lepton sector and to consider instead one of two suppression mechanisms: a $U(1)^3$ symmetry or multiple flavor scales.

We address first the case of a $U(1)^3$-symmetric strong sector. In section~\ref{sec:u1cube} we showed this scenario is phenomenologically viable, in particular it evades the bounds from $\mu \to e \gamma$
for $m_* \gtrsim 5 g_*$ TeV, which is associated with a relatively low degree of electroweak fine-tuning, $\xi \sim g^2_{*} v^2/m^2_{*} \lesssim 0.2 \%$.
Keeping this in mind, let us analyze the implications of the $B$-meson semi-leptonic constraints, for two limiting values of the PC parameters, already introduced in section~\ref{sec:u1cube}: 
\begin{itemize}

\item[(i)] {\it Left-right symmetry}: $\epsilon^{\ell}_i \sim \epsilon^{e}_i$.

In this limit   \eq{LLRR} implies that the $X^{\ell}$ matrix takes the form (for brevity, the elements indicated by a Ò$*$Ó are not shown explicitly because the matrix is symmetric)
\begin{equation}
|X^{\ell}| =
\left(
\begin{array}{ccc}
\frac{m_e}{m_{\mu}} & \frac{m_e}{m_{\mu}} & \frac{m_e}{m_{\mu}}\\
* & 1 & 1 \\
 *&* & \frac{m_\tau}{m_{\mu}}
\end{array}
\right) =
\left(
\begin{array}{ccc}
4.8 \cdot 10^{-3} & 4.8 \cdot 10^{-3} & 4.8 \cdot 10^{-3}\\ 
* & 1 & 1 \\
 *&* & 17
\end{array}
\right)~,
\end{equation}
so the bounds can be easily satisfied given the present experimental situation. Notice that the value of $X_{\tau \tau}^\ell$ is just a factor 2 below the current bound, therefore in this scenario we predict visible effects in $\mathcal{B}(B \to K^{(*)} \overline{\nu} \nu)$ at Belle II in the forthcoming years.
  
Also note that since the right leptons couple with similar strength to the strong sector, then $X^{e} = X^{\ell}$. Bounds on $X^{e}$ are however weaker or at most equal to those on $X^{\ell}$.

\item[(ii)] {\it Left anarchy}: $\epsilon^{\ell}_i \sim \epsilon^{\ell}_j $.

Left anarchy predicts $X^{\ell}_{ij} = \mathcal{O}(1)$, which is consistent with present bounds. A diligent comparison with (\ref{Xbounds}) shows that possible effects should be visible in flavor conserving processes involving electrons, $\mathcal{B} (B^+ \to X_s e^+ e^-)$ or $\mathcal{B} (B^+ \to K^+ e^+ e^-)$, as well as in LFV decays, $\mathcal{B}(B^+ \to K^+ e^{\pm} \mu^{\mp})$. In this scenario it is also important to check the size of operators with lepton singlets. Given a degree of left-handed compositeness $\epsilon^\ell$, we can predict $\epsilon^e_i$ from the mass relation (\ref{yuksym3}), {and then use \eq{LLRR} to arrive at $X^e_{ij} = \frac{2}{g_*^2 v^2 \epsilon^4_{\ell}} m_i m_j $}. Numerically:
\begin{equation}
|X^e| = \frac{1}{g^2_*} \left( \frac{2.2 \cdot 10^{-2}}{\epsilon_{\ell}} \right)^4
\left(
\begin{array}{ccc}
3.7 \cdot{10}^{-5} & 7.7 \cdot{10}^{-3} & 0.13 \\
* & 1.6 & 26 \\
 *&* & 440
\end{array}
\right)~.
\label{XeA}\end{equation}
The reference value for $\epsilon^\ell$ is the largest allowed by (\ref{46}), which corresponds to the conservative benchmark point $g_*=1$ and $m_*=10$ TeV. The comparison with (\ref{Xbounds}) then shows that 
already for the largest $\epsilon^\ell$, and up to the $g_*$ suppression, $B^+ \to K^+ \tau \tau$ could be within discovery reach, the LFV decay $B^+ \to K^+ \mu \tau$ constitutes an extra discovery channel, and the anomalous $B^+ \to K^+ \mu \mu$ should receive contributions also from a right-handed muon current.
\end{itemize}
We would like to note that presently the FCNC anomalies are well fitted either by a purely left-handed muon current, or by a vector muon current (for fit practitioners, $C_9 \neq 0$, $C_{10} = 0$). 
Incidentally, the range for $\epsilon^\ell$ predicted in both the left-right and left anarchy limits implies a similar degree of compositeness of the left- and right-handed muon, compatible with the vector current hypothesis. This pattern could be confirmed once new data will become available.

The predicted correlations among various semi-leptonic channels should allow to test the $U(1)^3$ symmetry in the near future, as well as to discriminate cases (i), (ii), or other possible values of the mixing parameters 
$\epsilon^\ell_i$.
Recall that neutrino mixing in the $U(1)^3$ scenario is independent from the hierarchies in $\epsilon^\ell_i$.

Let us add a final remark on the role of the $U(1)^3 \times CP$ symmetry for the $B$-meson FCNC anomalies.  
If such symmetry were to be respected also by the elementary-composite mixings (\ref{PC1}), 
as briefly discussed in section~\ref{sec:u1cube},
the constraints from LFV transitions ($\mu \to e \gamma$) and CP violation (electron EDM) would be significantly relaxed.
Therefore, lower (more natural) values of $m_*/g_*$ would be viable, as low as allowed by precision tests of the flavor-diagonal electroweak couplings of leptons, and four-lepton flavor-preserving operators.
Such a scenario is perfectly aligned with the $B$-anomalies, which currently do not show any sign of LFV. 
Under the exact symmetry hypothesis, the limits from LFV semi-leptonic processes are trivially satisfied, and only the diagonal entries of the matrices in (\ref{Xbounds}) provide relevant constraints.

We discuss now the multi-scale scenario of section~\ref{sec:multi}. We focus here on two cases, both of them consistent with experimental constraints from flavor and CP violation, whose main difference is in the compositeness scale of lepton doublets, with important consequences for the $B$-anomalies. 
\begin{itemize}

\item[(i)] {\it Left-right symmetry}: $m_{a*}^\ell=m_{a*}^e$.

In this case both left and right leptonic scales are correlated with the charged lepton Yukawa. The prediction for the $X^{\ell}$ matrix, for a Higgs scaling dimension $d_H = 2$, is given by (we are implicitly assuming here 
the unique dynamical scale in the quark sector is $m_*$)
\begin{equation}\label{Xldyn}
X^{\ell}_{ij} = \frac{m_i m_j}{m^2_{\mu}} \quad \to
\quad
X^{\ell} = 
\left(
\begin{array}{ccc}
2.3 \cdot{10}^{-5} & 4.8 \cdot{10}^{-3} & 8.1 \cdot{10}^{-2} \\
* & 1 & 17 \\
 *&* & 283
\end{array}
\right)~,
\end{equation}
which is in tension with the upper bound on $X^{\ell}_{\tau \tau}$ from $\mathcal{B} (B \to K^{*} \overline{\nu} \nu)$. Values of $d_H$ closer to the elementary limit $d_H = 1$ could alleviate such a tension, however they are at odds with keeping the Higgs-squared operator not relevant. We conclude that this scenario is not well-suited to explain the flavor anomalies.

\item[(ii)] {\it (Partial) left anarchy}: $m_{a*}^e, m_{1*}^\ell \gg m_{2*}^\ell \sim m_{3*}^\ell$.

The problem of the previous case can be easily fixed if the flavorful scales associated with the muon and tau lepton doublets are similar.
In this case it is natural to assume that the charged lepton masses are entirely reproduced by the hierarchies in the lepton singlet scales, for $d_H = 2$ we have $m_i \propto 1/m_{i*}^e$. This modifies the matrix $X^\ell$ with respect to (\ref{Xldyn}) in the $\tau$-row (and column), which now reads $X_{i \tau} = \left(m_e/m_\tau, 1, 1 \right)$, safely consistent with semi-leptonic processes. The $ee$ and $e\mu$ entries remain the same, since the breaking of the LFU in the electron-muon sector (hinted by the anomalies) is achieved by the hierarchy $m^{\ell}_{1*} \gg m^{\ell}_{2*}$.
\end{itemize}

Even if in the scenario (ii) the scales $m_{2*}^\ell \sim m_{3*}^\ell$, which from now on we simply denote as $m_*^\ell$, are not related to the charged lepton masses, this model is particularly attractive for what concerns 
the $B$-anomalies.
In fact, we can successfully extend this particular idea to the quark sector, along the lines of \cite{Panico:2016ull}.
For the down-quark sector we consider then
\begin{equation}
m_{a*}^d, m_{1*}^q \gg m_{2*}^q \gtrsim m_{3*}^q \sim m_*~,
\end{equation}
where we take the compositeness scale of the third generation quark doublet to be (approximately) the same as that of the Higgs, for considerations of electroweak naturalness (and in fact also $m_{3*}^u \sim m_*$ for the same reason).
The scale associated with the second generation quark doublet is kept low, relatively close to $m_*$, since, as we show below, it might play an important role in explaining the anomalies.
In this scenario we assume the down and up quark masses are reproduced via suitable hierarchies of the right scales, $m_{a*}^d$ and $m_{a*}^u$ respectively, and the CKM matrix is aligned with the left-handed down rotation, i.e.~$U_{\rm CKM} \sim U^{d}$ (for the same reasons discussed in section~\ref{sec:u1cube}, the mixing angles are determined by ratios of quark masses and such ratios are smaller in the down sector).

The bounds on a quark sector with multiples scales have been discussed at length in \cite{Panico:2016ull} and we report here their main results adapted to our analysis:
\begin{equation}\label{boundquark}
m_* \gtrsim 5 (x_t \, , \sqrt{g_* x_t})\, {\rm TeV} \, , \quad m_{2*}^q \gtrsim 240\, x_t \, {\rm TeV}~,
\end{equation}
with $x_t = \epsilon^q_3/\epsilon^u_3$ (recall $y_t \sim \epsilon^q_3 \epsilon^u_3 g_*$) and we have taken $\epsilon_2^q \sim \epsilon_3^q$ for simplicity. The first bound on $m_*$ comes from $K$- ($\epsilon_K$) and $B$-meson ($\Delta M_{B_{b,s}}$) mixing, while the second is from $Z \bar{b}b$. The bound on $m_{2*}^q$ is due to $\Delta M_K$.
%The first two bounds, on $m_*$, come from $K$- ($\epsilon_K$) and $B$-meson ($\Delta M_{B_{b,s}}$) mixing, while the last is from $\Delta M_K$.
Interestingly, these bounds are not too far from those in the lepton sector, discussed in section~\ref{sec:neutrMulti}. Given our breaking of LFU, $m_{1*}^\ell \gg m_*^\ell$, one finds that the strongest constraints arise from $\mu\to eee$ 
and $\mu\, Au \to e\, Au$. Following the discussion below (\ref{ellbounds}), we will assume a $P_{LR}$ symmetry acting on $\ell$ is at work, such that the bound from $\mu \to e$ conversion in nuclei is avoided. In this case
\begin{equation}\label{boundlepton}
m_*^\ell \gtrsim 15 g_* (\epsilon^\ell)^2 \, {\rm TeV}~.
\end{equation}
The size of the semi-leptonic operator giving rise to the anomaly, \eq{PCsemi-leptonic}, depends on the interplay between the scales above. Given the absolute size of the anomaly, the best case scenario is that in which (\ref{PCsemi-leptonic}) is generated at $m^\ell_* \gtrsim m_*$. 

Then, if $m^q_{2*} > m^\ell_*$, only semi-leptonic operators with third generation quark doublets $q_3$ are generated at $m^\ell_*$. 
After taking into account a rotation by $(U_{\rm CKM})_{32}$, one arrives at
\begin{equation}
\label{anom3}
\frac{1}{\Lambda^2_{NC}} \sim \lambda_C^2 (\epsilon^q_3)^2 (\epsilon^\ell)^2 \left( \frac{g_*}{m^\ell_{*}} \right)^2 \lesssim \frac{x_t}{g_*} \frac{1}{(\epsilon^\ell)^2} \left( \frac{1}{65\, {\rm TeV}} \right)^2~,
\end{equation}
where the inequality follows from (\ref{boundlepton}). Amusingly, the anomaly can be reproduced 
for suitable values of $x_t$, $g_*$, and $\epsilon^\ell$: comparing the bounds in (\ref{boundquark}) and (\ref{boundlepton}), our assumption that $m^\ell_* \gtrsim m_*$ simply requires $x_t/g_* \lesssim 3 (\epsilon^\ell)^2$, a condition that allows a relatively low $\Lambda_{NC} \gtrsim 37 \, {\rm TeV}$, a value consistent with the flavor anomalies given the $\mathcal{O}(1)$ accuracy of our estimates.
We should note that the scale of the anomaly is predicted to be largest when $m^\ell_*$ is close to $m_*$, which for e.g.~$x_t = 1$ and $g_* = 1$ is around $5 \, {\rm TeV}$. While such a low $m_*$ is welcome for naturalness considerations, the correspondingly low $m_*^\ell$ implies potentially large contributions to leptonic processes precisely measured at LEP. Fortunately in our construction problematic 4-fermion operators involving electrons are suppressed due to the hierarchy $m^{\ell}_{1*} \gg m^{\ell}_{*}$ while, due to $P_{LR}$, only the neutrino couplings to the $Z$, but not $Z \mu \bar \mu$, are modified, the deviation being of order $(\epsilon^\ell g_*v/m^{\ell}_{*})^2 \lesssim 10^{-3}$ and thus consistent with the measured $Z$ decay width to invisible.

If, {on the other hand,} $m^q_{2*} < m^\ell_*$, then semi-leptonic operators with second generation quark doublets $q_2$ are also generated at $m^\ell_*$. In this alternative case we find
\begin{equation}
\label{anom2}
\frac{1}{\Lambda^2_{NC}} \sim (\epsilon^q_3)^2 (\epsilon^\ell)^2 \left( \frac{g_*}{m^\ell_{*}} \right)^2 \lesssim \frac{x_t}{g_*} \frac{1}{(\epsilon^\ell)^2} \left( \frac{1}{15 \, {\rm TeV}} \right)^2~,
\end{equation}
where one needs to take into account that $m^q_{2*} < m^\ell_*$ is consistent with the bounds on $m_{2*}^q$ in (\ref{boundquark}) for $x_t/g_* \lesssim (\epsilon^\ell/4)^2 \le 1/16$. 
In this case the anomaly can be reproduced consistently with current constraints in both the quark and lepton sectors only if an accidentally large order-one factor is present in (\ref{anom2}) (a factor of 4  suffices).

The upshot of this example, which is born out of a consistent picture of flavor, is that if the flavor anomalies are confirmed, non-standard effects in both purely quark and leptonic processes ($\epsilon_K$, $\Delta M_{B_{b,s}}$, $\Delta M_K$, and $\mu \to eee $ or the electron EDM, respectively), should be close to the current sensitivity.

%%%%%%%%%%%%%%%%%%%%%%%%%%%%%%%%%%%%%%
\section{Conclusions}\label{end}
%%%%%%%%%%%%%%%%%%%%%%%%%%%%%%%%%%%%%%

In this paper we focused on the paradigm of Partial Compositeness (PC), a very appealing and predictive picture of charged-fermion flavor. In this context we believe the spectrum of model building possibilities has not been fully explored, and our work aims to close a few of the remaining important gaps. Our main original results concern
\begin{itemize}
\item neutrino masses and mixing;
\item scenarios with suppressed flavor violation in the lepton sector;
\item CP symmetry.
\end{itemize}

We pointed out that PC is very predictive in the neutrino sector. Anarchic realizations, in which the composite sector violates maximally flavor and CP symmetries, generate one of three possible neutrino mass textures, see \eq{Mclass}. 
Two textures are correlated with the degree of compositeness of the lepton doublets, that is forced to be of the same order for the second and third families. We argued there exists a natural mechanism to suppress neutrinoless $2\beta$ decay, and showed that, order-one PMNS phases are predicted in all textures. 
Concrete models that realize such textures have been identified and briefly discussed. These models encompass (a strongly-coupled version of) most well-known mechanisms of neutrino mass generation,
but they do not suffer from a naturalness problem even for a large lepton-number breaking scale.

We then carried out a thorough analysis of charged lepton flavor- and CP-violating processes, updating the constraints and presenting them in a model-independent fashion. 
Because of the absence in the near future of direct experimental access to energies beyond those probed by the LHC, the remarkable experimental sensitivity of indirect searches may turn out to be crucial to unravel where the scale of new physics might be.
Indeed, when considering PC, we find that anarchic scenarios require the typical new physics scale $m_*$ {and coupling $g_*$} to satisfy $m_*\gtrsim 100 g_* \, {\rm TeV}$ in order to be compatible with data: here the dominant constraint comes from the electron EDM, with a  weaker bound from $\mu-e$ transitions. 
This severe constraint implies a large hierarchy with the electroweak scale, $m_*^2/m_h^2\gg 1$, suggesting the presence either of a large fine-tuning of the Higgs mass, or an additional mechanism to address the leftover hierarchy problem. Retaining a low new physics scale makes the flavor puzzle a concrete issue, and this motivated us to look at more structured realizations of PC, departing from the minimal anarchic hypothesis.

The first non-anarchic scenario we considered is based on the assumption that the strong sector has a global $U(1)^3\times CP$ symmetry, generically broken by the mixing with the elementary sector. We proposed the flavor group $U(1)^3$ because this is the simplest symmetry compatible with the PC generation of the SM fermion mass hierarchy. The resulting pattern of flavor violation and neutrino masses is very different from that of anarchic PC. When married with the assumption that the composite sector respects CP, our construction is capable of passing the stringent constraint from the electron EDM and $\mu\to e\gamma$ with $m_* \gtrsim 7 g_* \, {\rm TeV}$, a remarkable improvement compared to anarchic scenarios. These observables remain the most sensitive probes of this scenario although, depending on the degree of compositeness of the lepton doublets, $\mu \to eee$ can also saturate the current experimental bound. In addition, in this framework the neutrino mass texture strongly depends on the $U(1)^3$ charges of the most relevant operator that violates lepton number. This may lead to interesting correlations among neutrino masses, mixing and phases,
and in particular {a massless active neutrino} is predicted in minimal models.

The second non-generic realization of PC relaxes the hypothesis that the composite sector is characterized by a unique $m_*$, assuming that it develops different mass scales for different flavors.
The resulting suppression of flavor and CP violation is very efficient, and all constraints can be satisfied with $m_*$ (here the scale of Higgs/top compositeness) of order a few TeV and higher compositeness scales for leptons. 
While processes mediated by dipole operators become out of reach, channels like $\mu \to eee$ could be close to discovery for example when $m_*^e \gg m_*^\ell \sim 200 g_* \, {\rm TeV}$.
Realistic neutrino masses can be obtained if the lepton-number breaking source is independent from the flavored scales $m_*^\ell$, or an approximate $U(1)^3$ symmetry selects a single dominant contribution to $m^\nu$, or a unique compositeness scale for the three lepton doublets is assumed.

Our results have also implications in the quark sector. First of all, we demonstrated that CP can be a good assumption in the composite sector, and specifically this hypothesis is well compatible with the generation of a CKM phase. The reason is that a source of CP violation is expected to be injected by the mixing with the elementary sector and, as a matter of fact, this is enough to ensure the correct size of the CKM phase.
A combination of $CP$ with an appropriate $U(1)^3$ may be used, as done for leptons, to relax the strong bounds from the neutron EDM. However, as opposed to the lepton sector, some tweaking of $\mathcal{O}(1)$ parameters is necessary to generate the observed quark mixing pattern.

As a final application of our findings, we analyzed the compatibility of PC and the current anomalies in semi-leptonic $B$-meson decays. We find that each of the above realizations predicts a characteristic pattern of violation of flavor and lepton flavor universality. Whether or not these are fully compatible with data depends however on the quark observables as well. Interestingly, we found that scenarios with multiple flavor scales can be consistent with all current constraints, and simultaneously introduce the breaking of lepton universality necessary to explain the anomalies in neutral-current $B$ decays.
{Besides, we argued that the option of a $U(1)^3\times CP$ symmetry respected also by the elementary-composite mixings turns out to be particularly interesting. In this extreme setting all lepton flavor violation is controlled by the mechanism of neutrino mass generation and the scale of lepton compositeness can be lowered down to $m_* \sim g_* \, {\rm TeV}$, which is optimal for naturalness, leaving as main signature the violation of flavor universality in flavor-conserving processes}.

Other anomalies currently present in lepton observables, such as the $\sim 4\sigma$ discrepancies in $b\to cl\bar\nu$ transitions and in the muon anomalous magnetic moment, seem to require new physics states {close or below the TeV}, as a necessary and not sufficient condition to be accommodated. Therefore, their explanation is highly model-dependent, as the spectrum of low-lying composite resonances cannot be characterized by the PC framework alone.

\section*{Acknowledgements}

We would like to thank K.~Agashe, A.~Azatov, R.~Coy, S.~Davidson, G.M.~Pruna, and A.~Weiler for discussions.
JS and LV would also like to express a special thanks to the Mainz Institute for Theoretical Physics (MITP) for its hospitality and support.
MF is supported by the European Union's Horizon 2020 research and innovation programme 
under the Marie Sklodowska-Curie grant agreements No 674896 and No 690575, and by the OCEVU Labex (ANR-11-LABX-0060) funded by the ``Investissements d'Avenir" French government program.
JS has been supported by the Collaborative Research Center SFB1258 and the DFG cluster of excellence EXC 153 ``Origin and Structure of the Universe''.
LV is supported by the Swiss National Science Foundation under the Sinergia network CRSII2-16081.

%%%%%%%%%%%%%%%%%%%%%%%%%%%%%%%%%%%%%%
%%%%%%%%%%%%%%%%%%%%%%%%%%%%%%%%%%%%%%
\appendix
%%%%%%%%%%%%%%%%%%%%%%%%%%%%%%%%%%%%%%
%%%%%%%%%%%%%%%%%%%%%%%%%%%%%%%%%%%%%%

%%%%%%%%%%%%%%%%%%%%%%%%%%%%%%%%%%%%%%
\section{Assumptions on the strong sector} 
\label{sec:ass}
%%%%%%%%%%%%%%%%%%%%%%%%%%%%%%%%%%%%%%

Our key assumptions can be stated as follows:
\begin{itemize}
\item[(a1)] The operators $O^\psi_a$ are part of a flavorful strongly-coupled conformal field theory (CFT). Such a CFT is described in terms of operators $O$ of a given scaling dimension $\Delta[O]$. To allow us to make concrete predictions we will assume that scaling dimensions algebraically sum, so that $\Delta[O_1O_2]=\Delta[O_1]+\Delta[O_2]$.
\item[(a2)] The CFT is perturbed by the couplings $\lambda^\psi$ defined in (\ref{PC1}), and other small couplings, e.g.~SM gauge couplings and possibly higher-dimensional operators, as those breaking lepton number, see section~\ref{NeutrinoSection}.
\item[(a3)] The strong sector develops a mass gap and composite resonances at low energies, much below $\Lambda_{\rm UV}$. The self-couplings $g_*$ and masses $m_*$ of these states are assumed to be all of the same order. In section \ref{sec:supp} we will allow the presence of more scales in the composite sector.
\item[(a4)] The CFT can have its own global symmetries, but these are interpreted as accidental, as in the SM. This means in particular that, unless motivated otherwise, such symmetries are broken by  the mixing (\ref{PC1}) with the SM, and by higher-dimensional operators like those in section~\ref{NeutrinoSection}. A CFT with no flavor symmetries will be called anarchic. More structured CFTs are discussed in section~\ref{sec:supp}.
\item[(a5)] UV interactions between the SM and the CFT are anarchic in flavor space, meaning that all couplings (in general in the form of matrices) have no special structure in the UV (unless enforced by gauge symmetries).
\end{itemize}
The hypothesis (a1) of conformality is necessary to stabilize a large hierarchy $m_*\ll\Lambda_{\rm UV}$ without introducing unnatural fine-tuning. The conformal symmetry is assumed to be abruptly broken at $\sim m_*$, where the quanta of the strong dynamics manifest themselves. 

The assumption (a3) of one coupling and one scale~\cite{Giudice:2007fh} allows us to estimate the coefficients appearing in the EFT at the scale $m_*$, up to unknown numbers of order unity, that we denote by $c$. Specifically, we postulate the low energy effective Lagrangian has the form dictated by naive dimensional analysis~\cite{Giudice:2007fh},
\ba\label{NDA}
{\cal L}_{\rm NDA}=\frac{m_*^4}{g_*^2}\left[\hat{\cal L}_0\left(\frac{g_* H}{m_*},\epsilon^\psi_{ai}\frac{g_*\psi_i}{m_*^{3/2}},\frac{D_\mu}{m_*}\right)+\frac{g_*^2}{16\pi^2}\hat{\cal L}_1\left(\frac{g_* H}{m_*},\epsilon^\psi_{ai}\frac{g_*\psi_i}{m_*^{3/2}},\frac{D_\mu}{m_*}\right)+{\cal O}\left(\frac{g_*^2}{16\pi^2}\right)^2\right],
\ea
with $\hat{\cal L}_n$ that depend on order-one coefficients $c_{a\cdots}$ that in general violate the composite flavor index $a$. In section \ref{sec:supp} we will either introduce symmetries in the strong sector, or allow the presence of more than one scale in the strong sector. Both changes have the effect of selecting a specific structure for the order one coefficients. 

Note that (\ref{NDA}) assumes the Higgs is maximally coupled to the CFT sector. In this sense we will consider scenarios with a composite Higgs. One may relax this assumption by replacing $g_*H/m_*\to g_*\epsilon_HH/m_*$ in (\ref{NDA}), thus allowing even the Higgs to be ``partially composite''.

Finally, we want to qualify what we mean by anarchy and hierarchy. We will often use the terms ``order one", ``anarchic", and ``non-hierarchical" throughout the paper, but the truth is that the real meaning depends on the context, and more precisely on the parameters $\epsilon_i^\psi$ involved. A number $c$ is said to be order unity (a matrix is said to have numbers of order one) whenever the pattern $\epsilon_1^\psi\ll\epsilon^\psi_2\ll\epsilon^\psi_3$ remains valid even if one of these $\epsilon_i^\psi$ is multiplied by $c$. This statement of course depends on the hierarchy in the $\epsilon_i^\psi$'s. For example, from table \ref{input} we see that any $|c|\lesssim10$ may be considered order unity when dealing with the up-type quarks, whereas in the down-quark sector a $c$ of order one must be literally close to one. An anarchic, or non-hierarchical, matrix is one whose entries have ratios of order unity. Our assumption (a5) of UV anarchy is equivalent to the statement that all flavor hierarchies follow from $\epsilon_1^\psi\ll\epsilon^\psi_2\ll\epsilon^\psi_3$.

%%%%%%%%%%%%%%%%%%%%%%%%%%%%%%%%%%%%%%
\section{{Derivation of the neutrino mass textures}} \label{sec:textures}
%%%%%%%%%%%%%%%%%%%%%%%%%%%%%%%%%%%%%%

We now show that, under the hypothesis that only operators with at most two fundamental fermions contribute significantly to the neutrino mass -- an assumption motivated below (\ref{scalingtilde}) --, there are only 
three classes of Majorana neutrino mass textures, shown in~(\ref{Mclass}), and that the corresponding $\tilde\epsilon$ matrices are anarchic.

Let us begin with operators of the form $\tilde \lambda O$, that have no fundamental fields. These are the simplest ones because they can only contribute to $m^\nu$ after the PC couplings (\ref{spur1}) are included, in particular $\lambda^\ell$. This is because $\tilde\lambda$ can carry only $U(1)_c$ charges, but no fundamental quantum numbers, i.e.~$\tilde \lambda \sim (1_0, 1_0, -c)$ under $[SU(3)_\ell\times U(1)_\ell]\times [SU(3)_e\times U(1)_e]\times U(1)_c$, where $c$ is the ``composite'' lepton number of the operator $O$. As a consequence, it should be immediately clear to the reader that these models belong to class 2M in (\ref{Mclass}). As a further remark, the relation between $\tilde\epsilon$ and $\tilde\lambda$ depends on the charge $c$, and is therefore model-dependent. Fortunately, this aspect plays no role in our analysis because the flavor structure is completely determined by $\epsilon^\ell$. Moreover, it is easy to see that $\tilde \epsilon \propto \tilde \lambda^n$, where $n \, c = 2$ (if this condition cannot be satisfied, $m_\nu$ will not be generated). The overall size of $\tilde\epsilon$ then depends on both the scaling dimension of $O$ and its $U(1)_c$ charge.

The impact of CFT deformations of the form $\tilde\lambda\psi O$ is also easy to understand. The first point to observe is that, once we choose a basis in which $\lambda^{\ell,e,q,u,d}$ are as in (\ref{PC3}), we cannot put $\tilde\lambda$ in triangular form as well. It then follows that $\tilde\lambda$ at the scale $m_*$ will be {\emph{anarchic}} in this field basis. 
To further proceed is useful to consider separately the case where $\psi = \ell$ and $\psi \neq \ell$. 
Let us start with the latter.
In this case the flavor indices of $\tilde \lambda$, associated with $\psi \neq \ell$, must be contracted with $\epsilon^{\psi \neq \ell}$ to yield a spurion with no fundamental quantum numbers, but only a definite $U(1)_c$ charge. 
The resulting spurion $\tilde \lambda^{\rm eff} \equiv \tilde \lambda \cdot \epsilon^{\psi \neq \ell \, \dagger} \sim (1_0, 1_0, -c^{\rm eff})$ is of the same form as the $\tilde \lambda O$ discussed above, thus these models belong to class 2M as well. 
The associated $\tilde \epsilon_{ab}$ will be suppressed by the entries in $\epsilon^{\psi \neq \ell}$, but such $\tilde \epsilon$ matrix will anyway be anarchic: the strong dynamics is, by assumption, flavor blind, being only sensitive to the $U(1)_c$ quantum number.

Certainly more interesting are  models where $\psi = \ell$, where $\tilde \lambda \sim (\bar 3_{-1}, 1_0, -c)$. 
By recalling that $\epsilon^\ell \sim (\bar{3}_{-1},1_0,1)$ and we need to induce $m^\nu \sim (6_{-2},1_0,0)$, it is easy to identify the three relevant $U(1)_c$ charge assignments: for $c = -3$, the contraction $\tilde \epsilon \propto \epsilon^\ell \cdot \tilde \lambda^\dagger$ realizes neutrino masses in class 2M; 
for $c = 1$, $\tilde \epsilon \propto \tilde \lambda$ leads to class 1M; for $c = 0$,  $\tilde \epsilon \propto \tilde \lambda \cdot \tilde \lambda$ belongs to class 0M.
This can be easily understood by noticing that the $U(1)_c$ charges of $\tilde \epsilon$ is $-2$, $-1$ and $0$ for classes 2M, 1M and 0M respectively, see (\ref{Mclass}).
Different values of $c$ may lead to less minimal models, where $\tilde\epsilon$ includes extra powers of $\epsilon^\ell \cdot \tilde \lambda^\dagger\sim (1_0,1_0,c)$, that reduce to the previous ones upon contraction of the indices.
Importantly, because the only relevant CFT quantum number is lepton number, the contractions above do not give rise to any flavor structure, meaning the $\tilde \epsilon$ matrices are anarchic. 
Finally note that, in general, when $\tilde \epsilon$ depends on some elementary spurion $\epsilon^\psi$, neutrino masses arise at one loop, with the fundamental fermion $\psi$ in the internal line.~\footnote{In general, the number of elementary loops needed to generate $m_\nu$ from a spurion $\tilde \lambda$ associated with $n_\ell$ lepton doublets and $n_\psi$ fermions $\psi \neq \ell$ is $n_{\rm loops} = |n_\psi n_{s}|+\min(|n_\ell n_{s}|,|2-n_\ell n_{s}|)$ where $n_{s}$ is the number of spurions required, i.e.~$m^\nu \propto \tilde \lambda^{n_{s}}$, and negative values of $n_{\ell, \psi}$ should be interpreted as the number of $\ell^\dagger, \psi^\dagger$ while negative $n_s$ means $m^\nu$ is proportional to powers of $\tilde \lambda^\dagger$.}

If one considers a bilinear deformation $\tilde\lambda\psi\psi' O$, similarly to the previous case the coupling $\tilde\lambda$ will generically be anarchic in the basis of (\ref{PC3}).
The spurion quantum numbers of $\tilde\lambda$ will be again fixed by the transformation properties of $\psi$ and $\psi'$. 
One important difference is that a bilinear deformation has scaling dimension $\Delta\gtrsim 5$, therefore
neutrino masses at $\mathcal{O}(\tilde \lambda^2)$
are too suppressed for $m_* \ll \LambdaL$, as the effective dimension of the CFT perturbation becomes $2 \Delta \gtrsim 10$.
Therefore one needs to realise $m^\nu\propto \tilde \lambda$.

In the case $\psi, \psi' \neq \ell$, the quantum numbers of $m_\nu$ can only be reproduced if the fundamental indices of $\tilde \lambda$ are contracted with $\epsilon^{\psi,\psi' \neq \ell}$, giving rise to an effective spurion $\tilde \lambda^{\rm eff} \equiv \epsilon^{\psi' \neq \ell \, \dagger} \cdot \tilde \lambda \cdot \epsilon^{\psi \neq \ell \, \dagger} \sim (1_0, 1_0, -c^{\rm eff})$. The resulting neutrino mass matrix belongs to class 2M, with $\tilde \epsilon$ anarchic. 
Similarly, whenever $\psi \neq \ell$ \emph{or} $\psi' \neq \ell$, generating neutrino masses requires a $\tilde \lambda^{\rm eff} \equiv \tilde \lambda \cdot \epsilon^{\psi \neq \ell \, \dagger} \sim (\bar 3_{-1}, 1_0, -c^{\rm eff})$, which, as explained above, gives rise to neutrino masses in classes 2M, 1M or 0M depending on $c^{\rm eff}$. 
Finally, let us discuss models where $\psi, \psi' = \ell$, with $\tilde \lambda_{3} \sim (3_{-2}, 1_0, -c)$ or $\tilde \lambda_{\bar 6} \sim (\bar 6_{-2}, 1_0, -c)$, the two possibilities arising from the tensor product $(\bar 3 \otimes \bar 3)_{-2}$. 
Note that the CFT deformation, even with the assumption of UV-anarchy, could naturally select either $\tilde \lambda_{3}$ or $\tilde \lambda_{\bar 6}$, depending on the gauge quantum numbers of $O$, either a weak singlet or triplet respectively, 
as explicitly shown in section~\ref{sec:Mmodels}. 
For this reason we analyze these two $U(1)_L$-violating perturbation separately. 
From $\tilde \lambda_{3}$ neutrino masses in class 1M are generated, with $\tilde \epsilon \propto \epsilon^{\ell \, \dagger} \times \tilde \lambda_3$. 
Importantly, the anti-symmetry of $\tilde \lambda_3$ implies that the resulting $\tilde \epsilon_{ai}$ is not fully anarchic, the third row ($a=3$) being suppressed by $\epsilon^\ell_2/\epsilon^\ell_3$ with respect to the others.
From $\tilde \lambda_{\bar 6}$ it should be evident that the resulting $m_\nu$ belongs to class 0M, with $\tilde \epsilon \propto \tilde \lambda_{\bar 6}$, anarchic. 
In both these cases $c = 0$ must be required for neutrino masses to be generated at leading order in $\tilde \lambda$.
We note again that, in the cases where $\tilde \epsilon = \tilde \epsilon (\epsilon^\psi,\epsilon^{\psi'})$, neutrino masses require loops of fundamental fermions.

Finally, let us briefly comment on $U(1)_L$-violating sources beyond those of type $\tilde\lambda O$, $\tilde\lambda\psi O$, $\tilde\lambda\psi\psi' O$. 
We therefore relax our assumption, by allowing CFT deformations with e.g. more than two elementary fields.
Let us consider e.g. $\Delta{\cal L}=\tilde\lambda_{ijk\tilde a} \ell^i \ell^j \ell^k O_{\tilde a, c=-1}$, with scaling dimension $\Delta \gtrsim 6$ (we assume $O$ has spin 1/2, and therefore dimension $\ge 3/2$ by unitarity). 
In this case the possible $SU(3)_\ell$ quantum numbers of $\tilde \lambda$ are $\bar 3 \otimes \bar 3 \otimes \bar 3 = 1 \oplus 8 \oplus 8 \oplus \overline{10}$.
Let us suppose for a moment that only an adjoint component is present, given by $(\tilde\lambda_{8})_{ijk} = (\tilde\lambda_{8})^l_i \epsilon_{ljk}+(\tilde\lambda_{8})^l_j \epsilon_{lik}$, symmetric under $i\leftrightarrow j$.
Such contraction is certainly curious, since it gives rise to $\tilde\epsilon_{ij} \simeq (\tilde\lambda_{8})_{ijk \tilde a} \epsilon_{ak}^{\ell \, *} c_{\tilde a a}(m_*/\LambdaL)^{\Delta_O+1/2} g_*/(16 \pi^2)$, 
and for $i=j=3$ the term with $k=3$ vanishes by antisymmetry. As a consequence,
the resulting neutrino mass matrix belongs to class 0M, but it has a suppressed 33 entry, $m^\nu_{33} \propto \epsilon^\ell_2$, while the other entries are proportional to $\epsilon^\ell_3$.
However, it is difficult to imagine how to select the above form for $\tilde\lambda$: in fact, gauge interactions can distinguish only between $O^{(2)}$ and $O^{(4)}$, where the subscript indicates the $SU(2)_L$ representation of the operator.
Choosing a single operator does not select the adjoint flavor contraction used above. This shows how difficult is to avoid anarchy in the entries of $\tilde\epsilon$.
Similar features are expected whenever one tries to depart from the assumptions leading to (\ref{Mclass}).

%%%%%%%%%%%%%%%%%%%%%%%%%%%%%%%%%%%%%%
\section{Dirac neutrinos}\label{dirac}
%%%%%%%%%%%%%%%%%%%%%%%%%%%%%%%%%%%%%%

When total lepton number $U(1)_L$ is conserved, the neutrinos cannot acquire Majorana masses. However, one can introduce an additional set of elementary chiral fermions, $N_{i}\equiv N_{Ri}$, singlet under electroweak interactions, with $L(N_i)=1$. 
In this case Dirac masses can arise in the EFT from the operator
\ba\label{Diracm}
y^\nu_{ij}\overline{\ell_i} N_{j}H ~.
\ea
The presence of sterile neutrinos implies the existence of an additional approximate accidental symmetry $U(n_N)$, with $n_N\geq2$ the number of $N_i$ families. This symmetry must be broken by couplings to the CFT, in such a way that the operator (\ref{Diracm}) can emerge in the low energy theory. Yet, if the neutrinos are to be (approximately) Dirac, an approximate $U(1)_{L=\ell+e+N+c}$ is still necessary, in order to suppress the operator (\ref{Weinberg}).

Following a logic similar to the one leading to \eq{Mclass} here we find two basic scenarios,
\ba\label{Dclass}
{\rm 2D}:&&y^\nu_{ij}= g_*\epsilon^{\ell*}_{ai}\epsilon^N_{bj} c^\nu_{ab}~,\no\\
{\rm 0D}:&&y^\nu_{ij}= g_*  \epsilon^\nu_{ij}~,
\ea
where $c^\nu_{a b}$ are numbers of order unity when the strong sector is anarchic.
In order for $y^\nu$ to have rank $\ge 2$, one needs a number of sterile neutrinos $n_N\ge 2$. In addition, in class 2D the sum over $b$ has to include at least two (three) terms to provide $y^\nu$ with rank two (three).
The neutrino mass matrix structure of class 2D is analogous to the charged lepton one (\ref{Yukawa}), thus both $U_\nu$ and $U_\ell$ are controlled by the hierarchies in $\epsilon^\ell$.
Indeed, it is easy to check that $y^\nu y^{\nu\dag}$ has the same flavor structure of class 2M in \eq{Mclass}, therefore \eq{ratiosNEW} holds with the corresponding discussion.
Anarchy of $\epsilon_{ij}$ in class 0D ensures that neutrino masses are of the same order and large mixing angles can be reproduced naturally, as in class 0M of \eq{Mclass}.
In particular, \eq{nop} holds, and the same comments we made there apply here as well.

We identify two minimal, representative models that lead to \eq{Dclass}:
\ba\label{Dirac}
\Delta{\cal L}&=&\lambda_{ai}^N \overline{O^N_{a,c=1}} N_{i}~~~~~~~~~~~~~~~~~~~~~\in~{\rm class~2D}, \no\\
&&\lambda_{ija} \overline{\ell_i} N_{j}O_{a, c=0}~~~~~~~~~~~~~~~~~\,\in~{\rm class~0D}.
\ea
The first is a straightforward generalization of (\ref{PC1}) and (\ref{Yukawa}) to the lepton sector.
In this case the smallness of the neutrino mass relatively to the charged lepton mass relies on $O^N$ being significantly more irrelevant than $O^e$, such that $\epsilon^N \simeq (\lambda^N/g_*) (m_*/\Lambda_{\rm UV})^{\Delta^N-5/2} \ll \epsilon^\ell$.
For the bilinear model in class 0D, first suggested in~\cite{Agashe:2008fe}, 
naturalness requires the operator to have dimension $\Delta\gtrsim 5$, thus in this case neutrino masses are naturally suppressed at low energies, since $\lambda(m_*) \simeq (m_*/\Lambda_{\rm UV})^{\Delta-4} \lambda$. 
We stress in passing that there is only one more operator breaking $U(n_N)$ and allowed to have scaling dimension close to five: $NNO_{c=-2}$. However, by $U(n_N)$ invariance it follows that such a CFT deformation can induce $m^\nu$ only if $\lambda^N \overline{O^N_{c=1}}N$ ($\lambda \overline{\ell} N O_{c=0}$) is also included, in which case the first (second) option in \eq{Dirac} would inevitably dominate the neutrino Yukawa matrix, as the new operator contributes only via a loop of $N$ fermions.

Let us briefly discuss the impact of a $U(1)^3$ symmetry of the strong sector on Dirac neutrino masses.
In class $0D$ the flavor structure of $y^\nu$ is independent from the strong sector flavor symmetry. On the contrary, in class $2D$ the $U(1)^3$ symmetry has important consequences.
Consider three sterile neutrinos $N_i$ and three spurions with different $U(1)^3$ charges: $\epsilon^N_{a=1\, i}\sim (-1,0,0)$, $\epsilon^N_{a=2\, i}\sim (0,-1,0)$, $\epsilon^N_{a=3\, i}\sim (0,0,-1)$.
In this case one has $c^\nu_{ab}= c^\nu_a \delta_{ab} $, and
the matrix $\epsilon^N_{ai}$ can be brought to the triangular form (\ref{PC3}) by a choice of basis for the $N_i$ fields.
As a consequence, $y^\nu$ takes the form of \eq{yukFlip}, with $\epsilon^e_i$ replaced by $\epsilon^N_i$, therefore the neutrino masses and mixing angles are controlled by the three quantities $y^\nu_i\equiv g_*\epsilon^\ell_i\epsilon^N_i$,
up to order one parameters. In order to reproduce oscillation data, we find that one needs either $0\le y^\nu_3 \lesssim y^\nu_1\sim y^\nu_2$, or $0\le y^\nu_2 \lesssim y^\nu_1\sim y^\nu_3$.
Note that $y^\nu_3$ (or $y^\nu_2$) can vanish, therefore only two spurions $\epsilon^N_{a=1,2\, i}$ ($\epsilon^N_{a=1,3\, i}$) are sufficient, and in this limit one neutrino would be massless.
On the other hand all three $y^\nu_i$ can be of the same order, and  $y^\nu_1$ can be smaller than the others by at most a factor of five or so. 
If only two sterile neutrinos exist, the matrix $y^\nu$ maintains the same flavor structure with the first column dropped, and the quantities $y^\nu_i$ should satisfy the same relations as in the case of three $N_i$.
Extensions to more than three sterile neutrinos, or more than three spurions $\epsilon^N_a$, do not lead to qualitatively different flavor patterns.

Regarding the impact of multiple scales on Dirac neutrino masses, let us first note that we could also introduce dynamical scales where the (three) sterile neutrinos decouple, $m_{a*}^N$. In class 2D the interplay of these with $m_{a*}^\ell$ determines the structure of $y^\nu$. Indeed, assuming $\epsilon^\ell \sim \epsilon^N$, the lightness of neutrinos compared to charged leptons immediately requires $m_{a*}^N \gg m_{a*}^\ell$. The neutrino Yukawa matrix takes the same form as in the charged lepton sector, \eq{yedyn}, with $m_i^\nu$ fixed entirely by $m_{i*}^N$. Reproducing the PMNS matrix then requires $m_1^\nu \lesssim m_2^\nu \sim m_3^\nu$, which is achieved via $m_{1*}^N \gtrsim m_{2*}^N \sim m_{3*}^N$. In class 0D instead the presence of flavorful scales does not have an impact on Dirac neutrino masses.

We end this section stressing that sizable violations of $U(1)_{L}$ can immediately bring us back to the Majorana models of section \ref{NeutrinoSection}. 
As a typical example one can consider adding a mass $m_N\equiv\LambdaL\leq\Lambda_{\rm UV}$ for $N$, corresponding to a specific source of $U(1)_{L}$ breaking. The physics below $m_N$ is now captured by the analysis of section
\ref{sec:Mmodels}. 
In particular, at scales $m_*<m_N$ we can integrate out $N$, thus obtaining an EFT involving the fields $\ell,e,q,u,d$ and appropriate $U(1)_{L}$-violating spurions.

%%%%%%%%%%%%%%%%%%%%%%%%%%%%%%%%%%%%%%
\section{Derivation of the constraints on the Wilson coefficients} \label{sec:bounds}
%%%%%%%%%%%%%%%%%%%%%%%%%%%%%%%%%%%%%%

In this appendix we go through the list of operators defined in table \ref{ops} and identify the most constraining observable for each of them. Two are the main assumptions:

(1) RG effects are ignored. This hypothesis introduces a small uncertainty of order a few percent and allows us to write $C(m_*)=C(\mu)$. 

(2) We constrain a single coefficient $C(\mu)$ at a time. This simplifying assumption neglects possible destructive and constructive interference among different contributions to the same rate, which can result in a change of order unity on the actual bounds.

Our choice of neglecting the RG is motivated by the fact that the present analysis inevitably suffers from an ${\cal O}(1)$ uncertainty in the estimate of the coefficients $c$ in table~\ref{ops}. RG effects are parametrically of order $\frac{\alpha}{4\pi}\ln(m_*/\mu)$ and hence much smaller. Importantly, within our picture the RG evolution does not generate new coefficients via mixing either. Indeed, our EFT is assumed to contain non-vanishing Wilson coefficients for {\emph{all}} operators already at $m_*$. Since these have natural size at the matching scale, as dictated by NDA, operator mixing will simply correct them by the same small factor $\frac{\alpha}{4\pi}\ln(m_*/\mu)$. In fact, the simplifying hypothesis that a single coefficient $C(\mu)$ dominates the observable affects our analysis more significantly, as this may result in an over- or under-estimation of our constraints by ${\cal O}(1)$ (still in line with our uncertainty of order unity from the NDA estimates).

Consistently with our assumptions (1) and (2), we will often include radiative effects when they correspond to the leading contribution of a single coefficient. Our Lagrangian is formally defined in the ${\overline{\rm MS}}$ scheme, so we will adopt this scheme also for radiative effects. 

\subsection{$Q_{e\gamma}$}
%%%%%%%%%%%%%%%%%%%%%%%%%%%%%%%%%%%%%%

The strongest bounds on this operator are from $\mu\to e\gamma$, the electron EDM $d_e$ and anomalous magnetic moment $a_e$, and then $\mu\to e$ in nuclei. In terms of the Wilson coefficient $C^{e\gamma}_{ij}$, we have
\ba
{\rm Br}\left(\mu\to e\gamma\right) 
&=&48\pi^2\frac{v^6}{\Lambda^4m_\mu^2} \left(|C^{e\gamma}_{12}|^2 + |C^{e\gamma}_{21}|^2\right),\\\no
d_e&=&\frac{2v}{\sqrt{2}\Lambda^2}{\rm Im}[C^{e\gamma}_{ee}]\\\no
\Delta a_e&=&\frac{4vm_e}{\sqrt{2}e\Lambda^2}{\rm Re}[C^{e\gamma}_{ee}].
\ea
Completely analogous expressions hold for the heavier leptons.
Imposing the experimental constraints in table \ref{boundsPDG} with $\Lambda=1$ TeV we find the results of table \ref{Ferruccio5}.

The dipole $Q_{e\gamma}$ contributes to ${\rm Br}\left(\mu\to eee\right)$ at tree level as well. However, this latter process has a rate parametrically suppressed by a factor of order $e^2/16\pi^2$ and the resulting bound is therefore less constraining. 

\subsection{$Q_{eH}$ }\label{QeH}
%%%%%%%%%%%%%%%%%%%%%%%%%%%%%%%%%%%%%%

The operator $Q_{eH}$ contributes at tree-level to three body decays $l_i \to l_j l_k \bar l_l$, via a Higgs exchange. 

More constrained are however loop contributions to $\mu\to e\gamma$, {here reported in the ${\overline{\rm MS}}$ scheme}. The latter are collected for example in \cite{Harnik:2012pb}. For the radiative decay of $l_i$ one finds (we do not make distinctions between chiralities, since they contribute equally)
\ba\label{1loop}
\left.C^{e\gamma}_{\rm IR}\right|_{\rm 1-loop}= \frac{\sqrt{2} e m^e_i}{8\pi^2v} \frac{1}{12m_h^2}  \frac{y^e_i}{\sqrt{2}} \frac{v^2C_{eH}}{\sqrt{2}}  \left(3\ln\frac{m_h^2}{m_i^2}-4\right)~,
\ea
and
\ba\label{2loop}
\left.C^{e\gamma}_{\rm IR}\right|_{\rm 2-loop}= \frac{\sqrt{2}e m^e_i}{8\pi^2 v}\frac{1}{m_h^2}\frac{v^2C_{eH}}{\sqrt{2}}\times0.055 \left(\frac{1.78~{\rm GeV}}{m_i^e}\right)~.
\ea
Note that within our assumptions the Wilson coefficients are interpreted as renormalized at $\sim m^e_i$. The 1-loop effect depends on the Yukawa coupling and the mass (needed for the necessary chirality-flip) of the decaying lepton. There is no need for a chirality flip in the 2-loop diagrams, which are therefore lepton independent. It turns out that for $\mu$ decays the one-loop is negligible compared to the 2-loop contribution, whereas for the $\tau$ the two contributions are of the same order. Overall we get
\ba\label{12loop}
C^{e\gamma}_{\rm IR}=C^{eH}\times\left\{
\begin{matrix}
6.0\times10^{-6}~~~~~~(\mu)\\
7.4\times10^{-6}~~~~~~(\tau)
\end{matrix}\right. ~.
\ea
With the help of \eq{12loop} we derive the bounds of table~\ref{Ferruccio5}.

\subsection{$Q_{\ell\ell,ee, \ell e}$}
%%%%%%%%%%%%%%%%%%%%%%%%%%%%%%%%%%%%%%

We begin observing that  
$$
Q_{ee}^{ijmn}=Q_{ee}^{mnij}=Q_{ee}^{inmj}=Q_{ee}^{mjin}~,\qquad Q_{\ell\ell}^{ijmn}=Q_{\ell\ell}^{mnij}~.
$$ 
In our convention, such identical operators are included only once in the effective Lagrangian. For example, for $Q_{ee}^{ijmn}$ we restrict the indexes to $i\ge m$ and $j\ge n$, with Wilson coefficient normalised as $C^{ee}_{ijmn}/\Lambda^2$.

All 4-lepton operators contribute at tree-level to three-body decays. One finds
\ba
{\rm Br}(l_i^-\to l_j^- l_j^-l_j^+)\!\!\!&=&\!\!\!\frac{m_{i}^5}{1536\pi^3\Lambda^4\Gamma_{i}}
\left({|C^{\ell e}_{jijj}|^2+|C^{\ell e}_{jjji}|^2+2|C^{\ell\ell}_{jijj}|^2+2|C^{ee}_{jijj}|^2} \right),
\no\\
{\rm Br}(l_i^-\to l_j^- l_k^+l_k^-)\!\!\!&=&\!\!\!\frac{m_i^5}{1536\pi^3\Lambda^4\Gamma_i}
\left({|C^{\ell e}_{jikk}|^2+|C^{\ell e}_{jkki}|^2+|C^{\ell e}_{kijk}|^2+|C^{\ell e}_{kkji}|^2
+|C^{\ell\ell}_{jikk}+ C^{\ell\ell}_{jkki}|^2+|C^{ee}_{jikk}|^2}\right),
\no\\
{\rm Br}(l_i^-\to l_j^+l_k^-l_k^-)\!\!\!&=&\!\!\!\frac{m_i^5}{1536\pi^3\Lambda^4\Gamma_i}
\left({|C^{\ell e}_{kikj}|^2+|C^{\ell e}_{kjki}|^2+2|C^{\ell\ell}_{kikj}|^2+2|C^{ee}_{kikj}|^2} \right),
\label{3body}
\ea
where the factors of $2$ arise combining a combinatoric $4$ in the Feynman rule and a $1/2$ in the phase space. These results agree with~\cite{Crivellin:2013hpa} except for the second line of (\ref{3body}), 
as some independent Wilson coefficients were identified with each other in \cite{Crivellin:2013hpa}.

The same operators also contribute to radiative decays at loop level. At one-loop the authors of~\cite{Crivellin:2013hpa} find (using ${\overline{\rm MS}}$)
\ba
(C^{e\gamma}_{\rm IR})_{ij,ji}=\sum_{k=e,\mu,\tau}\frac{e y^e_k}{16\pi^2}C^{\ell e}_{ikkj, jkki}~.
\ea
In table~\ref{Ferruccio6} we show these bounds assuming that a single Wilson coefficient is present at a time (a unique $k$).
The operators $Q_{\ell\ell,ee}$ first renormalize $C_{e\gamma}$ at two-loop order. This requires a chirality flip on one external lepton line, and an electroweak
gauge boson exchanged between the lepton loop and one external lepton line. We do not make an explicit calculation, but rather observe that the effect can be parametrized as
\ba
(C^{e\gamma}_{\rm IR})_{ij}=c\sum_{k=e,\mu,\tau}\frac{e y^e_{max(i,j)}}{16\pi^2}\frac{g^2}{16\pi^2}
\left\{C^{\ell\ell}_{ikkj,ijkk},C^{ee}_{ikkj}\right\} ~,
\ea
with $c$ a number of order one.
Using this rough estimate, we find that $\tau$ radiative decays lead to weak constraints, $|C^{\ell\ell,ee}|<{\cal O}(1)$. 
On the other hand, $\mu\to e\gamma$ sets an order of magnitude bound as shown in the table.

\subsection{$Q_{He,H\ell}$}
%%%%%%%%%%%%%%%%%%%%%%%%%%%%%%%%%%%%%%

The main effects of these operators are mediated by tree-level $Z^0$ exchange. In fact, at scales well below the $Z^0$ mass, $Q_{He,H\ell}$ contribute to several 4-fermion operators,
\ba
{\cal L}_{eff}^Z &=& + \dfrac{2}{\Lambda^2}\left[C^{He}_{ij} \overline{e_{Ri}}\gamma^\mu e_{Rj} + \left(C^{H\ell(1)}_{ij}+C^{H\ell(3)}_{ij}\right) \overline{e_{Li}}\gamma^\mu e_{Lj}
+ \left(C^{H\ell(1)}_{ij}-C^{H\ell(3)}_{ij}\right) \overline{\nu_{Li}}\gamma^\mu \nu_{Lj}\right] \no\\
&\times & {\displaystyle \sum\limits_{\psi = e_R,e_L,\nu_L,u_L,u_R,d_L,d_R}}\left[\, \overline{\psi_k}\left(T_{3\psi} - s_w^2 Q_\psi\right)\psi_k \,\right]~,
%\end{array}
\label{match1}\ea
where $T_3$ is the weak isospin, $Q$ the electric charge, and $s_w$ the sine of the weak mixing angle.
The four-lepton operators in (\ref{match1}) can induce the LFV decays  $l_i\to l_j l_k\bar l_l$. The branching ratios are obtained plugging the corresponding coefficients in (\ref{3body}). The resulting bounds are listed in table \ref{Ferruccio5}.
The two-quark, two-lepton operators in (\ref{match1}) can contribute to $\mu\to e$ conversion on nuclei, dominantly via the up and down quark vector currents. 
Using the formalism of \cite{Kitano:2002mt} to compute the rate for $\mu^-Au\to e^-Au$, we find a bound on $C^{He,H\ell}_{12}$ that is stronger than the one from $\mu\to 3e$, as shown in table \ref{Ferruccio5}.
 
The contribution of $Q_{He,H\ell}$ to radiative decays is of order
\ba
\frac{{\rm Br}(\mu\to e\gamma)}{{\rm Br}(\mu\to eee)}\sim\frac{e^2}{16\pi^2}\ll1.
\ea
As the experimental bounds on the branching ratios are just a factor of a few apart, it is clear that radiative decays are less constraining. Similarly, exotic $Z$ decays can also constrain these operators, but current limits are an order of magnitude weaker than those presented in table~\ref{Ferruccio5}.

%%%%%%%%%%%%%%%%%%%%%%%%%%%%%%%%%%%%%%
\section{{Extracting constraints on PC} \label{compApp}}
%%%%%%%%%%%%%%%%%%%%%%%%%%%%%%%%%%%%%

There are some technical subtleties in the derivation of the constraints on PC using~(\ref{NDA}) to directly extract the coefficients of the Warsaw basis operators defined in table~\ref{ops}.
The reason is that from eq.~(\ref{NDA}) we can also determine coefficients of operators that are {\emph{not}} independent from those of table \ref{ops}. Once these are removed via field redefinitions or the equations of motion, the coefficients of the operators in the table may receive contributions that are parametrically different from those estimated using~(\ref{NDA}) directly on the Warsaw basis.

We find there is only one instance in which this subtlety may be relevant. To appreciate this, note there is another class of flavor-violating operators we can write at dimension-6:
\ba\label{Fmunu}
g'\bar\ell^i_{L} \gamma^\mu \ell^j_{L} \partial_\nu B^\nu_\mu&=&g'^2\bar\ell^i_L \gamma^\mu \ell^j_L J^B_\mu\\\no
g'\bar e^i_{R} \gamma^\mu e^j_{R} \partial_\nu B^\nu_\mu&=&g'^2\bar e^i_R \gamma^\mu e^j_R J^B_\mu\\\no
g\bar\ell^i_L \gamma^\mu\tau^a \ell^j_L (D_\nu W^\nu_\mu)^a&=&g^2\bar\ell^i_L \gamma^\mu\tau^a \ell^j_L J^a_\mu,
\ea
where $J^{B,a}_\mu$ are the SM currents (including fermions and the Higgs doublet) of the hyper-charge and the $SU(2)$ gauge bosons respectively. The above equalities hold up to corrections due to operators of higher dimension. Now, the operators on the left-hand side of eq. (\ref{Fmunu}) have, according to~(\ref{NDA}), coefficients of order $\epsilon_i\epsilon_j/m_*^2$. However, because they are linear combinations of {$Q_{\ell \ell, ee, e\ell, H\ell, He}$}, they appear in our formalism as corrections of order $\delta C_{\ell\ell, ee, e\ell,H\ell,He}\sim{\epsilon^2g^2}/{m_*^2}$ to the estimates in our table. But these are parametrically different from the ones obtained using~(\ref{NDA}) to directly estimate the coefficient of {$Q_{\ell \ell, ee, e\ell, H\ell, He}$}!

\begin{table}[p]
\begin{center}
\resizebox{.7\hsize}{!}{$
\begin{array}{c|c|c|l}
\textrm{Operator} & \Lambda \,  (\textrm{TeV}) (C=1) & \textrm{PC bound } {(m_*=10\textrm{ TeV})} & \textrm{Observable} \\
\hline
(Q_{eW})_{12}, (Q_{eB})_{12} & 6.9 \times 10^4 & |c| \times \left(\frac{g_*}{4 \pi}\right)^2 \frac{\epsilon^{\ell}_1}{\epsilon^{\ell}_2} < 1.1 \times 10^{-4} & \mu \to e \gamma \\
(Q_{eW})_{21}, (Q_{eB})_{21} & 6.9 \times 10^4 & |c| \times \left(\frac{g_*}{4 \pi}\right)^2 \frac{\epsilon^{\ell}_2}{\epsilon^{\ell}_1} < 2.4 \times 10^{-2} & \mu \to e \gamma \\
(Q_{eH})_{12} & 1.7 \times 10^2 &  |c| \times \left(\frac{g_*}{4 \pi}\right)^2 \frac{\epsilon^{\ell}_1}{\epsilon^{\ell}_2} <  3.6 \times 10^{-2}  & \mu \to e \gamma \textrm{ [2-loop]} \\
(Q_{eH})_{21} & 1.7 \times 10^2 & |c| \times \left(\frac{g_*}{4 \pi}\right)^2 \frac{\epsilon^{\ell}_2}{\epsilon^{\ell}_1} <  7.5 & \mu \to e \gamma \textrm{ [2-loop]} \\
(Q^{(1)}_{H \ell})_{12}, (Q^{(3)}_{H \ell})_{12} & 4.5 \times 10^2 
& |c| \times \left(\frac{g_*}{4 \pi}\right)^2 \epsilon^{\ell}_1 \epsilon^{\ell}_2 < 3.1 \times 10^{-6}  
& \mu\, Au \to e\, Au\\ 
(Q_{H e})_{12}& 4.5 \times 10^2  
& |c| \times \frac{1}{ \epsilon^{\ell}_1 \epsilon^{\ell}_2} < 2.9 \times 10^5  
& \mu\, Au \to e\, Au\\
(Q_{\ell \ell})_{2111}& 2.1 \times 10^2 & |c|  \times \left(\frac{g_*}{4 \pi}\right)^2 (\epsilon^{\ell}_1)^3  \epsilon^{\ell}_2 < 1.5 \times 10^{-5} & \mu \to 3e  \\
(Q_{ee})_{2111}& 2.1 \times 10^2 & |c|  \times \left(\frac{g_*}{4 \pi}\right)^{-2} \frac{1}{(\epsilon^{\ell}_1)^3  \epsilon^{\ell}_2} < 2.4 \times 10^{19} & \mu \to 3e  \\
(Q_{\ell e})_{2111}& 1.7  \times 10^2 & |c| \times \frac{\epsilon^{\ell}_2}{\epsilon^{\ell}_1} < 3.8 \times 10^8 & \mu \to 3e   \\
(Q_{\ell e})_{1121}& 1.7  \times 10^2  & |c| \times \frac{\epsilon^{\ell}_1}{\epsilon^{\ell}_2} < 1.9 \times 10^6 & \mu \to 3e   \\
(Q_{\ell e})_{2221}& 77 & |c| \times \frac{\epsilon^{\ell}_2}{\epsilon^{\ell}_1} < 9.5 \times 10^6 &  \mu \to e \gamma \textrm{ [1-loop]}  \\
(Q_{\ell e})_{1222}& 77 & |c| \times \frac{\epsilon^{\ell}_1}{\epsilon^{\ell}_2} < 4.6 \times 10^4 & \mu \to e \gamma \textrm{ [1-loop]}  \\
(Q_{\ell e})_{2331}& 2.9 \times 10^2 & |c| \times \frac{\epsilon^{\ell}_2}{\epsilon^{\ell}_1} < 4.0 \times 10^4 &  \mu \to e \gamma \textrm{ [1-loop]}   \\
(Q_{\ell e})_{1332}& 2.9 \times 10^2 & |c| \times \frac{\epsilon^{\ell}_1}{\epsilon^{\ell}_2} < 1.9 \times 10^2 &  \mu \to e \gamma \textrm{ [1-loop]}   \\
\hline
(Q_{eW})_{13}, (Q_{eB})_{13} & 6.5 \times 10^2 & |c| \times \left(\frac{g_*}{4 \pi}\right)^2 \frac{\epsilon^{\ell}_1}{\epsilon^{\ell}_3} < 7.8 \times 10^{-2} & \tau \to e \gamma \\
(Q_{eW})_{31}, (Q_{eB})_{31} & 6.5 \times 10^2 & |c| \times \left(\frac{g_*}{4 \pi}\right)^2 \frac{\epsilon^{\ell}_3}{\epsilon^{\ell}_1} < 2.7 \times 10^{2} & \tau \to e \gamma \\
(Q_{eH})_{13} & 1.7 &  |c| \times \left(\frac{g_*}{4 \pi}\right)^2 \frac{\epsilon^{\ell}_1}{\epsilon^{\ell}_3} < 9.3 \times 10^{-1} & \tau \to e \gamma \textrm{ [2-loop]} \\
(Q_{eH})_{31} & 1.7 & |c| \times \left(\frac{g_*}{4 \pi}\right)^2 \frac{\epsilon^{\ell}_3}{\epsilon^{\ell}_1} < 3.2 \times 10^{3} & \tau \to e \gamma \textrm{ [2-loop]} \\
(Q^{(1)}_{H \ell})_{13}, (Q^{(3)}_{H \ell})_{13} & 8.4 & |c| \times \left(\frac{g_*}{4 \pi}\right)^2 \epsilon^{\ell}_1 \epsilon^{\ell}_3 < 8.9 \times 10^{-3} & \tau \to 3 e\\ 
(Q_{H e})_{13}& 8.2
& |c|  \frac{1}{ \epsilon^{\ell}_1 \epsilon^{\ell}_3} < 5.3 \times 10^7 & \tau \to 3 e\\ 
(Q_{\ell \ell})_{1311}& 10 & |c|  \times \left(\frac{g_*}{4 \pi}\right)^2 (\epsilon^{\ell}_1)^3  \epsilon^{\ell}_3 < 5.8 \times 10^{-3} & \tau \to 3e  \\
(Q_{ee})_{1311}& 10  & |c|  \times \left(\frac{g_*}{4 \pi}\right)^{-2} \frac{1}{(\epsilon^{\ell}_1)^3  \epsilon^{\ell}_3} < 5.6 \times 10^{20} & \tau \to 3e  \\
(Q_{\ell e})_{1311}& 8.8 & |c| \times \frac{\epsilon^{\ell}_3}{\epsilon^{\ell}_1} < 1.5 \times 10^{11} & \tau \to 3e \\
(Q_{\ell e})_{1113}& 8.8 & |c| \times \frac{\epsilon^{\ell}_1}{\epsilon^{\ell}_3} < 4.3 \times 10^{7} & \tau \to 3e\\
\hline
(Q_{eW})_{23}, (Q_{eB})_{23} & 6.1 \times 10^2 & |c| \times \left(\frac{g_*}{4 \pi}\right)^2 \frac{\epsilon^{\ell}_2}{\epsilon^{\ell}_3} < 8.7 \times 10^{-2} & \tau \to \mu \gamma \\
(Q_{eW})_{32}, (Q_{eB})_{32} & 6.1 \times 10^2 & |c| \times \left(\frac{g_*}{4 \pi}\right)^2 \frac{\epsilon^{\ell}_3}{\epsilon^{\ell}_2} < 1.5 & \tau \to \mu \gamma \\
(Q_{eH})_{23} & 1.6 & |c| \times \left(\frac{g_*}{4 \pi}\right)^2 \frac{\epsilon^{\ell}_2}{\epsilon^{\ell}_3} < 9.3 \times 10^{-1} & \tau \to \mu \gamma \textrm{ [2-loop]} \\
(Q_{eH})_{32} & 1.6 & |c| \times \left(\frac{g_*}{4 \pi}\right)^2 \frac{\epsilon^{\ell}_3}{\epsilon^{\ell}_2} < 10 & \tau \to \mu \gamma \textrm{ [2-loop]} \\
(Q^{(1)}_{H \ell})_{23}, (Q^{(3)}_{H \ell})_{23} & 8.8 
& |c| \times \left(\frac{g_*}{4 \pi}\right)^2 \epsilon^{\ell}_2 \epsilon^{\ell}_3 < 8.2 \times 10^{-3} & \tau \to 3 \mu\\ 
(Q_{H e})_{23}& 8.8 
& |c| \times \frac{1}{ \epsilon^{\ell}_2 \epsilon^{\ell}_3} < 2.2 \times 10^{5} & \tau \to 3 \mu\\ 
(Q_{\ell \ell})_{2322}& 11 & |c|  \times \left(\frac{g_*}{4 \pi}\right)^2 (\epsilon^{\ell}_2)^3  \epsilon^{\ell}_3 < 4.9 \times 10^{-3} & \tau \to 3\mu  \\
(Q_{ee})_{2322}& 11 & |c|  \times \left(\frac{g_*}{4 \pi}\right)^{-2} \frac{1}{(\epsilon^{\ell}_2)^3  \epsilon^{\ell}_3} < 5.4 \times 10^{13} & \tau \to 3\mu  \\
(Q_{\ell e})_{2322}& 9.5 & |c| \times \frac{\epsilon^{\ell}_3}{\epsilon^{\ell}_2} < 3.0 \times 10^6 & \tau \to 3\mu  \\
(Q_{\ell e})_{2322}& 9.5 & |c| \times \frac{\epsilon^{\ell}_2}{\epsilon^{\ell}_3} < 1.8 \times 10^5 & \tau \to 3\mu  \\
\hline
(Q_{eW})_{11}, (Q_{eB})_{11} & 1.4 \times 10^6 & \textrm{Im}(c) \times \left(\frac{g_*}{4 \pi}\right)^2 < 5.7 \times 10^{-5} & d_e \\
(Q_{\ell e})_{1221}& 1.6 \times 10^3 &  \textrm{Im}(c) \times \left(\frac{g_*}{4 \pi}\right)^2 < 2.4 \times 10^{4} & d_e \textrm{ [1-loop]} \\
(Q_{\ell e})_{1331}& 6.5 \times 10^{3} &  \textrm{Im}(c) \times \left(\frac{g_*}{4 \pi}\right)^2 < 8.3 \times 10 & d_e \textrm{ [1-loop]} \\
(Q_{eW})_{22}, (Q_{eB})_{22} & 11  & \textrm{Im}(c) \times \left(\frac{g_*}{4 \pi}\right)^2 < 4.6 \times 10^3 & d_{\mu} \\
(Q_{eW})_{33}, (Q_{eB})_{33} & 1.5  & \textrm{Im}(c) \times \left(\frac{g_*}{4 \pi}\right)^2  < 1.4 \times 10^4   & d_{\tau} \\
\end{array}
$}
\end{center}
\caption{\small{Bounds on Wilson coefficients and corresponding constraints on Anarchic PC.}\label{boundsall}}
\end{table}

In general this observation can have important implications, but fortunately not in anarchic PC. In fact, the contribution $\delta C_{H\ell,He}$ in (\ref{Fmunu}) is smaller than the one considered in table \ref{ops} by a factor $g^2/g_*^2\ll1$ or $g'^2/g_*^2\ll1$ and can thus be neglected. What about the new contribution to $\delta C_{\ell\ell, ee, e\ell}$ in (\ref{Fmunu})? Here the point is that in scenarios of PC the bound on the unknown factors of order unity coming from $Q_{\ell\ell, ee, e\ell}$ is typically weaker compared to that from $Q_{H\ell,He}$, so the overall $C_{\ell\ell, ee, e\ell}$ (including the new contribution $\delta C_{\ell\ell, ee, e\ell}$ from (\ref{Fmunu})) does not affect our analysis in practice. To see this, recall from (\ref{match1}) that at low scales $Q_{H\ell,He}$ become combinations of the 4-fermion operators $Q_{\ell\ell, ee, e\ell}$ with coefficients $\sim C_{eH,\ell H}\propto g_*^2\epsilon^2$, and that these are parametrically larger than $C_{\ell\ell, ee, e\ell}(m_*)\propto g_*^2\epsilon^4$ (see table~\ref{ops}) as well as $\delta C_{\ell\ell, ee, e\ell}\propto g^2\epsilon^2$ from (\ref{Fmunu}). Hence, when comparing experiments with the predictions of PC, the dominant constraints on 4-fermions operators actually translate into a bound on $Q_{H\ell,He}$, not on $Q_{\ell\ell, ee, e\ell}$. Note that these arguments heavily rely on our power-counting, that is, they hold as long as the coefficients $c$ in table \ref{ops} are of order one for all the operators. Without a concrete assumption on the UV, it would not be possible to neglect the operators on the left-hand side of (\ref{Fmunu}) nor, in general, compare different Wilson coefficients.
{We also checked that in the scenarios of section~\ref{sec:supp} and \ref{sec:pcanom} the operators (\ref{Fmunu}) do not lead to stronger constraints than those already discussed in the main text.}

A final comment regarding $Q_{eH}$ is in order. After having integrated out the Higgs at tree-level, this operator generates $Q_{\ell e}$ with a coefficient $\delta C_{\ell e}$ that is typically larger than the one induced by the heavy physics directly at the scale $m_*$. Indeed, $\Delta F=1$ processes are affected by a single insertion of $Q_{eH}$, whereas $\Delta F=2$ require two:
\ba
\frac{\delta C_{\ell e}}{C_{\ell e}}\sim \left\{
\begin{matrix}
\frac{g_*y_{\rm SM}v^2}{\epsilon_L\epsilon_Rm_h^2}\sim\frac{g_*^2v^2}{m_h^2}~~~~~~~~(\Delta F=1)\\
\frac{(g_*^2v^2)^2}{m_h^2m_*^2}~~~~~~~~~~~~~~~~~~~(\Delta F=2)
\end{matrix}
\right.
\ea
Since $g_*>1$ this means that $Q_{eH}$ effectively generates a larger contribution to the 4-fermion operator $Q_{\ell e}$. In the quark sector, the analogous $\Delta F=2$ effects are very dangerous, and are usually suppressed under reasonable assumptions about the UV physics \cite{Agashe:2009di}. Here we do not necessarily need this assumption. Indeed, in the context of lepton observables, 4-fermion transitions are not as constraining. The main constraint on $Q_{eH}$ comes from loop-induced contributions to radiative decays, but these are negligible compared to those directly arising from $Q_{e\gamma}$. 

We also provide a collection of the bounds on PC parameters in  table \ref{boundsall}.

%%%%%%%%%%%%%%%%%%%%%%%%%%%%%%%%%%%%%%
%%%%%%%%%%%%%%%%%%%%%%%%%%%%%%%%%%%%%%

\end{document}